%% file: mbs.tex
\documentstyle[12pt,jeep]{article}
\numberbysection
\begin{document}
\title{ Selection of the Ground State for Nonlinear Schr\"odinger Equations}
\author{
A. Soffer \thanks{Department of Mathematics, Rutgers University, New
Brunswick, NJ} \hspace{.05 in}
 and 
 M.I. Weinstein \thanks{Mathematical Sciences Research, Bell Laboratories - Lucent
 Technologies, Murray Hill, NJ 07974 
 }} 
\baselineskip=18pt
\maketitle
\newcommand{\ds}{\displaystyle}
\newcommand{\xtext}[1]{\mbox{#1}}
\newcommand{\ra}{\rightarrow}
\newcommand{\da}{\downarrow}
\newcommand{\wra}{\rightharpoonup}
\newcommand{\rf}{\widehat}
\newcommand{\nit}{\noindent}
\newcommand{\no}{\nonumber}
\newcommand{\be}{\begin{equation}}
\newcommand{\ee}{\end{equation}}
\newcommand{\ba}{\begin{eqnarray}}
\newcommand{\ea}{\end{eqnarray}}
\newcommand{\bom}{\mbox{$\Omega$}}
\newcommand{\Fee}{\mbox{$\Phi$}}
\newcommand{\del}{\mbox{$\nabla$}}
\newcommand{\delx}{\mbox{$\nabla_{x}$}}
\newcommand{\dely}{\mbox{$\nabla_{y}$}}
\newcommand{\ar}{\mbox{$\alpha$}}
\newcommand{\fee}{\mbox{$\varphi$}}
\newcommand{\dta}{\mbox{$\delta$}}
\newcommand{\lam}{\mbox{$\lambda$}}
\newcommand{\Lam}{\mbox{$\Lambda$}}
\newcommand{\eps}{\mbox{$\epsilon$}}
\newcommand{\gam}{\mbox{$\gamma$}}
\newcommand{\Gam}{\mbox{$\Gamma$}}
\newcommand{\al}{\mbox{$\alpha$}}
\newcommand{\bt}{\mbox{$\beta$}}
\newcommand{\va}{\mbox{$\bf a $}}
\newcommand{\bF}{\mbox{$\bf F $}}
\newcommand{\db}{\mbox{$\parallel$}}
\newcommand{\ov}{\overline}
\newcommand{\ud}{\underline}
\newcommand{\nW}{\left| W\right|_{\bf a}}
\newcommand{\D}{\partial}
\newcommand{\bu}{{\bf u}}
\newcommand{\bXz}{{\bf X}_0}
\newcommand{\bX}{{\bf X}}
\newcommand{\bY}{{\bf Y}}
\newcommand{\bN}{{\bf N}}
\newcommand{\bR}{{\bf R}}
\newcommand{\valpha}{\vec\alpha}
\newcommand{\vPsi}{\vec\Psi}
\newcommand{\talpha}{\tilde\alpha}
\newcommand{\tOmega}{\tilde\Omega}
\newcommand{\tetab}{\eta_b}
\newcommand{\tbeta}{\tilde\beta}
\newcommand{\tomega}{\tilde\omega}
\newcommand{\tA}{\tilde A}
\newcommand{\tE}{\tilde E}
\newcommand{\tLambda}{\tilde \Lambda}
\newcommand{\tGamma}{\tilde \Gamma}
\newcommand{\tchi}{\tilde\chi}
\newcommand{\tJ}{\tilde J}
\newcommand{\tn}{\tilde n}
\newcommand{\ts}{\tilde s}
\newcommand{\tp}{\tilde p}
\newcommand{\tpsi}{\tilde \psi}
\newcommand{\tTheta}{\tilde \Theta}
\newcommand{\eoY}{[\eta_0]_{\bf Y}}
\newcommand{\tG}{\tilde G}
\newcommand{\bT}{{\bf T}}
\newcommand{\bA}{{\bf A}}
\newcommand{\bAt}{{\bf A(t)}}
\newcommand{\bG}{{\bf G }}
\newcommand{\bH}{{\bf H }}
\newcommand{\bfe}{{\bf e }}
\newcommand{\bXYZ}{{\bf XYZ\ }}
\newcommand{\bTb}{\overline{\bf T}}
\newcommand{\cT}{{\cal T}}
\newcommand{\cO}{{\cal O}}
\newcommand{\cH}{{\cal H}}
\newcommand{\cI}{{\cal I}}
\newcommand{\cJ}{{\cal J}}
\newcommand{\cK}{{\cal K}}
\newcommand{\cN}{{\cal N}}
\newcommand{\cA}{{\cal A}}
\newcommand{\cB}{{\cal B}}
\newcommand{\cC}{{\cal C}}
\newcommand{\cD}{{\cal D}}
\newcommand{\cE}{{\cal E}}
\newcommand{\cF}{{\cal F}}
\newcommand{\cG}{{\cal G}}
\newcommand{\cQ}{{\cal Q}}
\newcommand{\cR}{{\cal R}}
\newcommand{\RR}{R^\#}
\newcommand{\SS}{S^\#}
\newcommand{\cS}{{\cal S}}
\newcommand{\cchi}{C_\chi}
\newcommand{\cTb}{\overline{\cal T}}
\newcommand{\Th}{\Theta}
\newcommand{\tH}{{\tilde H}}
\newcommand{\bx}{{\bf x}}
\newtheorem{theo}{Theorem}[section]
\newtheorem{defin}{Definition}[section]
\newtheorem{prop}{Proposition}[section]
\newtheorem{lem}{Lemma}[section]
\newtheorem{cor}{Corollary}[section]
\newtheorem{rmk}{Remark}[section]
\renewcommand{\theequation}{\arabic{section}.\arabic{equation}}
\def\R{{\rm \rlap{I}\,\bf R}}
\def\C{{\bf C} \llap{
 \vrule height 6.6pt width 0.6pt depth 0pt \hskip 4.7pt}}
\def\WB{W\tilde g_\Delta(H)}
\def\tg{\tilde g_\Delta (H)}
\def\tA{\tilde A}
\def\tmu{\tilde \mu}
\def\tf{\tilde f}
\def\3half{{3\over2}}
\def\half{{1\over2}}
\def\L2loc{L^2_{\rm loc}}
\def\khione{\chi_1}
\def\khitwo{\chi_2}
\def\Pc{{\bf P_c}}
\def\cO{{\cal O}}
\def\ta0{\tilde\alpha_0}
\def\tb1{\tilde\beta_1}
\def\t{\tau}
\def\nn{\nonumber}
\def\ol{\overline}
\def\bar{\overline}
\def\veps{\varepsilon}
\newcommand{\nm}[1]{\vert\vert {#1} \vert\vert}
\newcommand{\h}[1]{{#1}}
\def\la{\langle}
\def\ra{\rangle}
\def\jxs{\la x\ra^{\sigma}}
\def\jxms{\la x\ra^{-\sigma}}
\def\Jt{\la t\ra}
\def\jt0{\la t\ra_0}
\def\j#1{\la #1 \ra}
\def\ld{[\phi_d]_{LD}}
\def\Xnm#1{\nm{#1}_X}
\def\e#1{e^{#1}}
\def\i{\int^t_0}
\medskip
\date
\begin{abstract}
\nit 
We prove for a class of nonlinear Schr\"odinger systems (NLS)
having two nonlinear bound states that the (generic) large time
behavior is characterized by decay of the excited state, asymptotic
approach to the nonlinear ground state and dispersive
 radiation. Our analysis elucidates the mechanism through which
initial conditions which are very near the excited state branch evolve
into a (nonlinear) ground state, a phenomenon known as {\it ground state 
 selection}. 
 Key steps in the analysis are the introduction of a particular 
linearization and the derivation of a normal form which reflects the dynamics
on all time scales and yields, in particular, nonlinear Master equations.
 Then, a novel multiple time scale dynamic stability theory 
  is developed.
Consequently, we give a detailed description of the asymptotic behavior of the 
two bound state NLS for all small initial data. The methods are general 
 and 
 can be extended to treat NLS with more than two bound states
 and more general nonlinearities
including those of Hartree-Fock type.

\end{abstract}
\thispagestyle{empty}
\tableofcontents
%
%
%
\section{Introduction and statement of main results} 

In this paper we study the detailed dynamics of the nonlinear 
 Schr\"odinger equation with a potential (NLS)
\footnote{ In a similar way, one can treat nonlinearities $\lambda K[ |\phi|^2]$
 is a nonlinear term which can be local or nonlocal. Typical examples are:
 $K[|\phi|^2] = |\phi|^p$ and  $K[|\phi|^2] = {\cal K}\star|\phi|^2$,
 for some convolution kernel, ${\cal K}$.}:
\be
i\D_t\phi = H\phi + \lambda |\phi|^2\phi.\ 
\label{eq:nls}\ee
Here, $H = -\Delta + V(x)$ is a self-adjoint operator on $L^2(\R^3)$ and 
$\lambda$ is a coupling parameter, assumed real and of order one.
We assume that $V(x)$ is a smooth potential, 
 which decays sufficiently 
 rapidly as $|x|$ tends to infinity (short range). Finally, we assume
that the operator $H$ has no zero energy resonance \cite{JK,Mu}, a condition
which holds for generic $V$.

NLS is a Hamiltonian system with conserved Hamiltonian energy functional:
\be
\bH_{en}[\phi]\ =\ \int |\nabla\phi(x)|^2 + V(x)|\phi(x)|^2 + 
 {\lambda\over2} |\phi(x)|^4 \ dx 
\label{eq:Ham}
\ee
and additional conserved integral
\be
\bN[\phi] \ =\ \int |\phi(x)|^2\ dx
\label{eq:Ndef}\ee
These conserved integrals are continuous in the $H^1(\R^3)$ topology.
An extensive discussion of the well-posedness theory can be found in the 
\cite{Cazenave,Kato,SulemSulem}. In particular, NLS is 
  well-posed  globally in time in the space
 $H^1(\R^3)$, for initial data, $\phi_0$, 
 which is sufficiently small in $H^1$.  
\medskip

\nit Throughout this paper we shall assume that $\phi_0$ has sufficiently
 small $H^1$ norm. We shall use the notation:
\be \cE_0 \ \equiv\ \|\phi_0\|^2_{H^1}\label{eq:cE0def}\ee

For $V(x)$ which decays sufficiently rapidly as $|x|$ tends to infinity (short
 range potentials)  the spectrum of $H$ \cite{RS4} 
 consists of discrete spectrum, $\sigma_d(H)$, consisting of a  
 finite number of negative point eigenvalues, 
  and continuous spectrum, 
$\sigma_c(H)=[0,\infty)$.
 The dynamics of solutions
 for $\lambda=0$ (linear Schr\"odinger)
is very well understood. Let $\psi_{j*}$ and $E_{j*}$ denote bound states and 
bound state energies of the linear Schr\"odinger operator $H$:
\be
H\psi_{j*}\ =\ E_{j*}\psi_{j*},\ \
  \la\psi_{j*},\psi_{k*}\ra_{L^2}\ =\ \delta_{jk}. 
\label{eq:linearbs}\ee
Arbitrary initial conditions in an appropriate Hilbert
space, evolve as $t$ tends to infinity 
 into a time-quasiperiodic part consisting of a superposition of
time periodic and  spatially localized states with frequencies given by the 
eigenvalues and a dispersive or radiative part,
  which decays to zero as $t$ tends to infinity 
in appropriate spaces, {\it e.g.} $L^p,\ p>2$, $L^2(\la x\ra^{-\sigma}\ dx)$. 

 In order to be more precise, introduce $P_{c*}$,
 the orthogonal 
projection onto the continuous spectral subspace of $H$:
\be P_{c*}f\ =\ f\ -\ \sum_j\ \la\psi_{j*},f\ra\ \psi_{j*}. 
\label{eq:Pc*def}
\ee
The solution of the linear Schr\"odinger equation can be expressed as
\be
 e^{-iHt}\phi_0\ =\ 
 \sum_j\ \la\psi_{j*},\phi_0\ra\ \psi_{j*}e^{-iE_{j*}t}\ +\ e^{-iHt}P_{c*}\phi_0.
\label{eq:LSsolution}
\ee
The time decay of the continuous spectral part of the solution can be expressed,
  under suitable
smoothness, decay and genericity assumptions on $V(x)$, in terms of
{\it local decay estimates} \cite{JK,Mu}:
\be
\|\ \la x\ra^{-\sigma}  e^{-iHt}P_{c*}\phi_0\ \|_{L^2(R^3)}
\ \le\ C\la t\ra^{-{3\over2}}\ \|\ \la x\ra^{\sigma} \phi_0\ \|_{L^2(R^3)},
\label{eq:LSlocaldecay}
\ee
$\sigma\ge\sigma_0>0$, and $L^1-L^\infty$ {\it decay estimates} \cite{JSS,Y}
\be
\|\ e^{-iHt}P_{c*}\phi_0\ \|_{L^\infty(R^3)}
\ \le\ C |t|^{-{3\over2}}\ \|\ \phi_0\ \|_{L^1(R^3)}.
\label{eq:LSLpLq} 
\ee

For $\lambda\ne0$ the bound states of the linear problem persist, 
 and bifurcate 
from the linear states at zero amplitude into branches of {\it 
 nonlinear bound states}
 \cite{RW88}. Of interest to us is the detailed dynamics of the nonlinear
problem on short, intermediate
and long time scales and, in particular, the manner
in which the nonlinear bound states participate in the dynamics. In \cite{RW88}
 variational methods were used to establish the existence and  
orbital Lyapunov stability of bound states which are local minimizers 
 of $\bH_{en}$ subject to fixed $\bN$; see also 
 \cite{MIW86,GSS87}. This result says that 
initial data which is close, modulo a phase adjustment, in $H^1$ 
 to the ground state remains $H^1$ close to a phase adjusted 
 ground state 
for all time. The $H^1$ norm is closely related to the conserved
 Hamiltonian energy
of the system and  
   is insensitive to dispersive phenomena. Therefore,
 the detailed dynamics is not addressed by this result. For example, 
could the large time dynamics consist of a nonlinear ground state
plus a small nonlinear excited state part? The main result of this paper
implies that this cannot occur. 

For the nonlinear problem, 
 the simplest question to consider is the case
where $H$ has only one simple eigenvalue and the norm of the solution is small.
 The detailed dynamics was studied in 
 \cite{SW88,SW9092,PW95,Weder}.
  Small norm initial data
are shown to evolve into an asymptotic nonlinear ground state and a radiative
 decaying part.

In this paper we study the multibound state variant of this question. 
 We consider
the specific case where $H$ has two simple eigenvalues, $E_{0*}$ and $E_{1*}$.
The linear Schr\"odinger equation then has two time-periodic 
 solutions $\psi_{0*}e^{-iE_{0*}t}$ and $\psi_{1*}e^{-iE_{1*}t}$, with
 $H\psi_{j*}=E_{j*}\psi_{j*},\ \psi_{j*}\in L^2$. Therefore \cite{RW88},
  NLS has
two branches of nonlinear bound states bifurcating from the zero state 
at the eigenvalues of $H$, $\Psi_{\alpha_0}e^{-iE_0t}$ and  
 $\Psi_{\alpha_1}e^{-iE_1t}$, with $\Psi_{\alpha_j}\in L^2$ satisfying 
\be H\Psi_{\alpha_j}+\lambda|\Psi_{\alpha_j}|^2
 \Psi_{\alpha_j} = E_j\Psi_{\alpha_j}. 
\label{eq:Psieqn}
\ee
 Here, $\alpha_j$ denotes a coordinate along 
the $j^{\rm th}$ nonlinear bound state branch and  
\be E_j = E_{j*} + \cO(|\alpha_j|^2).\nn\ee

In contrast to the linear behavior (\ref{eq:LSsolution})
 our main result
  is the following:
\begin{theo}\label{theo:maintheorem}

Consider NLS with $V(x)$ a short range potential supporting two bound 
states as described above. Furthermore, assume that the linear Schr\"odinger
 operator, $H$, has no zero energy resonance.

\nit(1) 
 Assume the initial data, $\phi(0)$,
 is small in the norm defined by:
\be
[\phi(0)]_X\ \equiv\  \|\la x\ra^\sigma\phi(0)\|_{H^k},
\label{eq:triplenorm}
\ee
where $k>2$ and $\sigma>0$ are sufficiently large.
  Let $\phi(t)$ solve the initial value 
problem for  NLS. 

Assume the 
 (generically satisfied)
  {\it nonlinear Fermi golden rule} resonance
condition\footnote{The operator $f\mapsto\delta(H-\omega_*)f$ 
projects $f$ onto the generalized eigenfunction of $H$ with generalized
eigenvalue $\omega_*$. The expression in (\ref{eq:omegastar}) is finite  
by local decay estimates (\ref{eq:LSlocaldecay}); 
 see {\it e.g.} \cite{SW98}.}
\be 
\Gamma_{\omega_*} \equiv \lambda^2
\pi\left\la\psi_{0*}\psi_{1*}^2,\delta(H-\omega_*)\psi_{0*}\psi_{1*}^2
 \right\ra>0\label{eq:omegastar1}\ee
holds, where 
\be \omega_*=2E_{1*}-E_{0*}>0. \label{eq:omegastardef}\ee

\nit Then, as $t\to\infty$ 
\be
\phi(t)\ \rightarrow\  e^{-i\omega_j(t)}\Psi_{\alpha_j(\infty)}\ + \ 
 e^{i\Delta t}\phi_+, \label{eq:asymp}\ee 
in $L^2$, where {\em either} $j=0$ or $j=1$. The phase $\omega_j$ satisfies
\be
\omega_j(t) = \omega_j^\infty t + \cO(\log t) 
\label{eq:log-corr}\ee

\nit Here, $\Psi_{\alpha_j(\infty)}$ is a nonlinear bound state (section
 \ref{sec:bs}), with frequency $E_j(\infty)$ near $E_{j*}$.
When $j=0$, the solution is asymptotic to
a nonlinear ground state, while in the case $j=1$ the solution
 is asymptotic to a nonlinear excited state.

\nit(2)
 More specifically, 
we have the following expansion of the solution $\phi(t)$:
\ba
\phi(t) &=& e^{-i\tomega_0(t)}\Psi_{\alpha_0(t)} +
 e^{-i\tomega_1(t)}\Psi_{\alpha_1(t)}\nn\\ 
 &+& \pi_1 e^{-i\cH_0(\infty)t}P_c\tilde\phi_+(t) + R_{loc}(t) + R_{nloc}(t),
\nn\ea
where as $t\to\infty$
\be \tomega_j(t) - \omega_j(t) \to 0, \alpha_j(t)-\alpha_j(\infty)\to0
\nn\ee
and such that for each initial state, $\phi(0)$,
\ba
|\alpha_0(\infty)|\cdot |\alpha_1(\infty)|\
 &=&\ 0,\nn\\ 
 \tilde\phi_+(t)\ &\to& \  \phi_+\ \ {\rm in }\ L^2,\nn\\
\| R_{loc}(\cdot,t)\|_2 \ &=&\ \cO(t^{-{1\over2}})\nn\\
\| R_{nloc}(\cdot,t)\|_\infty\ &=&\ \cO(t^{-{1\over2}}).\nn
\ea
Here, $\cH_0(\infty)$ is a small spatially 
 localized perturbation of the operator
$$\sigma_3(-\Delta + V(x))$$
and $P_c=P_c(\cH_0(\infty))$, 
  the projection onto its continuous spectral part. Finally, 
 $\pi_1$ maps the vector
 $(z_1,z_2)$ to $(z_1,0)$.

\end{theo}
\bigskip

\begin{rmk}

\smallskip

\nit (A) 
 Theorem \ref{theo:maintheorem} implies the absence of small norm 
  time - quasiperiodic solutions for this class of nonlinear Schr\"odinger 
equations \cite{sigalcmp}. Intuitively, one can explain why
one expects only a {\it  pure state} in the limit $t\to\infty$ and
how the condition on $\omega_*=2E_{1*}-E_{0*}$ arises. Our intuition is based
on viewing the nonlinearity as linear time-dependent potential;
 see also \cite{sigalcmp,sigjapan}.
  An approximate superposition of a nonlinear ground
 state and excited state $\phi \sim
 \Psi_{\alpha_0}e^{-iE_0t} + \Psi_{\alpha_1}e^{-iE_1t} $ can be viewed as
  defining a self-consistent time-dependent potential:
\ba W(x,t) &=&
 \lambda|\Psi_{\alpha_0}e^{-iE_0t} + \Psi_{\alpha_1}e^{-iE_1t}|^2,\nn\\
 &=& \lambda|\Psi_{\alpha_0} + \Psi_{\alpha_1}e^{-i(E_1-E_0)t}|^2\nn\\
 &\sim& \lambda |\Psi_{\alpha_0}|^2 +
 2\lambda \cos((E_1-E_0)t + \gamma)\Psi_{|\alpha_0|}\Psi_{|\alpha_1|}
\label{eq:selfpotential}
\ea
for $|\alpha_1|<<|\alpha_0|$.
As shown in \cite{SW98jsp}, \cite{KW01}, \cite{KW02}
  data with initial conditions given
by the unperturbed excited state decay exponentially on a time scale of
order $\tau \sim \cO(|\alpha_0\alpha_1|^{-2})$ provided the forcing frequency,
 $E_1-E_0\sim E_{1*}-E_{0*}> -E_{1*}$ or $\omega_*=2E_{1*}-E_{0*}>0$.

\smallskip

\nit (B)
 Theorem \ref{theo:maintheorem}
  implies asymptotic stability and selection of the ground state
for generic small data.  Part 1 of Theorem \ref{theo:maintheorem} 
 implies a form of asymptotic completeness. 

\nit (C) Since we control the decay of solutions in $W^{k,\infty}$, our results
 imply global existence of small solutions in $H^s$ for all $s$ sufficiently 
 large. 

\nit (D)  The asymptotic state where $|\alpha_1(\infty)|\ne0$ (and therefore 
 $|\alpha_0(\infty)|=0$) is non-generic. This can be seen by linearization about 
the excited state. 
 The linearized operator, $\cH_1$, is a localized perturbation
of an operator having embedded eigenvalues in its continuous spectrum, 
 under our 
 hypothesis $\omega_*\equiv2E_{1*}-E_{0*}>0$.
 The connection between embedded eigenvalues in the continuous spectrum
 of an appropriate
linear operator and 
 the non-persistence of localized time periodic states
and between embedded eigenvalues in the continuous spectrum
 of an appropriate
linear operator was explored first in \cite{sigalcmp,sigjapan}.
 It is well known that embedded eigenvalues 
 in the continuous spectrum are unstable to generic
 perturbations; see, for example,
 \cite{CFKS,Grillakis,SW98}. 
 In this case, the embedded eigenvalues are perturbed to  complex
eigenvalues, with corresponding eigenstates whose evolution is 
  exponentially {\em growing} with time, under the condition 
 (\ref{eq:omegastar}). The perturbation to the linear operator with embedded
 eigenvalues is however both nongeneric (in that it comes from 
linearization of a Hamiltonian nonlinear term about a critical point
of the energy) and breaks self-adjointness 
with respect to the standard $L^2$ inner product. A second order perturbation
 theory calculation shows that if $\omega_*\equiv 2E_{1*}-E_{0*}>0$ 
 generically the embedded eigenvalue perturbs
to an exponential instability \cite{SW95notes}. This suggests the existence 
of an unstable manifold of solutions for the {\em nonlinear} equation. 
 The existence of such non-generic solutions 
 of NLS with $E_1(\infty)\ne0$ for the full nonlinear flow has recently 
been demonstrated \cite{TY01b}.

\nit (E) Theorem \ref{theo:maintheorem} is stated for the case of 
 two nonlinear
bound state branches. The technique  of proof, however, can be used to 
consider the more general case. We expect results which are analogous to 
those of our main theorem, but more complicated due to the presence of:
 direct bound state - bound state interactions, bound state - continuum 
interactions {\bf and} bound state - bound state interactions mediated 
by the continuum. Multimode Hamiltonian systems have been considered 
in the context of linear time almost periodic perturbations have been
 studied in \cite{KW02}.  
\end{rmk}

\nit{\bf Relation to other work}
 
\nit We now wish to further put our results in context.
Research on nonlinear scattering in the presence of bound states has
followed two related lines.

\nit{\bf A. Nonlinear dispersive waves in systems with defects, potentials,...}

\nit Our analysis centers around nonlinear
bound states which bifurcate from linear bound states of the operator, $H$,
obtained by linearizing about the zero solution. These bound states exist
at all sufficiently small amplitudes (measured in any $H^s$ norm, $s\ge0$).
The behavior of the ``bifurcation diagram'' for larger amplitudes depends
in a detailed way on the details of the nonlinearity, the spatial dimension
and the norm \cite{RW88}. Such nonlinear bound states are also called 
nonlinear defect modes, nonlinear localized modes or nonlinear pinned modes.
They are localized about or ``pinned'' to the support of the potential, 
 $V$, and  
 arise due to a  local deviation  from translation invariance
or a ``defect'' in the homogeneous background which acts as 
an attractive potential well. 
To get a more refined picture of the dynamics
than in the $H^1$ theory, one must consider the linearized evolution
  about the family of 
nonlinear ground states. This linearized operator has continuous   
and discrete spectral parts inherited from the linear bound state
  spectral structure. 
 In particular, the discrete spectrum contains an eigenvalue at zero
 corresponding to the ground state
and a pair of
eigenvalues (located symmetrically about zero) 
 corresponding to the excited state. Thus, at linear order a solution
infinitesimally close to the ground state formally appears to be 
 quasiperiodic in time - a ground state plus a small excited state oscillation.
However, at higher order in perturbation theory one finds nonlinear resonant
coupling of the neutral oscillatory 
 modes to the continuum and as a result these slowly
damp to zero; generically, 
 for very large time energy splits between the ground state and  dispersive
parts of the solution. This mechanism for relaxation to the ground state was
earlier considered 
 for the nonlinear Klein-Gordon wave equation with a potential,
where the decay of ``breather-like'' solutions was studied \cite{SW99}. 
In this work, small norm solutions relax to the zero solution via resonant
energy transfer out of the bound state to radiation modes and dispersive radiation
of energy to infinity;
  the zero solution plays the role of the ground state.
 Results concerning special classes of initial data are 
considered in work of S. Cuccagna \cite{Cucc2} and T.-P. Tsai and H.-T. Yau
 \cite{TY00,TY01a}. Similar results to those of Theorem \ref{theo:maintheorem} 
 are considered in \cite{TY02}.

\nit {\bf B. Nonlinear dispersive translation invariant equations:} 
 A closely related line of research focuses on the translation invariant
 nonlinear Schr\"odinger equation. Here the equation is (\ref{eq:nls}) with
  $V$ taken to be  identically zero nonlinear coupling  parameter
 $\lambda<0$. In this case, the equation has solitary
wave solutions, obtainable by minimization of ${\bf \cH_{\rm en} }[\phi]$ 
subject to ${\bf \cN}[\phi]=\cN_0$. For NLS in dimension $n$ with cubic 
 nonlinearity replaced by the general power nonlinearity $|\phi|^{p-1}\phi$,
we have that if $p<1+{4\over n}$, the foregoing variational problem has a
unique (up to translation) radially symmetric ground state solution for 
any $\cN_0>0$. In the case when $V(x)\ne0$ is a potential supporting 
bound states, the small $\cN_0$ solutions agree with the  bifurcating bound
states discussed above \cite{RW88}. 
 As pointed out earlier, constrained energy minimizers are
 $H^1$ orbitally Lyapunov stable \cite{CL82,MIW86}.
An interesting feature of  solitary waves in the translation invariant case
 is the
presence spurious neutral oscillations. These are sometimes called 
 {\it internal modes} \cite{Kivshar}. To explain this, consider the 
 linearization about a 
ground state solitary wave ($p<1+{4\over n}$). 
 Due to the underlying symmetric group of the 
equation (translation invariance, phase invariance, Galilean invariance,..)
 this linearization has a (generalized) zero eigenvalue of multiplicity 
related to the dimension of equation's symmetry group. 
In the non-integrable cases ($n=1, \ p\ne3$ and $n\ge2$) the linearization
has additional neutral modes. These neutral modes approach
zero as $p$ approaches $1+{4\over n}$; 
 the dimension of the zero subspace jumps by two 
at $p=1+{4\over n}$, the critical case, corresponding to the larger group of 
symmetries and the existence of a pseudo-conformal invariant \cite{MIW85}.
  V.S. Buslaev
 and G. Perel'man \cite{BP2} considered the problem in one space dimension
 and showed that nonlinear resonance of these ``internal modes'' with the 
continuum is responsible for their damping on long time scales and the
 asymptotic stability of solitons. See also the recent work of V.S. Buslaev and 
 C. Sulem \cite{BS03}. Their analysis was restricted to one space 
dimension only in their use of explicit eigenfunction expansion methods 
to obtain the required local energy decay estimates. 
 S. Cuccagna \cite{Cucc1,Cucc2}
 extended 
their results to more general nonlinearities and general space dimensions. 
In his analysis, the required dispersive estimates are obtained by adapting
K. Yajima's \cite{Y} approach in which  the wave operators, which conjugate
the linearized operator on its continuous spectral part to the constant
coefficient ``free'' dispersive evolution, are  shown to be bounded on $W^{k,p}$
spaces. This method was also used in \cite{SW99}. 

Another feature, common to problems of type {\bf A.} and {\bf B.} is the use
of the method of normal forms. In the context of nonlinear scattering, 
 normal form ideas were used to obtain the local behavior
  in a neighborhood of a soliton in 
 \cite{BP2} and for the decaying breather-like state in \cite{SW99}. 
 In contrast 
 to the normal form for finite dimensional Hamiltonian systems, resonant interaction
with the {\it continuous} spectrum gives rise to a more general normal form
 which captures internal damping, due to  energy transfer  out of certain discrete
modes to the continuum modes; 
see the discussion in the introduction to \cite{SW99}. In the present work,
 we derive a 
 {\it nonlinear master equation},
coupled equations for the renormalized (up to near identity
transformations on the complex discrete mode amplitdues) discrete mode square
amplitudes ("mode powers"),
 which governs generic dynamics on large intermediate and very
 long time scales.  Normal forms of this type, expected to be valid for 
very long times, were derived and studied 
 in the local analysis
about the steady and `` wobbling '' kink-like solutions of discrete 
 nonlinear wave equations in \cite{KeWe00}. 
A key feature of the normal form of the current work is our analysis  
of its behavior on different time scales and the analysis
 of its transitional behavior across time scales, for general initial data.
   

Finally, we point out that there 
 are  many important areas of application which motivate the 
 for study of the class of
 models we treated in this paper. We mention two.
  At the most fundmanental level the nonlinear
Schr\"odinger equation arises as a mean field limit model governing the
interaction of a very large number of weakly interacting bosons \cite{Hepp},
\cite{Spohn}, \cite{LSY99}, \cite{FTY}. At a macroscopic level, it has
been shown that equations of this type arise as the equation governing
 the evolution of the envelope of the
 electric field of a light pulse propagating in a medium
with defects. See, for example,
 \cite{GWH01,GHW01,GSW01}.

\bigskip
\noindent{\bf Acknowledgments:} This work was supported in part by grants
 from the National Science Foundation.

\section{Structure of the proof}

We now sketch our analysis. Certain notation is defined in the appendix of 
section \ref{sec:notation}.  In analogy with the approach
 introduced in \cite{SW88},\cite{SW9092} in the one bound state case, we 
represent the solution in terms of the dynamics of the bound state part,
 described through the evolution of the {\it collective coordinates} 
 $\alpha_0(t)$ and $\alpha_1(t)$, and a remainder $\phi_2$,
  whose dynamics is controlled by a  dispersive equation. 
In particular we have
\be
\phi(t,x) = 
e^{-i\int_0^t E_0(s) ds - i\tTheta(t)}\left(\  \Psi_{\alpha_0(t)}
 + \Psi_{\alpha_1(t)} + \phi_2(t,x)\ \right)
\label{eq:fansatz}
\ee
 
We substitute (\ref{eq:fansatz}) into NLS and use the nonlinear equations 
 (\ref{eq:Psieqn}) for
 $\Psi_{\alpha_j}$ 
 to simplify. Anticipating the decay of the excited state, we center
the dynamics about the ground state. 
 We therefore obtain for  $\Phi_2 \equiv (\phi_2,\ol{\phi_2})^T$
 the equation:
\be
i\D_t\Phi_2 = {\cal H}_0(t)\Phi_2 + 
 {\cal G}(t,x, \Phi_2;\D_t\valpha(t),\D_t\valpha,
 \D_t\tTheta(t))
\label{eq:Phieqn}
\ee
where, ${\cal H}_0(t)$ denotes the matrix operator which is the linearization 
about the time-dependent nonlinear ground state $\Psi_{\alpha_0(t)}$. 
 The idea is that
in order for $\phi_2(t,x)$ to decay dispersively to zero 
 we must choose $\alpha_0(t)$
and $\alpha_1(t)$ to evolve in such a way as to remove all secular resonance
 terms from ${\cal G}$. Thus we require,
\be P_c({\cal H}_0(t))\Phi_2(t) = 0,\label{eq:PcPhid0}
\ee where $P_c({\cal H}_0)$ denotes the continuous spectral projection 
of ${\cal H}_0$; see also condition (\ref{eq:phi1inM1}).
Since the discrete subspace of ${\cal H}_0(t)$ is four dimensional
 (consisting of
  a generalized null space of dimension two plus two oscillating 
neutral modes) (\ref{eq:PcPhid0}) is equivalent to four 
 orthogonality conditions implying to four  
 differential equations for $\alpha_0,\alpha_1$ 
and their complex conjugates. These equations are coupled to the 
 dispersive partial differential equation for $\Phi_2$.  
At this stage we have that NLS is equivalent to a dynamical system
consisting of a finite dimensional part (\ref{eq:a0eqn}-\ref{eq:a1eqn}),
governing $\valpha_j=(\alpha_j,\ol{\alpha_j}),\ j=0,1$,  
 coupled to an infinite dimensional
dispersive part governing $\Phi_2$:
\ba
i\D_t\valpha &=& {\cal A}(t)\valpha + \vec F_\alpha\nn\\
i\D_t\Phi_2 &=& {\cH_0}(t)\Phi_2+ \vec F_\phi
\label{eq:model}
\ea
We expect ${\cal A}(t)$ and $\cH_0(t)$ to have limits as $t\to\pm\infty$.
Our strategy is to fix $T>0$ arbitrarily large, and to study the dynamics
on the interval $[0,T]$. In this we follow the strategy of \cite{BP2,Cucc2}.
 We shall rewrite (\ref{eq:model}) as:

\ba
i\D_t\valpha &=& {\cal A}(T)\valpha + ({\cal A}(t)-{\cal A}(T))\valpha
 + \vec F_\alpha\nn\\
i\D_t\Phi_2 &=& {\cal H}(T)\Phi_2+ ({\cH_0}(t)-{\cH_0}(T))\Phi_2+ \vec F_\phi
\label{eq:model1}
\ea
and implement a perturbative analysis about the 
 {\it time-independent} reference linear, respectively, matrix and differential,
 operators ${\cal A}(T)$ and ${\cH_0}(T)$.

More specifically, we analyze the dynamics of (\ref{eq:model1}) by
 using (1) the eigenvalues of ${\cal A}(T)$ to calculate the key
resonant terms and (2) together with the dispersive  estimates of
 $e^{-i\cH_0(T) t}P_c(T)$ \cite{Cucc1}.

Note also  that $P_c(T)\Phi_2(t)\ne \Phi_2(t)$ because
$\Phi_2(t)\in\ {\rm Range}\ P_c(t)\ \ne\ {\rm Range}\ P_c(T)$.
  We therefore decompose $\Phi_2$
as:
\be
\Phi_2 = {\rm disc(t;T)}\ +\ \eta,\ \ \eta = P_c(T)\eta\nn\ee
where disc$(t;T)$ lies in the discrete spectral subspace of $\cH_0(T)$
and show that ${\rm disc(t;T)}$ can be controlled in terms of $\eta$.

 The expected generic behavior of this system is that $\alpha_1$ and 
 $\Phi_2$ decay with a rate $t^{-{1\over2}}$. This slow rate actually
leads to an equation for $\alpha_1$ with the character 
 $\D_t\alpha_1 \sim ( {1\over t}\rho  - \D_t\tTheta ) \alpha_1 + $ integrable
 in $t$; see (\ref{eq:tTheta}).
Thus, $\tTheta$ is chosen to satisfy $\D_t\tTheta \sim {1\over t}\rho$
ensuring that $\alpha_1$ has a limit. In this way a logarithmic correction 
to the standard phase arises; see (\ref{eq:log-corr}).  

Next we explicitly factor out the rapid oscillations from $\alpha_1$
 and show that, after a near identity
 change of variables $(\alpha_0,\alpha_1)\mapsto (\talpha_0, \tbeta_1)$,
 that the modified ground and excited state amplitudes satisfy the 
system:
\ba
i\D_t\ta0\ &=&\ (c_{1022}+i\Gamma_{\omega})\
 |\tb1|^4\ta0\ +\
 F_\alpha[\ta0,\tb1,\eta,t]\nn\\
i\D_t\tb1\ &=&\ (c_{1121}-2i\Gamma_{\omega})
 |\ta0|^2|\tb1|^2\tb1\
 +\ F_\beta[\ta0,\tb1,\eta,t];
\label{eq:nf-intro}\ea
see Proposition \ref{prop:normalform}.
It follows that a  {\it Nonlinear Master Equation}
governs 
 $P_j=|\talpha_j|^2$, the power in the $j^{th}$ mode: 
\ba
{dP_0\over dt} &=& 2\Gamma P_1^2 P_0 +  R_0(t)\nn\\
{dP_1\over dt} &=& -4\Gamma P_1^2 P_0 +  R_1(t).\label{eq:mastereqn}
\ea
Coupling to the dispersive part, $\Phi_2$, is through the source terms
$R_0$ and $R_1$. The expression "master equation" is used since the role
played by (\ref{eq:mastereqn}) is analogous to the role of master equations
in the quantum theory of open systems \cite{Davies76}.

 A novel multiscale
  Lyapunov argument is implemented in section
 \ref{sec:stab} characterizing
the behavior of the system (\ref{eq:mastereqn}) coupled to that of 
 the dispersive part on short, intermediate and long time scales.
 We consider  the system (\ref{eq:mastereqn}) on three time intervals: 
  $I_0=[0,t_0]$ (initial phase)\  $I_1=[t_0,t_1]$ (embryonic phase) 
 and 
  $I_2=[t_1,\infty)$ (selection of the ground state).

For $t\ge t_0$, the terms $R_0(t)$ and $R_1(t)$ are shown 
 (Proposition \ref{prop:ODE}) to have the form
\ba
R_0(t)\ &\sim&\ \frac{b_0(t_0,\cE_0)}{\la t\ra^2}\ +\ \rho_0(\cE_0,t)P_0P_1^2 
\label{eq:R0rough}\\
R_1(t)\ &\sim&\ \frac{b_1(t_0,\cE_0)}{\la t\ra^2}\ +\ 
            \delta_m(t)\sqrt{P_0}P_1^m\ +\ \rho_1(\cE_0,t)P_0P_1^2,
\label{eq:R1rough}
\ea 
where 
\be b_0=\cO(\la t_0\ra^{-1})\ ,\ \ b_1=\cO(\la t_0\ra^{-{1\over2}})\
\label{eq:bjrough}
\ee
 $R_0(t)$ and $R_1(t)$ have parts which are local in time and nonlocal in 
 time. The handling of nonlocal terms is explained in  
section \ref{sec:stab}. 

We set (Proposition \ref{prop:propB})
\be
Q_0\ =\ P_0\ -\ \frac{\tilde{b_0}}{\la t\ra},\ \ 
Q_1\ =\ P_1\ +\ \frac{\tilde{b_1}}{\la t\ra}\label{eq:Q0Q1rough}
\ee
where $\tilde{b_0}$ and $\tilde{b_1}$ are positive and 
 satisfy (\ref{eq:bjrough}) as well.
\bigskip

\nit{\bf Discussion of time scales}

Estimates  (\ref{eq:R0rough}-\ref{eq:R1rough}) and the definitions
 (\ref{eq:Q0Q1rough}) imply an effective finite-dimensional 
 reduction to a system of 
equations for the ``effective mode powers'': $Q_0(t)$ and $Q_1(t)$, 
whose character on different time scales dictates the 
 full infinite dimensional  dynamics, in a manner analogous to role
of a center manifold reduction of a dissipative system \cite{Carr}.

\nit\underline{\it Initial phase --  $t\in I_0=[0,t_0]$:}
  Here, $I_0$ is the maximal interval on which 
 $Q_0(t)\le0.$ If $t_0=\infty$, then $P_0(t)=\cO(\la t\ra^{-2})$ and the 
ground state decays to zero. 
 In this case, we show in section \ref{sec:nongeneric} that the excited
state amplitude has a limit as well (with may or  may not be zero). This case 
is nongeneric.

\nit\underline{\it Embryonic phase -- $t\in I_1=[t_0,t_1]$:} 
 If $t_0<\infty$, then for $t>t_0$:
\ba
\frac{dQ_0}{dt}&\ge& 2\Gamma' Q_0Q_1^2\nn\\
\frac{dQ_1}{dt}&\le& -4\Gamma' Q_0Q_1^2\ +\ \cO(\sqrt{Q_0}Q_1^m). 
 \ m\ge4\label{eq:Q0Q1I1}
\ea
Therefore, $Q_0$ is montonically increasing; {\it the ground state grows}.
  Furthermore, if $Q_0$ is small
relative to $Q_1$, then
\be \frac{Q_0}{Q_1}\ {\rm is\  montonically\  increasing}, \nn\ee 
in fact exponentially increasing; {\it 
 the ground state grows rapidly relative to the 
 excited state.}

\nit\underline{\it Selection of the ground state $t\in I_2=[t_1,\infty)$:}
 There exists a time $t=t_1$, $t_0\le t_1<\infty$ at which 
 the $\cO(\sqrt{Q_0}Q_1^m)$ term in (\ref{eq:Q0Q1I1}) is dominated
by the leading ("dissipative") term. For $t\ge t_1$ we have
\ba
\frac{dQ_0}{dt} &\ge& 2\Gamma' Q_0Q_1^2\nn\\
\frac{dQ_1}{dt} &\le& -4\Gamma' Q_0Q_1^2.  
 \label{eq:Q0Q1I2}
\ea
It follows that $Q_0(t)\to Q_0(\infty)>0$  and $Q_1(t)\to0$
 as $t\to\infty$; {\it the ground state is selected.}


%
\section{ Linear and nonlinear bound states}\label{sec:bs}

In this section we introduce bound states of the linear ($\lambda=0$)
 and nonlinear ($\lambda\ne0$) 
 Schr\"odinger equation (\ref{eq:nls}).
\bigskip

\noindent{\bf Bound states of the unperturbed problem}

Let $H= -\Delta + V(x)$. We assume that $V(x)$ is smooth and
sufficiently rapidly decaying, so that $H$ defines a self-adjoint
operator in $L^2$. Additionally, we assume that the spectrum of $H$
consists of continuous spectrum extending from $0$ to positive infinity
and two discrete negative eigenvalues, each of multiplicity one.
\be
\sigma(H)\ =\ \{E_{0*}, E_{1*}\}\ \cup\ [0,\infty )
\label{eq:specH}\nn\ee
Therefore, there exist eigenstates, $\psi_{j*}\in {\cal D}(H),\ j=0,1$ such that
\be H\psi_{j*} = E_{j*} \psi_{j*}. \label{eq:linearevp}\ee

We also introduce spectral projections onto the discrete eigenstates and 
continuous spectral part of $H$, respectively:
\ba P_{j*}f\ &\equiv&\ \la\psi_{j*},f\ra\psi_{j*}, \ j=0,1
\nn\\
    P_{c*}\ &\equiv &  I - P_{0*}- P_{1*}\nn\ea 
     
\bigskip

\noindent{\bf Nonlinear bound states}

We seek solutions of (\ref{eq:nls}) of the form
\be \phi = e^{-iE_jt}\Psi_{E_j}\nn\ee
Substitution into (\ref{eq:nls}) yields
\be
H \Psi_{E_j} + \lambda |\Psi_{E_j}|^2 \Psi_{E_j} = E_j \psi_{E_j}
\label{eq:nonlinearevp}\ee
We introduce a bifurcation parameter, $\alpha_j$,
  for the $j^{th}$ nonlinear bound state branch and define
\be \valpha_j = (\alpha_j,\bar\alpha_j)\label{eq:boldalpha}\ee
\medskip

\begin{prop} \cite{RW88} \label{prop:bifurcate}
For each $j=0,1$ we have a one parameter family, 
 $\Psi_{\alpha_j}=\Psi_j$, of   
bound states depending on the complex parameter 
$\alpha_j=|\alpha_j|e^{i\gamma_j}$ and defined 
 for $|\alpha_j|$ sufficiently small:
\ba
  \Psi_j(x)&\equiv&\ 
 \ \alpha_j\psi_j(x;|\alpha_j|^2) 
   = e^{i\gamma_j}|\alpha_j|\psi_j(x;|\alpha_j|^2)\nn\\ 
  &=& \alpha_j\left(\ \psi_{j*}(x) + 
 \lambda |\alpha_j|^2\psi_j^{(1)}(x;|\alpha_j|^2)\ \right)\nn\\
 &=& \alpha_j\left(\ \psi_{j*}(x) +
 \lambda |\alpha_j|^2\psi_j^{(1)}(x;0) + \cO(\lambda^2|\alpha_j|^4\right).
 \label{eq:jbranch}
\ea
Here, 
\ba
 \psi_j^{(1)}(x;0) &=& -(H-E_{0*})^{-1}(I-P_{E_{j*}})\psi_{j*}^3
 \label{eq:psij10}\\
E_j &\equiv&\ E_j(\valpha_j)\ \equiv\  
 E_{j*} + |\alpha_j|^2 E_j^{(1)}(|\alpha_j|^2)\nn\\
 &=& E_{j*} + \lambda\int\psi_{0*}^4 dx\  |\alpha_j|^2 \ 
 +\ {\cal O}(|\alpha_j|^4).
\label{eq:Ej_expand}
\ea
The mapping $\valpha_j\mapsto \left(E_j(\valpha_j),\Psi_j(\cdot;\valpha_j)
 \right)$
 is smooth.
\end{prop}
 The proof uses standard bifurcation 
theory \cite{Nirenberg}, which is based on the implicit function
 theorem. The analysis extends to the case of nonlocal nonlinearities.
A variational approach can also be used to construct nonlinear bound 
 states. Variational approaches, though more global, do not directly
 yield the information we require concerning smooth variation 
 with respect to parameters.

\begin{rmk} $\Psi_j$ depends on $\alpha_j$ and $\bar\alpha_j$.
We shall  compute derivatives of the nonlinear bound states
$\Psi_{\alpha}$ with respect to $\alpha_j$ and $\ol{\alpha_j}$
and use the notation:
\be \vec\nabla_j = \left(\D_{\alpha_j},\D_{\ol{\alpha_j}}\right)
 \equiv (\D_j,\ol{\D_j}).
 \label{eq:nablaj}\ee
\end{rmk}

In what follows we shall ``modulate'' these bound states. That is, we shall
allow $\valpha$ to vary with time. For convenience, we shall use the notation:
\ba
\Psi_j(t,x)\ &=&\ \Psi_{\alpha_j(t)}(x), 
\nn\\
E_j(t)\ &=&\ E_j(|\alpha_j(t)|^2)\nn\ea
\section{Linearization about the ground state}\label{sec:linearization}

Let $\Psi$ denote a nonlinear bound state (ground state or excited state)
 of (\ref{eq:nls}); see section \ref{sec:bs}. Then,
\be
H\Psi\ +\ \lambda|\Psi|^2\Psi\ =\ E\ \Psi.
\label{eq:boundstateeqn}
\ee
We first derive the linear stability problem.
Let
\be
\phi\ =\ \left( \Psi\ +\ p \right) e^{-iEt},
\label{eq:lin_pert}
\ee
where $p$ denotes the perturbation about $\Psi$. Substituting  
(\ref{eq:lin_pert}) into (\ref{eq:nls}) and neglecting all terms which 
are nonlinear in $p$ and $\ol{p}$, we obtain the {\it linearized perturbation
equation}
\be
i\D_t p \ =\ \left(\ H-E+2\lambda |\Psi|^2\ \right)\ p +\ \lambda \Psi^2\ \ol{p} 
\label{eq:lin_pert_eqn}
\ee
Since $\ol{p}$ appears explicitly in (\ref{eq:lin_pert_eqn}) it is natural
to consider the system for 
\be
\vec p\ =\ 
\left(\begin{array}{c} p\\ \ol{p} \end{array}\right):
\label{eq:vec_p_def}
\ee
\be
i\D_t\vec p\ =\ {\cal H}\vec p,
\label{eq:vec_p_eqn}\ee
where
\be
{\cal H}\ =\ \sigma_3
 \left(\begin{array}{cc} H-E+2\lambda |\Psi|^2 & \lambda \Psi^2 \\ 
 \lambda \ol{\Psi}^2 & H-E+2\lambda |\Psi|^2  \end{array}\right),
\label{eq:H_def}\ee
where 
\be
\sigma_3 = \left(\begin{array}{cc} 1 & 0 \\ 0 & -1\end{array}\right).
\label{eq:sigma3}
\ee

Later in this paper we shall refer specifically to the linearization about 
 a ``curve" of  bound states $(\Psi_j(t), E_j(t))$ and will denote by 
 ${\cal H}_j(t)$ the operator (\ref{eq:H_def}) with $E$ replaced 
 by $E_j(t)$ and 
 $\Psi$ replaced by $\Psi_j(t)$.
Our main focus will be on the operator family
\be
 \cH_0(t)\ =\ \cH_{E_0(t),\Psi_0(t)}.
\nn\ee

The nonlinear bound state $\Psi$ is {\it linearly spectrally stable}
if the spectrum of ${\cal H}$, $\sigma({\cal H})$, is a subset of the 
 real line
\footnote{
The spectrum of $\cH$ has the symmetries one expects for Hamiltonian systems.
The mappings $\lambda\mapsto -\lambda$ and $\lambda\mapsto\bar\lambda$ send 
 points in the spectrum to points in the spectrum.
Note that if $\xi$ is an eigenvector of ${\cal H}$ with eigenvalue $\mu$ then
 $\ol{\sigma_1\xi}$ is an eigenvector of ${\cal H}$ with eigenvalue $-\mu$.
Therefore, $\xi_{-\mu}= \ol{\sigma_1\xi_\mu}$. }. 
 $\Psi$ is {\it linearly dynamically stable} if, in an appropriate
 space, all solutions of the initial value problem for (\ref{eq:vec_p_eqn})
 are bounded in time. That is, in some norm $e^{-i\cH_0t}$ is a bounded 
 operator. Linear dynamical stability of the ground state $\Psi_0$ follows
from \cite{MIW85}. 
 For this result and the necessary stronger dispersive estimates on 
 $e^{-i\cH_0t}$ \cite{Cucc1}, we require information on the 
 discrete spectrum of ${\cal H}_0$ and the corresponding spectral subspaces.

Before stating these results we observe that the operator $e^{-i\cH_0t}$ 
 can
be expressed in terms of the operator treated explicitly in \cite{MIW85}, 
 \cite{Cucc1}. 
 To see this,
  express the ground state as $\Psi_0 = |\Psi_0| e^{i\gamma}$ ($|\Psi_0|>0$), 
 where $\gamma=\arg\alpha$ is a constant. 
 Set $p=e^{i\gamma}q$.
Then, by (\ref{eq:lin_pert_eqn}) we have:

\be
i\D_t q \ =\ \left(\ H-E_0+2\lambda |\Psi_0|^2\ \right)\ q +\ 
 \lambda \Psi_0^2\ \ol{q}
\label{eq:lin_pert_eqn1}
\ee
Now let $q=u+iv$, where $u$ and $v$ are real. Then,
\be
\D_t \left(\begin{array}{c} u \\ v\end{array}\right)\ = \ J\tilde H_0
 \left(\begin{array}{c} u \\ v\end{array}\right), \nonumber\ee
where
\be
J= \left(\begin{array}{cc} 0 & -1 \\ 1 & 0 \end{array}\right) 
\ {\rm and}\ 
 \tilde H_0=\left(\begin{array}{cc} H_0-E_0-\lambda |\Psi_0|^2  & 0 \\ 
                0 & H_0-E_0-3\lambda |\Psi_0|^2  \end{array}\right)
\label{eq:J_tH0_def}
\ee
Note that 
\ba p(t)\ &=&\ \pi_1\vec p(t)\ =\ \pi_1 e^{-i{\cal H}_0t}\ \vec p_0\nonumber\\
 &=&\ e^{i\gamma}\left(\ \pi_1e^{J{\tilde H}_0t}\ \Re q_0\ +\ 
\ i\pi_2 e^{J{\tilde H}_0t}\ \Im q_0\right),
\ea
where $\pi_1(z_1,z_2)=(z_1,0)$ and  $\pi_2(z_1,z_2)=(0,z_2)$.
Therefore, we have:
\begin{prop}
Estimates on $e^{-i{\cal H}_0t}$ are equivalent to those for
 $e^{J{\tilde H}_0t}$
 and are independent of $\gamma$.
\end{prop}

We now turn to a detailed discussion of the spectral properties of $\cH_0$.
\begin{prop}\label{prop:Hspectrum}
Consider ${\cal H}_0$, the linearization about the ground state. Let 
 $\alpha_0$ be sufficiently small.

\noindent (1) $\sigma({\cal H}_0)$ is a subset of the real line.

\noindent (2) $\sigma_{\rm discrete}({\cal H}_0)\ = \ \{-\mu, 0,\mu\}$,
 where $0<\mu<|E_0|$.

\noindent (3) Zero is a generalized eigenvalue of ${\cal H}_0$.
 The generalized null space, $N_g({\cal H}_0)$, is given by 
\be
N_g({\cal H}_0)\ =\  
{\rm span}\left\{
 \sigma_3\left(\begin{array}{c} \Psi_0\\ \ol{\Psi_0} \end{array}\right)\ ,
 \ \left(\begin{array}{c} \D_{E_0}\Psi_0\\ \ol{\D_{E_0}\Psi_0} \end{array}\right)
\right\}
\label{eq:NgH}\ee

\noindent (4) $\pm\mu$ are simple eigenvalues. We denote their corresponding
 eigenfunctions by $\xi_\mu$ and $\xi_{-\mu}$. For $|\alpha_0|$ small 
we have the expansion:
\ba \mu\ &=&\  E_1-E_0 + {\cal O}(|\alpha_0|^2)\label{eq:muexpand}\\
    \xi_\mu\ &=&\ \left(\begin{array}{c} 1 \\ 0\end{array}\right)
\psi_{1*} + 
\left(\begin{array}{c} |\alpha_0|^2c_1(|\alpha_0|^2) \\ 
                       \overline{\alpha_0}^2c_2(|\alpha_0|^2)
 \end{array}\right)
\label{eq:ximu_def}\\
    \xi_{-\mu}\ &=&\ \overline{\sigma_1\xi_\mu}, 
\label{eq:ximinusmu_def}
\ea
where $c_1(a)$ and $c_2(a)$ are real analytic functions in $a$.

\noindent (5) $\sigma({\cal H}_0) - \sigma_{\rm discrete}({\cal H}_0)\ =\ 
 (-\infty, E_0] \cup [-E_0,\infty)$.
\end{prop}
The proof of Proposition \ref{prop:Hspectrum} is in the appendix of 
 section \ref{sec:appB}.
\bigskip

Note that   if $\omega$ is an eigenvalue of ${\cal H}$:
\be \sigma_3{\cal L}g\ \equiv {\cal H}g\ =\ \omega g \label{eq:evpH}\ee
then $\omega$ is an eigenvalue of ${\cal H}^*$ with corresponding eigenfunction
 $\sigma_3 g$:
\be
{\cal L}\sigma_3 (\sigma_3 g) = {\cal H}^* (\sigma_3 g) = \omega\sigma_3 g.
\label{eq:evpH*}\ee
Therefore, we have
\begin{prop}\label{prop:H*spectrum}
\smallskip

\noindent (1) $\sigma({\cal H}_0^*)\ =\ \sigma({\cal H}_0)$

\noindent (2) 
\be
N_g({\cal H}_0^*)\ =\
{\rm span}\left\{
 \left(\begin{array}{c} \Psi_0\\ \ol{\Psi_0} \end{array}\right)\ ,
 \
 \sigma_3\left(\begin{array}{c} \D_{E_0}\Psi_0\\ \ol{\D_{E_0}\Psi_0} \end{array}
 \right)
\right\}
\label{eq:NgH*}\ee

\noindent (3) 
\be N({\cal H}_0^*\mp \mu) \ =\ {\rm span}
 \left\{\sigma_3 \xi_\mu, \sigma_3\xi_{-\mu}  \right\}
 \equiv \left\{\zeta_\mu, \zeta_{-\mu} \right\}
\ee
Here, 
\be \zeta_\mu=\sigma_3\xi_\mu\ {\rm and}\ \zeta_{-\mu} = -\sigma_3\xi_{-\mu},
\label{eq:zetamu_def}
\ee
where this choice of $\zeta_{-\mu}$ is taken so that 
$\la \zeta_{-\mu},\xi_{-\mu}\ra = 1$ in the $|\alpha_0| \downarrow 0$ limit. 
\end{prop}
\medskip

\nit {\bf
 Nondegenerate basis for the discrete subspaces of $\cH_0$ and 
 $\cH_0^*$}

For fixed $E\ne E_0$, the basis of $N_g({\cal H}_0^*)$ displayed in 
 (\ref{eq:NgH*}) is a natural basis 
due to its direct connection to the symmetries of NLS. 
 However, this basis is degenerate
 and singular in the limit $E\to E_{0*}$, as we shall now see. 

Consider the basis of $N_g({\cal H}_0)$ displayed in (\ref{eq:NgH}).
Beginning with the first element of this basis, explicitly we have:
\be
\sigma_3\left( \begin{array}{c}\Psi_0\\ \ol{\Psi_0} \end{array}\right) 
= \sigma_3\left( \begin{array}{c}\alpha_0\psi_0(\cdot;|\alpha_0|^2)\\ 
                 \ol{\alpha_0} \psi_0(\cdot;|\alpha_0|^2) \end{array}\right) 
= \alpha_0 \sigma_3 G_0\left(\begin{array}{c} 1\\ 1\end{array}\right) F_0,
\nn\ee 
where  
\be \alpha_0 = |\alpha_0|e^{i\gamma_0},\ 
 \ {\bf G_0} = \left( \begin{array}{cc}e^{i\gamma_0} & 0\\ 0 & e^{-i\gamma_0} 
            \end{array}\right),
\label{eq:G0_def}
\ee
and 
\ba F_0 \equiv \psi_0(\cdot;|\alpha_0|^2)
 &=& \psi_{0*}+|\alpha_0|^2\chi(\cdot,|\alpha_0|^2)\nn\\
 &=& \psi_{0*}+|\alpha_0|^2\chi(\cdot,0) + \chi_4^{(0)}.
\label{eq:F0_def}
\ea
Here and subsequently we use the notation $\chi(\cdot,p)$
  to denote a generic 
real-valued localized function of $x$ with smooth dependence on a parameter, $p$,
 and $\chi_k^{(j)}$ is localized in $x$ and $\cO(|\alpha_j|^k)$.

  A nonsingular element of the 
 $N_g({\cal H}_0)$ is obtained by dividing out $\rho_0$. We therefore define 
\be \xi_{01} \equiv \sigma_3 G_0\left(\begin{array}{c} 1\\ 1\end{array}\right) 
           F_0(\rho_0^2)\label{eq:xi01_def}
\ee

We now turn to the second element of the basis displayed in (\ref{eq:NgH}).
First note that
\ba {\D\over\D |\alpha_0|}\Psi_0 &=&
 e^{i\gamma_0}
  {\D\over\D |\alpha_0|}\left( |\alpha_0|\psi_0(\cdot;|\alpha_0|^2)\right)\nn\\
 &=& e^{i\gamma_0}\left[ \psi_0 +
       |\alpha_0|^2\chi(\cdot;|\alpha_0|^2)\right]
\nn\ea

Differentiation of (\ref{eq:nonlinearevp}) with respect to $|\alpha_0|$ 
yields: 
\be
(H-E)\D_{|\alpha_0|}\Psi_0 + 2|\Psi_0|^2\D_{|\alpha_0|}\Psi_0 + 
 \Psi_0^2\overline{\D_{|\alpha_0|}\Psi_0} = (\D_{|\alpha_0|}E)\Psi_0
\label{eq:Psi0primeeqn}
\ee
Taken together with the complex conjugate of (\ref{eq:Psi0primeeqn}), this 
yields, after multiplication by $\sigma_3$:
\ba
\sigma_3{\cal H} 
  G_0 \left(\begin{array}{c} 1\\ 1\end{array}\right)F_0' &=& 
 (\D_{|\alpha_0|}E)\ 
 \sigma_3 G_0
    \left(\begin{array}{c} 1\\ 1\end{array}\right) |\alpha_0|\psi_0\nn\\ 
 &=& |\alpha_0| \xi_{01}
\label{eq:taken}
\ea

We define
\be \xi_{02} \equiv  G_0 \left(\begin{array}{c} 1\\ 1\end{array}\right)
           F_0',\label{eq:xi02_def}
\ee
where  
\be
F_0' = 
     {\D\over\D |\alpha_0|}\left( |\alpha_0|\psi_0(\cdot;|\alpha_0|^2)\right)
  = \psi_{0*} + |\alpha_0|^2\chi(x,|\alpha_0|^2).
\label{eq:F0prime_def}
\ee
By the above calculation $\xi_{02}$ lies in the null space of
  $(\sigma_3{\cal H})^2$.
Therefore, the pair of vectors: $\xi_{01}$ and $\xi_{02}$ spans 
 $N_g({\cal H}_0)$ and is nonsingular as $E_0\to E_{0*}$. By a previous
remark:
\be \zeta_{01} \equiv \sigma_3\xi_{02}\ {\rm and}\ \zeta_{02} \equiv
                          \sigma_3\xi_{01}
\label{eq:zeta01}
\ee
form a nonsingular basis for ${\cal H}_0^*$.
 This choice of basis will facilitate a {\it uniform} description
 of the dynamics in a neighborhood of the origin.

The above construction and Proposition \ref{prop:Hspectrum} imply the following
basis for the discrete subspaces of $\cH_0$ and $\cH_0^*$.
\begin{prop}\label{prop:xis}
\ba
N_g({\cal H}_0)\ &=&\ {\rm span} \{ \xi_{01}, \xi_{02} \} 
\label{eq:NgH1}\\
N_g({\cal H}^*_0)\ &=&\ {\rm span} \{ \zeta_{01}, \zeta_{02} \}
\label{eq:NgH1*}\\
N(\cH_0\mp\mu)\ &=&\ {\rm span} \{ \xi_\mu, \xi_{-\mu}\}
\label{eq:NHpmmu}\\ 
N(\cH_0^*\mp\mu)\ &=&\ {\rm span} \{ \zeta_\mu, \zeta_{-\mu}\} 
 = \{ \sigma_3\xi_\mu, -\sigma_3\xi_{-\mu}\}\label{eq:NH*pmmu}\\
\la \zeta_a, \xi_b\ra &=& C_{ab}\delta_{ab} + {\cal O}(|\alpha_0|^2),
\label{eq:normalization}
\ea
where $a$ and $b$ vary over the set $\{(01),(02),\mu,-\mu\}$.
For $\alpha_0$ small we have the expansions
\ba
\xi_{01} &=& \sigma_3G_0\left(\begin{array}{c}1 \\ 1\end{array}\right)F_0
 = \sigma_3G_0\left(\begin{array}{c}1 \\ 1\end{array}\right)
 \left(\psi_{0*}+ |\alpha_0|^2\chi^{(0)}_0\right)\nn\\
\xi_{02} &=& G_0\left(\begin{array}{c}1 \\ 1\end{array}\right)F_0'
 = G_0\left(\begin{array}{c}1 \\ 1\end{array}\right)
 \left(\psi_{0*}+ |\alpha_0|^2\chi^{(0)}_0\right)\nn\\
\xi_\mu &=& \left(\begin{array}{c}1 \\ 0\end{array}\right)
 \left( \psi_{1*} + |\alpha_0|^2\chi_0^{(0)}\right)
 + \ol{\alpha_0}^2\left(\begin{array}{c}0 \\ 1\end{array}\right)\chi_0^{(0)}
\label{eq:xi_exps}
\ea
and $\xi_{-\mu} =\ol{\sigma_1\xi_\mu}$.
\end{prop}

Finally, we shall find it useful to note that 
\be
\zeta_{01} = \sigma_3\xi_{02} = \xi_{01} + |\alpha_0|^2\chi(x;|\alpha_0|^2).
\label{eq:useful}\ee
\medskip

\nit{\bf Estimates for the linearized evolution operator}\label{sec:linevolution}

\begin{theo}\label{theo:lin_stab} (Linear dynamical stability \cite{MIW85})
Let
\be {\cal M}\ \equiv\ N_g({\cal H}_0^*)^\perp\ \cap \left(H^1\times H^1\right).
\label{eq:Mdef}\ee
There exists $C>0$ such that for any $f\in {\cal M}$
\be || e^{-i{\cal H}_0t } f ||_{H^1}\ \le\ C||f||_{H^1}\label{eq:H1est}\ee
\end{theo}

\begin{theo} \label{theo:lin_asymp_stab} (Dispersive estimates) \cite{Cucc1}
Let
\be {\cal M}_1\ \equiv\ \left[
 N_g({\cal H}_0^*)\oplus N(H^*\mp\mu)\right]^\perp 
\label{eq:M1def}\ee
and $P_c$ the associated continuous spectral projection.
For any $q\ge2$ there exists $C_{1,q}>0$ such that
\be || e^{-i{\cal H}_0t }P_c f ||_{L^q}\ \le\ C_{1,q}\ t^{-{n\over2}+{n\over q}}
 ||f||_{L^p},\label{eq:Lpqest}\ee
where $p^{-1}+q^{-1}=1$.
\end{theo}
\begin{rmk}
We shall in later sections use the notation ${\cal M}(t)$ and ${\cal M}_1(t)$ to
denote corresponding time-dependent subspaces relative to the time dependent
 operator ${\cal H}_0^*(t)$.
\end{rmk}

\begin{theo} (Local decay estimate) \label{theo:local_decay}
Let $\sigma$ be sufficiently large. 
 Let $\omega\in \{\ \nu\in\R:\ |\nu|> |E_0|\ \}$, 
 the interior of the continuous spectrum of
 $\cH_0$. Then, for $t>0$
\ba
||\la x\ra^{-\sigma}
  e^{-i{\cal H}_0t} P_c \left({\cal H}_0-\omega -i0\right)^{-l}
\la x\ra^{-\sigma}||_{\cB(L^2)}\ &\le& C\la t\ra^{-{3\over2}},\ \ l=0,1
\label{eq:ldest}\\
||\la x\ra^{-\sigma} {\tilde\cH_0}^k
  e^{-i{\cal H}_0t} P_c 
\la x\ra^{-\sigma}||_{\cB(H^{2k},L^2)}\ 
 &\le& C\la t\ra^{-{3\over2} -k},\ 
\label{eq:H0kldest}
\ea
where ${\tilde\cH_0}=\cH_0+E_0\sigma_3$.
For $t<0$, the same estimates hold with $-i0$ replaced by $+i0$
 in (\ref{eq:ldest}).
\end{theo}
Theorem \ref{theo:local_decay} follows from the technique of \cite{JK}
and results in \cite{Cucc2}; see also \cite{JSS}. 
 In particular, for a Schr\"odinger operator
$H_0=-\Delta+V$, one has the  local decay estimate 
\be
||\la x\ra^{-\sigma} 
  e^{-iH_0t} P_c(H_0)
\la x\ra^{-\sigma}||_{\cB(L^2)}\
 \le C\la t\ra^{-{3\over2}},\
\label{eq:JKldest}
\ee
where $\sigma$ is positive and  sufficiently large.
The key to the proof is an analysis of the resolvent, $(H_0-\lambda)^{-1}$
near $\lambda=0$. One uses the spectral theorem 
\be e^{-itH_0}P_c(H_0)\ =\ \int_0^\infty\ e^{-it\lambda} E'(\lambda)d\lambda
\nn\ee
and an explicit expansion of $E'(\lambda)$, 
which can be expressed in terms of the imaginary part of the resolvent.
The expansion of the resolvent is valid in the space~ 
 $\cB(L^2(\la x\ra^{\sigma}dx),L^2(\la x\ra^{-\sigma}dx)$
  for $\sigma$ sufficiently large.
 Time decay is arbitrarily fast for functions with
  spectral support away from $\lambda=0$,
while the $t^{-{3\over2}}$ decay results from the behavior near $\lambda=0$.  
 The analogue of 
(\ref{eq:H0kldest}) is an estimate for $e^{-itH_0}H_0^kP_c(H_0)$,
which follows from the expansion of \cite{JK} applied to the formula:
\be e^{-itH_0}H_0^kP_c(H_0)\ =\ \int_0^\infty\ e^{-it\lambda} \lambda^k\ 
 E'(\lambda)d\lambda.
\nn\ee 
Finally, this result can be obtained in the matrix case using the 
approach of \cite{Cucc2}. 

\section{Decomposition and Modulation Equations}\label{sec:dec-mod}

Consider the nonlinear Schr\"odinger equation
\be
i\D_t\phi\ =\ H\phi\ +\ \lambda |\phi|^2\phi.
\label{eq:NLS2}\ee

In the regime of low energy we decompose the solution of NLS 
 in the following form:
\be
\phi(t)\ =\ e^{-i\Theta(t)}\left[ \Psi_0(t) +\Psi_1(t) +\phi_2(t)\right].
\label{eq:decomp}
\ee
Here, $\Psi_0(t)$ and $\Psi_1(t)$ represent motion along the ground state
and excited state manifolds of equilibria and $\phi_2$ 
 is a decaying correction
 term, lying in an appropriate {\it dispersive subspace}. The phase,
 $\Theta$ is divided into two parts:
\be \Theta(t) = \Theta_0(t) + \tTheta(t),\label{eq:Theta_def}\ee
where 
\be \Theta_0(t)=\int_0^t\ E_0(s)\ ds.
 \label{eq:Theta_0_def}\ee
Thus, $\D_t\Theta_0(t) = E_0(t)$  
 is the modulated ground state energy, and $\tTheta(t)$ is 
a ``long range'' logarithmic correction, which is to be derived below

We begin by setting
\be
\phi\ =\ e^{-i\Theta(t)}\left[ \Psi_0(t) + \phi_1\right].
\label{eq:firststep}
\ee
Substitution of (\ref{eq:firststep}) into (\ref{eq:NLS2}) yields the 
following equation for $\phi_1$:

\ba
i\D_t\phi_1\ &=&\ \left(H-E_0(t)\right)\phi_1\ +\ 
 2\lambda |\Psi_0(t)|^2\phi_1\ +\ \lambda\Psi_0^2(t)\ol{\phi_1}\nn\\
&+&\ \left(-\D_t\tTheta+2\lambda|\phi_1|^2\right)\Psi_0(t)\ 
 +\ \lambda\ol{\Psi_0(t)}\phi_1^2\ +
\ \lambda |\phi_1|^2\phi_1 - \D_t\tTheta(t)\phi_1\nn\\
&-& i\D_t\Psi_0(t) 
\label{eq:phi1_eqn}\ea

Next, we decompose $\phi_1$ into a part along the excited state manifold and a
correction term:
\be \phi_1\ \equiv\ \Psi_1(t)\ +\ \phi_2.\label{eq:phi1_expand}\ee
 We have, using (\ref{eq:phi1_eqn}),
\ba
i\D_t\phi_2 &=& \left( H-E_0(t) \right)\phi_2 + 2\lambda |\Psi_0(t)|^2\phi_2
 + \lambda \Psi_0^2(t)\ol{\phi_2} - \D_t\tTheta(t)\phi_2\nn\\
&\ -& \left( E_{01}(t) +\D_t\tTheta(t) + i\D_t \right)\Psi_1(t)\ + \ 2\lambda |\Psi_0(t)|^2\Psi_1(t)
\ +\ \lambda \Psi_0(t)^2\ol{\Psi_1(t)}\nn\\
 &+&\ \left(-\D_t\tTheta(t) + 2\lambda \Psi_0(t)|\phi_1|^2\right)\Psi_0(t)
 + \lambda \ol{\Psi_0(t)}\phi_1^2\nn\\
&\ +&  \lambda\left( |\phi_1|^2 \phi_1 -
 |\Psi_1(t)|^2\Psi_1(t)\right)\nn\\
& -&\ i \D_t\Psi_0(t) 
\label{eq:phi2_eqn}
\ea
Since equation (\ref{eq:phi2_eqn}) involves $\ol{\phi_2}$ it is natural to
consider the system governing $\phi_2$ and $\ol{\phi_2}$.
Let
\be \Phi_2\ =\ \left(\begin{array}{c}\phi_2\\ \ol{\phi_2}\end{array}\right)
\label{eq:Phi2_def}\ee
and introduce the matrix linear operator:
\be
{\cal H}_0(t)\ \equiv\ \sigma_3
 \left(\begin{array}{cc} H-E_0(t)+2\lambda|\Psi_0(t)|^2 & \lambda\Psi^2_0(t)\\
  \lambda\ol{\Psi}^2_0(t) & H-E_0(t)+2\lambda|\Psi_0(t)|^2 \end{array}\right).
\label{eq:H0_def}\ee
Then we have (recall that $\phi_1 = \Psi_1 +\phi_2$) 
\ba
i\D_t\Phi_2\ &=&\ 
{\cal H}_0(t)\Phi_2\ + \bF -i \D_t\left(\begin{array}{c}
 \Psi_0(t)\\ c.c. \end{array}\right) 
- \left(\ ( E_{01}(1)+\D_t\tTheta(t) ) \sigma_3 + i\D_t\ \right)\ 
 \left(\begin{array}{c}
 \Psi_1(t)\\ c.c. \end{array}\right),\nn\\
&& \label{eq:Phi2_eqn}
\ea
where
\ba
\bF &\equiv&\ 
- \D_t\tTheta(t)\sigma_3
 \left(\begin{array}{c}\Psi_0\\ c.c.\end{array}\right)\  
-\ \D_t\tTheta(t)\sigma_3
\left(\begin{array}{c}\phi_2\\ c.c.\end{array}\right)\
\nn\\ 
&&+ \lambda\sigma_3\left(\begin{array}{c}2|\Psi_0|^2\Psi_1+\Psi_0^2\ol{\Psi_1}
 \\
 c.c. \end{array}\right)\nn\\
&+& \lambda\sigma_3\left(\begin{array}{c}2\Psi_0|\phi_1|^2 +
 \ol{\Psi_0}\phi_1^2\\
 c.c. \end{array}\right)
\ +\ \lambda\sigma_3\left(\begin{array}{c}|\phi_1|^2\phi_1
 - |\Psi_1|^2\Psi_1\\
 c.c. \end{array}\right)
\label{eq:bFdef}
\ea
\medskip

\subsection{ Modulation Equations} \label{sec:modulationeqns}

Motivated by the results of Theorem \ref{theo:lin_asymp_stab}
  on dispersive decay, we shall require that
\be \Phi_2(t)\in {\cal M}_1(t),\label{eq:phi1inM1}\ee
where ${\cal M}_1$ is defined in (\ref{eq:M1def}). Equivalently,
 \be P_c(\cH_0(t))\Phi_2(t) = \Phi_2(t),\nn\ee  
where $P_c(\cH_0(t))$ denotes the continuous spectral projection of 
$\cH_0(t)$. By Proposition
 \ref{prop:H*spectrum} this imposes four orthogonality conditions on 
 $\Phi_2(t)$:
\be \la \sigma_3\xi_a(t),\Phi_2(t) \ra = 0,\label{eq:orthogonality}\ee
where $a\in\{(01), (02), \mu,-\mu\}$.
We impose (\ref{eq:orthogonality}) at $t=0$ and now derive 
{\it modulation equations} for the coordinates $\alpha_0(t)$ and $\alpha_1(t)$
ensuring that (\ref{eq:orthogonality}) 
  persists for all $t\ne0$.
 
To derive the modulations we first take the inner product of 
(\ref{eq:Phi2_eqn})
with the adjoint vectors $\sigma_3\xi_a$ to obtain the identity: 
\ba
&& \left\langle \sigma_3\xi_a\ ,\  
 i\D_t\left(\begin{array}{c} \Psi_0 
                        \\ c.c. 
        \end{array}\right)
 \right\rangle
+ \left\langle \sigma_3\xi_a,\left(\ (E_{01}+\D_t\tTheta)\sigma_3 +i\D_t\right)
              \left(\begin{array}{c} \Psi_1\\ c.c.
                    \end{array}\right)\right\rangle\nn\\
&&=\ 
\la \cH_0^*(t)\sigma_3\xi_a,\Phi_2\ra + \la\sigma_3\xi_a,\bF\ra
\ + i\langle \D_t(\sigma_3\xi_a),\Phi_2\rangle 
 -i\D_t\la\sigma_3\xi_a,\Phi_2\ra
\label{eq:identity}
\ea

The initial data for NLS is decomposed so that 
$\la\sigma_3\xi_a(t),\Phi_2(t)\ra=0$ for $t=0$. In order for 
this condition to persist for all time it is necessary and sufficient that
the last term in (\ref{eq:identity}) vanish, or equivalently:

\begin{prop}\label{prop:ModulationEqns} 
The condition that $P_c(\cH_0(t))\Phi_2(t)=\Phi_2(t)$ is equivalent to the
following modulation equations for the coordinates $\alpha_0$ and $\alpha_1$
 which specify the dynamics along the ground state and excited 
 state manifolds  of equilibria:
\ba
&&\left\langle \sigma_3\xi_a\ ,\ 
      i\D_t\vPsi_0\right\rangle
 \ +\ \left\langle \sigma_3\xi_a\ ,\ 
 \left( \sigma_3 (E_{01}+ \D_t\tTheta)+i\D_t \right)
   \vPsi_1 
   \right\rangle, 
\nn\\
&&= 
-\lambda \left\langle \xi_a,
\left(\begin{array}{c}2|\Psi_0|^2\Psi_1+\Psi_0^2\ol{\Psi_1}\\
 c.c. \end{array}\right)\right\rangle\nn\\
&& +\ \left\la \xi_a,
 \left(\ -\D_t\tTheta(t)+2\lambda|\phi_1|^2\ \right)
 \left(\begin{array}{c}\Psi_0\\ c.c.\end{array}\right)\right\ra
\nn\\
&&+ \lambda\left\la \xi_a, \left(\begin{array}{c} \ol{\Psi_0}\phi_1^2\\
 c.c. \end{array}\right)\right\ra
 +\ \lambda \left\langle \xi_a,
\left(\begin{array}{c}|\phi_1|^2\phi_1
- |\Psi_1|^2\Psi_1\\ c.c. \end{array}\right)\right\rangle\nn\\  
&& - \D_t\tTheta(t)\la \xi_a,\Phi_2\ra +\ i\la \D_t(\sigma_3\xi_a),\Phi_2\ra.
\label{eq:modeqns}
\ea
where $a\in\{(01), (02), \mu,-\mu\}$ and 
 $\vec\Psi_j\equiv (\Psi_j,\ol{\Psi_j}),\ j=0,1$.
\end{prop}

\begin{rmk} 
\smallskip

\nit (A)  Note that the term $\la \cH_0^*(t)\sigma_3\xi_a,\Phi_2\ra$
in (\ref{eq:identity}) vanishes by the orthogonality constraint and because 
 $\cH_0^*(t)$ maps the discrete
 subspace into itself.
It therefore does not appear in 
 (\ref{eq:modeqns}).

\nit (B) The last term in (\ref{eq:modeqns}) is present
 due to the time-dependence of the
eigenvectors $\xi_a(t)$. An important simplification
of this ``commutator term'', which we require for $a=(01)$,
is carried out in the appendix of section \ref{sec:commutator}.
\end{rmk}

Initial data for the system (\ref{eq:modeqns}),
 (\ref{eq:Phi2_eqn}), governing $\alpha_0, \alpha_1$ and 
$\Phi_2$ are obtained as follows. Given data $\phi_0$ for NLS,
we find $\alpha_0$ so as to minimize 
\be \| \phi_0-\Psi_{\alpha_0}\|_2;\nn\ee
see \cite{SW9092}.
This $\alpha_0$ is used to define the initial Hamiltonian $\cH_0(0)$.
Now decompose $\phi_0$ using the biorthogonal decomposition 
associated with $\cH_0(0)$. This specifies $\alpha_0(0), \alpha_1(0)$
and $\Phi_2(0)$. 
\bigskip

\subsection{Conservation laws and 'a priori bounds}\label{sec:bounds}

In this subsection we obtain bounds on $\alpha_0$, $\alpha_1$ and $\phi_2$
using the conservation laws of NLS, noted in the introduction.

By the $L^2$ conservation law: $\bN[\phi]=\bN[\phi_0]$ we have
\ba
 h(t)\ &\equiv&\ |\Psi_0(t)|^2 + |\Psi_1(t)|^2 + \int\ |\phi_2(t,x)|^2\ dx
\nn\\
&=&\ \bN[\phi_0]-2\Re\int\Psi_0\ol{\Psi_1} - 2\Re\int\Psi_0\ol{\phi_2}
 - \Re\int\Psi_1\ol{\phi_2}
\label{eq:con1}
\ea 
Note that $\int\ \Psi_0\ol{\Psi_1}\ =\ \cO(|\alpha_0|\ |\alpha_1|^2)
 + \cO(|\alpha_0|^2\ |\alpha_1|)$. Furthermore,  we have
the orthogonality relation (\ref{eq:orthogonality}) with $a=(01)$,
 $\la\sigma_3\xi_{01},\Phi_2\ra=0$  or equivalently
 $\Re\int\Psi_0\ol{\phi_2}\ =\ 0$. Therefore 
 (\ref{eq:con1}) is equivalent to
\be
\left(\ 1\ +\ \cO(h(t))\ \right)\ h(t)
\ =\ \bN[\phi_0]\ 
\ -\ \Re\int\Psi_1\ol{\phi_2}
\label{eq:con2}
\ee 
The orthogonality relation (\ref{eq:orthogonality}) with $a=\mu$ or
 $\la\sigma_3\xi_\mu,\Phi_2\ra=0$ implies 
 $\Re\int\Psi_1\ol{\phi_2}\ =\
  \cO\left(|\alpha_1|^3\ +\ |\alpha_1|\ |\alpha_0|^2)\right)\|\phi_0\|_2$ 
 and therefore if $\bN[\phi_0]$ is sufficiently small
\be
|\Psi_0(t)|^2 + |\Psi_1(t)|^2 + \int\ |\phi_2(t,x)|^2\ dx\ \le\ C\bN[\phi_0]
\label{eq:L2componentbound}
\ee
In a similar way, the conservation law $\bH_{en}[\phi(t)]=\bH_{en}[\phi_0]$
 implies, if $\cE_0$ is sufficiently small,  the bound:
\be
\int\ |\nabla\phi_2(t,x)|^2\ dx\ \le\ C\cE_0 
\label{eq:phi2H1bound}
\ee
\section{ Toward a normal form - algebraic reductions and frequency analysis}
We now embark on a detailed calculation leading to a form
of this system, which though equivalent, is of a form to which
normal form methods can be easily applied. 

\subsection{Modulation equations}

Equations (\ref{eq:modeqns}) are a coupled system for $\alpha_0$,
 $\alpha_1$ and their complex conjugates. It is natural to write the system
as one which is nearly diagonal. This can be done by taking 
 appropriate linear combinations
of the equations in (\ref{eq:modeqns}). 

The equation which essentially determines  $\alpha_0$ can be found by 
 adding the two equations obtained from (\ref{eq:modeqns})
by setting $a=(01)$ and $a=(02)$. This gives:
\ba
&& \la \sigma_3(\xi_{01}+\xi_{02})\ ,\
i\D_t\vPsi_0\ra
 \ +\ \la \sigma_3(\xi_{01}+\xi_{02})\ ,\
 \left( \sigma_3(E_{01}+\D_t\tTheta)+i\D_t \right)
 \vPsi_1\ra
\nn\\
&&= 
-\lambda \left\langle \xi_{01}+\xi_{02},
\left(\begin{array}{c}2|\Psi_0|^2\Psi_1+\Psi_0^2\ol{\Psi_1}\\
 c.c. \end{array}\right)\right\rangle\nn\\
&& +\ \left\la \xi_{01}+\xi_{02},
 \left(\ -\D_t\tTheta(t)+2\lambda|\phi_1|^2\ \right)
 \left(\begin{array}{c}\Psi_0\\ c.c.\end{array}\right)\right\ra
\nn\\
&&+ \lambda\left\la \xi_{01}+\xi_{02},
  \left(\begin{array}{c} \ol{\Psi_0}\phi_1^2\\
 c.c. \end{array}\right)\right\ra
 +\ \lambda \left\langle \xi_{01}+\xi_{02},
\left(\begin{array}{c}|\phi_1|^2\phi_1
- |\Psi_1|^2\Psi_1\\ c.c. \end{array}\right)\right\rangle\nn\\
&&- \D_t\tTheta(t)\la \xi_{01}+\xi_{02},\Phi_2\ra 
 + i\la \D_t(\sigma_3(\xi_{01}+\xi_{02})),\Phi_2\ra.
\label{eq:me1}
\ea
The difference of the $a=(01)$ and $a=(02)$
equations is the complex conjugate of the equation (\ref{eq:me1}).
\bigskip

The equation which essentially determines  $\alpha_1$ is equation 
 (\ref{eq:modeqns}) with  $a=\mu$:
\ba
&& \la \sigma_3\xi_\mu\ ,\
      i\D_t\vPsi_0\ra
 \ +\ \la \sigma_3\xi_\mu\ ,\
 \left( \sigma_3(E_{01}+\D_t\tTheta) + i\D_t \right)
      \vPsi_1\ra
\nn\\
&&=\ 
-\lambda \left\langle \xi_\mu,
\left(\begin{array}{c}2|\Psi_0|^2\Psi_1+\Psi_0^2\ol{\Psi_1}\\
 c.c. \end{array}\right)\right\rangle\nn\\
&& +\ \left\la \xi_\mu,
 \left(\ -\D_t\tTheta(t)+2\lambda|\phi_1|^2\ \right)
 \left(\begin{array}{c}\Psi_0\\ c.c.\end{array}\right)\right\ra
\nn\\
&&+ \lambda\left\la \xi_\mu, \left(\begin{array}{c} \ol{\Psi_0}\phi_1^2\\
 c.c. \end{array}\right)\right\ra
 +\ \lambda \left\langle \xi_\mu,
\left(\begin{array}{c}|\phi_1|^2\phi_1
- |\Psi_1|^2\Psi_1\\ c.c. \end{array}\right)\right\rangle\nn\\
&& - \D_t\tTheta(t)\la \xi_\mu,\Phi_2\ra +\  i\la \D_t(\sigma_3\xi_\mu),\Phi_2\ra.\
\label{eq:me2}
\ea
The equation corresponding to $a=-\mu$ is the complex conjugate of this 
 equation.

Since
\be
\D_t \left(\begin{array}{c} \Psi_k\\ \ol{\Psi_k} \end{array}\right) =\
\left(\begin{array}{c} \D_k\Psi_k \\ \ol{ \ol{\D_k}\Psi_k }
 \end{array}\right) \D_t\alpha_k
\ +\ \left(\begin{array}{c} \ol{\D_k}\Psi_k \\ \ol{\D_k\Psi_k}
 \end{array}\right)\ \D_t\ol{\alpha_k},
\label{eq:dd}
\ee
we see that the equations for $\valpha_j=(\alpha_0,\ol{\alpha_0})^T,\
 j=0,1$ can be expressed in the form:
\ba
 i{\bf M_{00}}
 \D_t\left(\begin{array}{c}\alpha_0\\ \ol{\alpha_0}\end{array}\right)
+i{\bf M_{01}}
 \D_t\left(\begin{array}{c}\alpha_1\\ \ol{\alpha_1}\end{array}\right)
+ (E_{01}+\D_t\tTheta){\bf N_{01}}
         \left(\begin{array}{c}\alpha_1\\ \ol{\alpha_1}\end{array}\right)
&=& \bF_0\label{eq:me1a}\\
 i{\bf M_{10}}
 \D_t\left(\begin{array}{c}\alpha_0\\ \ol{\alpha_0}\end{array}\right)
+i{\bf M_{11}}
 \D_t\left(\begin{array}{c}\alpha_1\\ \ol{\alpha_1}\end{array}\right)
+(E_{01}+\D_t\tTheta){\bf N_{11}}
\left(\begin{array}{c}\alpha_1\\ \ol{\alpha_1}\end{array}\right)
&=& \bF_1
\label{eq:me2a}\ea
Here, for $k=0,1$:
\ba
{\bf M_{0k}} &=&
{\bf G_0} \left(\begin{array}{cc}
\left\la\sigma_3(\xi_{01}+\xi_{02}),
 \left(\begin{array}{c} \D_k\Psi_k \\ \ol{ \ol{\D_k}\Psi_k }\end{array}\right)
 \right\ra
&  \left\la\sigma_3(\xi_{01}+\xi_{02}),
  \left(\begin{array}{c}
 \ol{\D_k}\Psi_k \\ \ol{\D_k\Psi_k}\end{array}\right)\right\ra\\
 \left\la\sigma_3\ol{(\xi_{01}+\xi_{02})},
 \left(\begin{array}{c} \ol{\ol{\D_k}\Psi_k} \\ \D_k\Psi_k
 \end{array}\right)\right\ra &
\left\la\sigma_3\ol{(\xi_{01}+\xi_{02})},
 \left(\begin{array}{c}\ol{\D_k\Psi_k}\\ \ol{\D_k}\Psi_k\end{array}\right)
 \right\ra
\end{array}\right)\label{eq:M0k}\\
\nn\\
{\bf N_{01}} &=&
{\bf G_0}\left(\begin{array}{cc}
\left\la\sigma_3(\xi_{01}+\xi_{02}),
 \left(\begin{array}{c}\psi_1\\ 0\end{array}\right)\right\ra &
\left\la\sigma_3(\xi_{01}+\xi_{02}),\left(\begin{array}{c}0\\
 \psi_1\end{array}\right)
\right\ra\\
\left\la\sigma_3\ol{(\xi_{01}+\xi_{02})},
 \left(\begin{array}{c}0\\ \psi_1\end{array}\right)\right\ra &
\left\la\sigma_3\ol{(\xi_{01}+\xi_{02})},
 \left(\begin{array}{c}\psi_1\\ 0\end{array}\right)
\right\ra
\end{array}\right)
\label{eq:N01}\\
\nn\\
{\bf M_{1k} } &=&
{\bf G_0}\left(\begin{array}{cc}
\left\la\sigma_3\xi_\mu,
 \left(\begin{array}{c} \D_k\Psi_k \\ \ol{ \ol{\D_k}\Psi_k }\end{array}\right)
 \right\ra
& \left\la \sigma_3\xi_\mu,
 \left(\begin{array}{c} \ol{\D_k}\Psi_k \\ \ol{\D_k\Psi_k}\end{array}\right)
 \right\ra
 \\
 \left\la\sigma_3\ol{\xi_\mu},
 \left(\begin{array}{c} \ol{\ol{\D_k}\Psi_k} \\ \D_k\Psi_k
 \end{array}\right)\right\ra &
\left\la\sigma_3\ol{\xi_\mu},
 \left(\begin{array}{c}\ol{\D_k\Psi_k}\\ \ol{\D_k}\Psi_k
\end{array}\right)\right\ra
 \end{array}\right),\label{eq:M1k}\\
\nn\\
 {\bf N_{11} } &=&
{\bf G_0}\left(\begin{array}{cc}
\left\la\sigma_3\xi_\mu,\left(\begin{array}{c}\psi_1\\ 0\end{array}\right)
\right\ra &
\left\la\sigma_3\xi_\mu,\left(\begin{array}{c}0\\ \psi_1\end{array}\right)
\right\ra\\
\left\la\sigma_3\ol{\xi_\mu},
 \left(\begin{array}{c}0\\ \psi_1\end{array}\right)\right\ra &
\left\la\sigma_3\ol{\xi_\mu},
 \left(\begin{array}{c}\psi_1\\ 0\end{array}\right)\right\ra
\end{array}\right)
\label{eq:N11}
\ea

\subsection{Algebraic reductions and determination of $\tTheta(t)$}
To express the modulation  equations (\ref{eq:modeqns}) in a tractable 
form we shall make use of a number of notations and  relations
 which now list for convenience; see also the section \ref{sec:notation}. 

\medskip

\nit $\chi_k^{(j)}$ denotes a spatially localized  function of order $|\alpha_j|^k$,
 as $|\alpha_j|\to0$. 

\nit $\cO_k^{(j)}$ denotes a quantity which is of order
 $|\alpha_j|^k$ as $|\alpha_j|\to0$.  Both $\chi_k^{(j)}$  and $\cO_k^{(j)}$
 are invariant under the map 
$\alpha_j\mapsto\alpha_je^{i\gamma}$. 

\nit $\cO_k^{(0,1)} = \cO_{k_1}^{(0)}\cO_{k_2}^{(1)},\ \ k=k_1+k_2$.

\ba
\phi_1 &=& \Psi_1 + \phi_2\nn\\
\alpha_0 &=& |\alpha_0| e^{i\gamma_0},\ \ol{\alpha_0}e^{2i\gamma_0} = \alpha_0
 \nn\\
 \left(\D_k\Psi_k, \ol{\D_k}\Psi_k\right) &=& 
 \left(\psi_{k*}+|\alpha_k|^2\chi_0^{(k)}, \alpha_k^2\chi_0^{(k)}\right),
 \ k=0,1\nn\\
\xi_{01}+\xi_{02} &=& 
 G_0\left(\begin{array}{c} F_0+F_0'\\ F_0-F_0'\end{array}\right)
 = \left(\begin{array}{c}
2e^{i\gamma_0}(\psi_{0*} + \chi_2^{(0)}) \\ e^{-i\gamma_0}\chi_2^{(0)}
\end{array}\right)\nn\\
\ \xi_\mu &=& 
 \left(\begin{array}{c} \psi_{1*}+\chi_2^{(0)}\\ \ol{\alpha_0}^2\chi_0^{(0)}
 \end{array}\right)
\nn\\
F_0+F_0' &=& 2\psi_{0*}+|\alpha_0|^2\chi(\cdot;|\alpha_0|^2) =
  2(\psi_{0*} + \chi_2^{(0)})\nn\\
  F_0-F_0' &=& |\alpha_0|^2\chi(\cdot;|\alpha_0|^2)=\chi_2^{(0)}.
 \nn\\
 \la \psi_{0*},\psi_0^{(1)}(\cdot;0)\ra &=& 0,\label{eq:ids}
\ea

Using (\ref{eq:ids}) in (\ref{eq:me1}) we get:
\ba
&&2i(1+\cO_4^{(0)})\D_t\alpha_0 + i\alpha_0^2\cO_0^{(0)}\D_t\ol{\alpha_0}
+ \la \sigma_3(\xi_{01}+\xi_{02})\ ,\
 \left(\ \sigma_3(E_{01}+\D_t\tTheta)+i\D_t\ \right)
           \vPsi_1 \ra
 \nn\\
&&= -2\left( \D_t\tTheta \la F_0',\Psi_0\ra-2\lambda\la F_0',|\phi_1|^2\Psi_0\ra
 \right)\nn\\
&&+\la F_0+F_0',2\lambda|\Psi_0|^2\Psi_1+\lambda\Psi_0^2\ol{\Psi_1}
 +\lambda|\phi_1|^2\phi_1 -\lambda|\Psi_1|^2\Psi_1\ra
-e^{2i\gamma_0}\la F_0-F_0', c.c.\ra\nn\\
&& + ie^{2i\gamma_0}\la \D_t[\sigma_3(\xi_{01}+\xi_{02})],\Phi_2\ra
\label{eq:a0-1}
\ea
\nit{\bf Determination of $\tTheta(t)$}

We anticipate  that 
 generically $|\phi_1|\sim |\alpha_1| \sim t^{-{1\over2}}$ for $t$ very large.
$\alpha_0$ will have a limit as $t\to\pm\infty$
if $\D_t\alpha_0$ is integrable. We ensure this by  
 choosing $\D_t\tTheta$ to cancel 
the terms which are of order $t^{-1}$ and {\it nonoscillatory}.
Thus, we choose $\tTheta$ to satisfy
\be
\D_t\tTheta \la F_0',\Psi_0\ra - 2\lambda\la F_0',|\phi_1|^2\Psi_0\ra = 0.
\label{eq:tTheta}
\ee
To leading order this gives:
\be
\D_t\tTheta \sim 2\lambda\la \psi_{0*}^2,|\phi_1|^2\ra.
\label{eq:tTheta-lead}
\ee
In this way, a logarithmic correction to the standard phase, $\int_0^tE_0(s)ds$
 arises.

Equations (\ref{eq:a0-1}), (\ref{eq:tTheta}) together with Proposition
 \ref{prop:commutators} imply
\ba
&&2i(1+\cO_4^{(0)})\D_t\alpha_0 + i\alpha_0^2\cO_0^{(0)}\D_t\ol{\alpha_0}
+ (E_{01}+\D_t\tTheta)\la\xi_{01}+\xi_{02}\ ,\vPsi_1\ra +
 \la \sigma_3(\xi_{01}+\xi_{02}),i\D_t\vPsi_1 \ra
\nn\\ 
&&= -\D_t\tTheta\la \chi_0^{(0)},\phi_2\ra + \D_t\tTheta \alpha_0^2
 \la \chi_0^{(0)}, \ol{\phi_2}\ra\nn\\
&& + \lambda\cO_0^{(0)}(|\alpha_0|^2\alpha_1+\alpha_0^2\ol{\alpha_1})
 +\lambda\la \chi_0^{(0)},|\phi_1|^2\phi_1 -|\Psi_1|^2\Psi_1\ra
-\lambda\alpha_0^2 \la \chi_0^{(0)}, |\phi_1|^2\ol{\phi_1} -
 |\Psi_1|^2\ol{\Psi_1}\ra\nn\\
&& -\D_t|\alpha_0|^2\left[\ \la \chi_0^{(0)}, \phi_2\ra + 
 e^{2i\gamma_0}\la \chi_0^{(0)},\ol{\phi_2}\ra\ \right] +
 i |\alpha_0|^2 \D_t\gamma_0
\left[\ \la \chi_0^{(0)}, \phi_2\ra + 
 e^{2i\gamma_0}\la \chi_0^{(0)},\ol{\phi_2}\ra\ \right] 
 \label{eq:a0-1a}
\ea
\bigskip

We now turn equation (\ref{eq:me2}). 
Using (\ref{eq:ids}) equation (\ref{eq:me2}) can be written as:
\ba
&&(1+\cO_2^{(0,1)})\ i\D_t\alpha_1 +
(\cO(\alpha_0^2)+\cO(\alpha_1^2) )\ i\D_t\ol{\alpha_1}\nn\\
&&+ (\ (1+\cO_2^{(0,1)})\alpha_1 + \alpha_0^2\ol{\alpha_1}\cO_2^{(0,1)}\ )
 (E_{01}+\D_t\tTheta) + \la\sigma_3\xi_\mu,i\D_t\vPsi_0\ra\nn\\
&& = -\D_t\tTheta(\ \la\chi,\phi_2\ra + \alpha_0^2\la\chi,\ol{\phi_2}\ra\ )
-\D_t\tTheta \cO_0^{(0,1)}\alpha_0\nn\\
&& -\lambda\cO_2^{(0,1)}(|\alpha_0|^2\alpha_1+\alpha_0^2\ol{\alpha_1})
+\lambda \cO_0^{(0)}\alpha_0\la\chi,|\phi_1|^2\ra\nn\\
&&+\lambda \ol{\alpha_0}\la\chi,\phi_1^2\ra +
 \lambda \alpha_0^3\la\chi,\ol{\phi_1}^2\ra\nn\\
&&+\lambda \la\chi,|\phi_1|^2\phi_1-|\Psi_1|^2\Psi_1\ra
+ \lambda\alpha_0^2\la\chi,|\phi_1|^2\ol{\phi_1}-|\Psi_1|^2\ol{\Psi_1}\ra
\nn\\
&& + i\la \D_t(\sigma_3\xi_\mu), \Phi_2\ra
\label{eq:a1-1a}
\ea
We next write the systems for $\valpha_0$ and $\valpha_1$.
\ba
&& i{\bf M_{00}}\D_t\valpha_0 + i{\bf M_{01}}\D_t\valpha_1 
 + (E_{01}+\D_t\tTheta)\sigma_3{\bf N_{01}}\valpha_1\nn\\
&&\ \  =\  \lambda\cO_0^{(0)}\sigma_3\left(\begin{array}{c}
                       |\alpha_0|^2\alpha_1+\alpha_0^2\ol{\alpha_1}\\ c.c.
                         \end{array}\right)\nn\\
&&\ \ + \lambda\sigma_3\left(\begin{array}{c}
\la\chi_0^{(0)},|\phi_1|^2\phi_1-|\Psi_1|^2\Psi_1\ra 
 + \alpha_0^2\la\chi_0^{(0)},
 |\phi_1|^2\ol{\phi_1}-|\Psi_1|^2\ol{\Psi_1}\ra \\ c.c.
                         \end{array}\right)\nn\\
&&\ \ -\D_t\tTheta
\left(\begin{array}{c}
\la\chi_0^{(0)},\phi_2\ra +\alpha_0^2\la\chi_0^{(0)},\ol{\phi_2}\ra
\\ c.c.  \end{array}\right)
\nn\\
&&\ \ - \D_t|\alpha_0|^2\sigma_3
\left(\begin{array}{c}
\la\chi_0^{(0)},\phi_2\ra + e^{2i\gamma_0}\la\chi_0^{(0)},\ol{\phi_2}\ra 
                         \\ c.c.
                         \end{array}\right)\nn\\
&&\  \ +i |\alpha_0|^2\D_t\gamma_0
\left(\begin{array}{c}
\la\chi_0^{(0)},\phi_2\ra + e^{2i\gamma_0}\la\chi_0^{(0)},\ol{\phi_2}\ra 
                         \\ c.c.
                         \end{array}\right)
\label{eq:a0-sys}\\
&& i{\bf M_{11}}\D_t\valpha_1 + i{\bf M_{10}}\D_t\valpha_0 
+ (E_{01}+\D_t\tTheta)\sigma_3{\bf N_{11}}\valpha_1\nn\\
&&\ \ =  \lambda\cO_0^{(0,1)}\sigma_3\left(\begin{array}{c}
                       |\alpha_0|^2\alpha_1+\alpha_0^2\ol{\alpha_1}\\ c.c.
                         \end{array}\right)\nn\\
&&\ \ + \lambda\sigma_3\left(\begin{array}{c}
 \la \chi,|\phi_1|^2\phi_1-|\Psi_1|^2\Psi_1\ra +\alpha_0^2
      \la \chi,|\phi_1|^2\ol{\phi_1}-|\Psi_1|^2\ol{\Psi_1}\ra\\ c.c.
           \end{array}\right)\nn\\
&&\ \ +\lambda\sigma_3
\left(\begin{array}{c}
\alpha_0\la \chi,|\phi_1|^2\ra +\ol{\alpha_0}\la \chi,\phi_1^2\ra
      +\alpha_0^3\la \chi,\ol{\phi_1}^2\ra\\ c.c.
           \end{array}\right)\nn\\
&&\ \ - \D_t\tTheta\ \sigma_3
\left(\begin{array}{c}
\la\chi,\phi_2\ra+\alpha_0^2\la\chi,\ol{\phi_2}\ra\\ c.c.\end{array}\right)
+ i\left(\begin{array}{c} \la\D_t(\sigma_3\xi_\mu),\Phi_2\ra\\ c.c.
         \end{array}\right)
\label{eq:a1-sys}
\ea

The matrices ${\bf M_{jk}},\ {\bf N_{j1}},\  j,k=0,1$ are displayed in
 equations (\ref{eq:M0k}-\ref{eq:N11}).
  For the (generic) case where we expect $|\alpha_0|$
to approach a nonzero limit and $\alpha_1$ and $\phi_2$ to decay to zero,
we shall use the expansion: 

\ba
{\bf M_{00}}
 &=& \left(\begin{array}{cc}
     2(1+\cO_4^{(0)}) & \alpha_0^2\cO_0^{(0)}
\\   \ol{\alpha_0}^2\cO_0^{(0)} & 2(1+\cO_4^{(0)})
         \end{array}
   \right)\nn\\
{\bf M_{11}}
 &=& \left(\begin{array}{cc}
     1+\cO_2^{(0,1)} & \cO(\alpha_0^2)+ \cO(\alpha_1^2) 
\\    \cO(\ol{\alpha_0}^2)+ \cO(\ol{\alpha_1}^2) & 1+\cO_2^{(0)}
         \end{array}
   \right)\nn\\
\ea
The matrices ${\bf M_{10}}$ and ${\bf M_{10}}$ are higher order in $\alpha_0$
and satisfy:
\be
{\bf M_{10}},\ {\bf M_{01}}
 = \left(\begin{array}{cc}
     \cO_2^{(0)} & \alpha_0^2\cO_0^{(0)}
\\   \ol{\alpha_0}^2\cO_0^{(0)} & \cO_2^{(0)}
         \end{array}
   \right)\label{eq:M10M01}
\ee
The matrices ${\bf N_{01}}$ and ${\bf N_{11}}$
satisfy:
\ba
{\bf N_{01}}
 &=& \left(\begin{array}{cc}
     \cO_2^{(0)}+\cO_2^{(1)} & \alpha_0^2(\cO_2^{(0)}+\cO_2^{(1)})
\\   \ol{\alpha_0}^2(\cO_2^{(0)}+\cO_2^{(1)}) & \cO_2^{(0)}+\cO_2^{(1)} 
         \end{array}
   \right)\nn\\
{\bf N_{11}}
 &=& \left(\begin{array}{cc}
     1+\cO_2^{(0)} & \alpha_0^2\cO_0^{(0,1)}
\\   \ol{\alpha_0}^2\cO_0^{(0,1)} & 1+\cO_2^{(0)}
         \end{array}
   \right)\nn
\ea
\bigskip

\nit{\bf Simplifications to equations (\ref{eq:me1}-\ref{eq:me2})}

\nit (1) Since ${\bf M_{10}}$ and ${\bf M_{01}}$ are higher order in $\alpha_0$
we can eliminate $\D_t\alpha_1$ from (\ref{eq:a0-sys}) and 
 $\D_t\alpha_0$ from (\ref{eq:a1-sys}).

\nit (2) Note also, that "commutator terms" with factors like
  $\D_t|\alpha_0|^2$ or $\D_t\alpha_0^2$ can be eliminated via redefinition
of the near identity 
 matrix ${\bf M_{00}}$ through incorporation of a higher order correction.

\nit (3)  
We can eliminate the term proportional to $|\alpha_0|^2\D_t\gamma_0$
as follows.
Consider the last two terms of (\ref{eq:a0-sys}). Since $\D_t|\alpha_0|^2
 = \alpha_0\D_t\ol{\alpha_0} + \alpha_0\D_t\ol{\alpha_0}$ we can incorporate
the second to last term of (\ref{eq:a0-sys}) as a higher order
 correction to the near identity matrix $\bf{M_{00}}$, $\tilde{\bf{M_{00}}}$.
Our goal is now to eliminate the eliminate $\D_t\gamma_0$ from the equation.
The system now be written as form:
\be
 i\D_t\valpha_0
 = \tilde{\bf{M_{00}}}^{-1}\left(\  ...\ +\
 i \cO(|\alpha_0|^2)\D_t\gamma_0\right)
\label{eq:hey1}\ee
The first component has the form:
\ba &&i\D_t\alpha_0 = { 1^{\rm st}\ {\rm component}\ {\rm of\ the\ vector:}}
 \nn\\
&& \tilde{\bf{M_{00}}}^{-1}(...) +
 i\cO(|\alpha_0|^2) \D_t\gamma_0 \tilde{\bf{M_{00}}}^{-1}
\label{eq:hey2}
\ea
Since $\alpha_0=|\alpha_0|e^{i\gamma_0}$ we have
\ba && i\D_t |\alpha_0| - |\alpha_0|\D_t\gamma_0
 = e^{-i\gamma_0}\times
 { 1^{\rm st}\ {\rm component}\ {\rm of\ the\ vector:}}\nn\\
&& \tilde{\bf{M_{00}}}^{-1}(...) +
 i\cO(|\alpha_0|)\ |\alpha_0| \D_t\gamma_0 \tilde{\bf{M_{00}}}^{-1}
\label{eq:hey3}
\ea
By taking the real part of (\ref{eq:hey3}), for $|\alpha_0|$ small,
 we can solve for $|\alpha_0|\D_t\gamma_0$.
This enables us to eliminate it from equation (\ref{eq:a0-sys}) as a higher 
 order term.
\bigskip

Implementation of these simplifications leads to the following
\begin{prop}
\ba
 i\D_t\valpha_0 &=& 
 {\bf M_{00}^\#}\left[
 -(E_{01}+\D_t\tTheta)\sigma_3{\bf N_{01}}\valpha_1\right.\nn\\
&&\  +\lambda\cO_0^{(0)}\sigma_3\left(\begin{array}{c}
                       |\alpha_0|^2\alpha_1+\alpha_0^2\ol{\alpha_1}\\ c.c.
                         \end{array}\right)\nn\\
&&\ \ + \lambda\sigma_3\left(\begin{array}{c}
\la\chi_0^{(0)},|\phi_1|^2\phi_1-|\Psi_1|^2\Psi_1\ra
 + \alpha_0^2\la\chi_0^{(0)},
 |\phi_1|^2\ol{\phi_1}-|\Psi_1|^2\ol{\Psi_1}\ra \\ c.c.
                         \end{array}\right)\nn\\
&&\ \left. -\D_t\tTheta
\left(\begin{array}{c}
\la\chi_0^{(0)},\phi_2\ra +\alpha_0^2\la\chi_0^{(0)},\ol{\phi_2}\ra
\\ c.c.  \end{array}\right) \ \right]
\nn\\
\label{eq:a0eqn}\\
i\D_t\valpha_1 &=& {\bf A}(t)\valpha_1\nn\\
&&\ \ +{\bf M_{11}^\#}\left[\ \lambda\sigma_3\left(\begin{array}{c}
 \la \chi,|\phi_1|^2\phi_1-|\Psi_1|^2\Psi_1\ra +\alpha_0^2
      \la \chi,|\phi_1|^2\ol{\phi_1}-|\Psi_1|^2\ol{\Psi_1}\ra\\ c.c.
           \end{array}\right)\ \right.\nn\\
&&\ \ +\lambda\sigma_3
\left(\begin{array}{c}
\alpha_0\la \chi,|\phi_1|^2\ra +\ol{\alpha_0}\la \chi,\phi_1^2\ra
      +\alpha_0^3\la \chi,\ol{\phi_1}^2\ra\\ c.c.
           \end{array}\right)\nn\\
&&\ \left. - \D_t\tTheta\ \sigma_3
\left(\begin{array}{c}
\la\chi,\phi_2\ra+\alpha_0^2\la\chi,\ol{\phi_2}\ra\\ c.c.\end{array}\right)
\ \right]
\label{eq:a1eqn}
\ea
Here, 
\ba
\bA\ &=&\ \left(\begin{array}{cc}
 E_{10}+ \cO_0^{(0,1)}|\alpha_0|^2 &\cO_0^{(0,1)}\alpha_0^2\\
-\cO_0^{(0,1)}\ol{\alpha_0}^2 & -E_{10}-\cO_0^{(0,1)}|\alpha_0|^2
\end{array}\right)\nn\\
\label{eq:Adef}\\
\bA_{22} &=& -\bA_{11},\ \bA_{21}=-\ol{\bA_{12}},\ {\rm and}\nn\\
\phi_1 &=& \Psi_1 + \phi_2\nn
\ea
${\bf M_{00}^\#}$ and ${\bf M_{11}^\#}$ are near-identity matrices,
whose deviations from the identity give rise to higher order terms which 
are subordinate to the leading order behavior obtained in the analysis which
follows.

\nit Finally, 
 equations (\ref{eq:a0eqn}) and (\ref{eq:a1eqn}) are coupled to the equation
for $\Phi_2$ given by (\ref{eq:Phi2_eqn}).
\end{prop}

 
\subsection{Peeling off the rapid oscillations of  $\alpha_1$}
\label{subsection:Peel}

Fix $T>0$ and large. We rewrite (\ref{eq:a1eqn}) centered about 
the ground state at time $t=T$:
\be
i\D_t\valpha_1 =\bA(T) \valpha_1 + \left(\bA(t)-\bA(T)\right)\valpha_1 +
\vec F,
\label{eq:alpha1_eqnE}
\ee
where
\be
\bA(T)\ =\ \left(\begin{array}{cc}
 E_{01}(T)+ \cO_0^{(0,1)}(T)|\alpha_0(T)|^2 &\cO_0^{(0,1)}(T)\alpha_0^2(T)\\
-\cO_0^{(0,1)}(T)\ol{\alpha_0}^2(T) & -E_{01}(T)-\cO_0^{(0,1)}(T)|\alpha_0(T)|^2
\end{array}\right)
\label{eq:AT}
\ee

Since $\bA(T)$ is a constant coefficient matrix it is a simple matter
to obtain the fundamental matrix.
\begin{prop}\label{prop:xdef}
The system
\be
i\D_t\vec\alpha_1\ =\ \bA(T)\vec\alpha_1\label{eq:homogalpha1}\ee
has a fundamental solution matrix:
\ba
&& \bX(t)\ =\ \left(\begin{array}{cc} c_{11}^+ & c_{12}^-\\
                                   c_{21}^+ & c_{22}^- \end{array}\right)
         \left(\begin{array}{cc} e^{-i\lambda_+t} & 0\\
                                 0   & e^{-i\lambda_-t}\end{array}\right)
\nn\\
&=& \left(\begin{array}{cc}
 1  & \cO_0^{(0,1)}(T)\alpha_0(T)^2E_{10}^{-1}(T) \\
\cO_0^{(0,1)}(T)\ol{\alpha_0(T)}^2 E_{10}^{-1}(T) & 1
\end{array}\right)
  \left(\begin{array}{cc} e^{-i\lambda_+t} & 0\\
    0   & e^{-i\lambda_-t}\end{array}\right)
\nn\\
\label{eq:Xdef}
\ea
The eigenfrequencies, $\lambda_\pm(T)$ are given by $\lambda_+(T)$
and $\lambda_-(T)=-\lambda_+(T)$, where:
\be \lambda_+(T)\ =\ E_{10}(T)+
 \cO_0^{(0,1)}|\alpha_0(T)|^2,\
\label{eq:lambda_def}
\ee
where $E_{10}=E_1-E_0$, and provided $|\alpha_0(T)|^2/E_{10}(T)$
  is sufficiently small.
\end{prop}
\medskip

We use the fundamental matrix, $X(t)$, to define a change of variables:
\be
\vec\alpha_1\ \equiv
             \ \bX(t)\vec\beta\ =\ \bX(t)\left(\begin{array}{c}\beta_1\\
                                                          \ol{\beta_1}
                                           \end{array}\right)
\label{eq:beta_def}\ee
Therefore,
\be
\alpha_1 = e^{-i\lambda_+t}\beta_1 + \cO_0^{(0,1)}\alpha_0^2(T)E_{10}^{-1}(T)
       e^{-i\lambda_-t}\ol{\beta_1}.
\label{eq:a1-b1}\ee
Then, $\vec\beta$ satisfies
\be
i\D_t\vec\beta\ = \ \bX^{-1}(t)\vec F(\bX(t)\vec\beta(t),t)
 \label{eq:vecbeta_eqn}\ee
Note that since the linear in $\alpha_1$ terms have been removed by the change
of variables (\ref{eq:beta_def}), $\D_t\vec\beta_1\sim \cO(|\beta_1|^2)$.

Note that by (\ref{eq:Xdef})
\be
\bX^{-1}(t)\ =\
        \left(\begin{array}{cc} e^{i\lambda_+t} & 0\\
                                0 & e^{i\lambda_-t}\end{array}\right)
\left(\begin{array}{cc} 1+\cO_0^{(0,1)}(T)|\alpha_0(T)|^4 E_{10}^{-2}(T) &
                               -\cO_0^{(0,1)}(T)\alpha_0^2(T)E_{10}^{-1}(T)\\
                   -\cO_0^{(0,1)}(T)\ol{\alpha_0(T)}^2E_{10}^{-1}(T) &
               1+\cO_0^{(0,1)}(T)|\alpha_0(T)|^2E_{10}^{-2}(T)
                     \end{array}\right)
\label{eq:Xinverse}\ee

Written out in detail, from (\ref{eq:vecbeta_eqn}) and the various definitions,
we have
\begin{prop}
The equation for $\beta_1$ has the form
\ba
i\D_t\beta_1\ &=&\ \lambda e^{i\lambda_+t}
\la \chi_0, 
|e^{-i\lambda_+t}\beta_1\psi_{1*}+\phi_2|^2\ra
 \alpha_0\nn\\
&+&\ \lambda e^{i\lambda_+t}\langle \chi_0,
 (e^{-i\lambda_+t}\beta_1\psi_{1*}+\phi_2)^2\rangle\ol{\alpha_0}\nn\\
&+&\ \lambda e^{i\lambda_+t}\la \chi_0, (e^{i\lambda_+t}\ol{\beta_1}\psi_{1*}+
 \ol{\phi_2})^2\ra\alpha_0^3\nn\\
&+&\ \lambda\langle\chi_0 ,
  \left[ |e^{-i\lambda_+t}\beta_1\psi_{1*}+\phi_2|^2
        (e^{-i\lambda_+t}\beta_1\psi_{1*}+\phi_2) -
              e^{-i\lambda_+t}|\beta_1|^2\beta_1\psi_{1*}^3\right]\ 
 \rangle\nn\\
&+& {\cal R}_\beta, \ {\rm where}
 \label{eq:beta_eqn2}\\
\cR_\beta &=& \bX^{-1}(t)\left[ (\bA(t)-\bA(T))\bX(t)\vec\beta_1+\cR_1\right]
\label{eq:Rbeta_def}
\ea
\end{prop}

\begin{rmk}
\be
\bX^{-1}(t)[ \bA(t) - \bA(T) ] \bX(t) = 
 {\rm \ Real \ Symmetric\ Diagonal}\ {\bf S}(t) + e^{-2i\lambda_+t} {\bf B}(t)
\nn\ee
Therefore, the nonoscillatory part involving {\bf S(t)} does not effect the 
evolution of $|\beta_1|^2$.
\end{rmk}

\subsection{Expansion of $\phi_2$}

The next step is to get an appropriate expansion of $\phi_2$,
 which upon substitution into (\ref{eq:beta_eqn2}) can be used
 to isolate the key
resonant terms in the $\beta_1$ equation.
  The analogous steps are then repeated for the $\alpha_0$ equation. Finally,
a near-identity change of variables is constructed which maps the system for
$\alpha_0$ and $\beta_1$ to a new system (a {\it normal form} plus corrections)
 for which the dynamical behavior is more transparent.

$\Phi_2$ solves equation (\ref{eq:Phi2_eqn}). We shall require dispersive
decay estimates for $\Phi_2$ and these are most naturally obtained relative to
a time independent Hamiltonian.
 Since $\alpha_0(t)$ is expected to tend to a limit as
 $t\to\infty $ and since we are fixing a time interval $[0,T]$, it is natural
to use as {\it reference Hamiltonian}, the operator ${\cal H}_0(T)$. We now make
 use of the linear spectral theory of section 3 and decompose $\Phi_2$ into a
 part lying in the ``discrete subspace of ${\cal H}_0(T)$'':
\be
N_g({\cal H}_0(T))\oplus N({\cal H}_0(T)-\mu(T)) \oplus N({\cal H}_0(T)+\mu(T))
\label{eq:discrete_subspace}\ee
and a part lying 
 in ${\cal M}_1(T)$, the ``dispersive subspace of ${\cal H}_0(T)$'';
 see (\ref{eq:M1def}).

Let \cite{BP2,Cucc2} 
\be \Phi_2\ =\ k\ +\ n\ +\ \eta,\label{eq:Phi2kneta}\ee
   where
\ba
k &\equiv&
  \sum_{\xi_a\in N_g({\cal H}_0(T))}\langle\sigma_3\xi_a(T),\Phi_2\rangle\xi_a
 \label{eq:k_def}\\
n &\equiv&
  \sum_{\xi_b\in N({\cal H}_0(T)\mp\mu(T))}\langle\sigma_3\xi_b(T),\Phi_2\rangle
\xi_b
 \label{eq:n_def}\\
\eta &\equiv& P_c(T)\Phi_2\ =\ \Phi_2-k-n
\label{eq:eta_def}
\ea
Since
\ba \la\sigma_3\xi_a(t),\Phi_2(t)\ra&=&0,\ \xi_a\in N_g({\cal H}_0(t))\nn\\
\la\sigma_3\xi_b(t),\Phi_2(t)\ra&=&0,\ \xi_b\in N_g({\cal H}_0(t)\mp\mu(t)),
\nn\ea
$\xi(t)$ may be replaced by $\xi(t)-\xi(T)$ in the definitions of $k$ and $n$.
Inserting the expansion (\ref{eq:Phi2kneta}) into (\ref{eq:k_def})
 and (\ref{eq:n_def}) and defining
\ba p_{\rm null}(t,T) &=& \sum_{\xi_a\in N_g({\cal H}_0(T))}
 \langle\sigma_3(\xi_a(T)-\xi_a(t)),\cdot\rangle\xi_a\label{eq:pnull_def}\\
p_{\rm neut}(t,T) &=& \sum_{\xi_b\in N({\cal H}_0(T)\mp\mu(T))}
 \langle\sigma_3(\xi_b(T)-\xi_b(t)),\cdot\rangle\xi_a\label{eq:pneutral_def}
\ea
we have that $k$ and $n$ may be expressed in terms of $\eta$ as follows:
Then,
\be
\left[\ I\ -\ \left(\begin{array}{cc} p_{\rm null}(t,T) & p_{\rm null}(t,T)\\
                                       p_{\rm neut}(t,T) & p_{\rm neut}(t,T)
                     \end{array}\right)\ \right]
        \left(\begin{array}{c} k \\ n\end{array}\right)\ =\
\left(\begin{array}{c} p_{\rm null}(t,T)\eta \\ p_{\rm neut}(t,T)\eta
   \end{array}\right).
\label{eq:knofeta}
\ee
Therefore, we have
\begin{prop}\label{prop:phi2kneta}
There exists $\varepsilon_0>0$ such that if $|\alpha_0(t)-\alpha_0(T)|
 <\varepsilon_0$ then the relation (\ref{eq:knofeta}) can be inverted and
we have
\be \Phi_2\ =\ \Phi_2[\eta]\ =\ k[\eta]+n[\eta]+\eta,\nn\ee
where $\Phi_2$ is linear in $\eta$ and continuous in the weighted
 (local decay)  norm
  of $f\mapsto \|\la x\ra^{-\sigma}f\|_2$.
\end{prop}
The statement about continuity in the weighted norm
follows from the spatial localization of the generalized eigenfunctions.

\begin{rmk}\label{rmk:knalpha}
 Note that since $|\xi_a(T)-\xi_a(t)|\le
 C\cE_0^{1\over2}|\alpha_0(T)-\alpha_0(t)|,$
we have the simple estimate:
\be |k|,\ |n|\ \le\ C |\alpha_0(T)-\alpha_0(t)|\  
 ||\la x\ra^{-\sigma}\eta(t)||_2.\nn\ee
Anticipating that for $t$ very large,
  $|\alpha_0(T)-\alpha_0(t)|\sim t^{-{1\over2}}$ and
$||\la x\ra^{-\sigma} \Phi_2(t)||_2\sim  t^{-{1\over2}}$, it follows that
$|k|,\ |n| \sim t^{-1}$. Thus,  $|k|$ and $|n|$,  
 are expected to decay faster than
 $\eta$.
\end{rmk}
Finally, $\eta$ satisfies the following evolution equation obtained from
(\ref{eq:Phi2_eqn}) by explicitly introducing the reference Hamiltonian,
 ${\cal H}_0(T)$, and applying the projection  $P_c(T)$ to the equation.
\ba
i\D_t\eta\ &=&\ {\cal H}_0(T)\eta\nn\\
&+& \left( {\cal H}_0(t)-{\cal H}_0(T) \right)\eta +
  \left( E_1(t)-E_0(t) \right)P_c(T)\sigma_3
             \left(\begin{array}{c}\Psi_1(t)\\ c.c.\end{array}\right)
\nn\\
&+&\ \lambda P_c(T)\sigma_3
 \left(\begin{array}{c}2|\Psi_0|^2\Psi_1+\Psi_0^2\ol{\Psi_1} \\
 c.c. \end{array}\right)\nn\\
&+& \lambda P_c(T) \sigma_3\left(\begin{array}{c}2\Psi_0|\Psi_1+
 \pi_1\Phi_2[\eta]|^2 +
 \ol{\Psi_0}(\Psi_1+\pi_1\Phi_2[\eta])^2\\
 c.c. \end{array}\right)\nn\\
&+&\ \lambda P_c(T)\sigma_3\left(\begin{array}{c}|\Psi_1+\pi_1\Phi_2[\eta]|^2
 (\Psi_1+\pi_1\Phi_2[\eta]) -
 |\Psi_1|^2\Psi_1\\
 c.c. \end{array}\right)\nn\\
&-&\ i P_c(T)\left(\begin{array}{c}
 \vec\nabla_1\Psi_1(t)\cdot\D_t\vec\alpha_1(t) +
 \vec\nabla_0\Psi_0(t)\cdot\D_t\vec\alpha_0(t)\\
 c.c. \end{array}\right)
\label{eq:eta_eqn}
\ea

\section{Normal Form and Master Equations}\label{sec:nfandmaster}

In section 4 we decomposed the solution, $\phi$, in terms of coordinates
 $\alpha_0(t)$ and $\alpha_1(t)$ along manifolds of nonlinear bound states
and $\phi_2$, a correction which lies in a time-dependent subspace, ${\cal M}_1(t)$,
 of continuum modes. $\phi_2(t)$ was then decomposed into its discrete ($k$ and $n$)
 and continuous ($\eta$) components with respect to a time \underline{independent}
 Hamiltonian, ${\cal H}_0(T)$. We also observed that $k$ and $n$ are determined  
 by and are expected to be more rapidly decaying than  $\eta$ (Proposition 
 \ref{prop:phi2kneta} and Remark \ref{rmk:knalpha}).
  Therefore, the evolution of $\phi$ is determined  
 by $\alpha_0$, $\alpha_1$ and 
 $\eta$. Finally, the fast oscillations of $\alpha_1$ are removed by the introduction
 of $\beta_1\ =\ [ X(t)\vec\alpha_1 ]_1\sim e^{-i\lambda_+t}\alpha_1$. 

We now seek a form of the system for $\alpha_0$ and $\beta_1$ from which the large 
time dynamics can be deduced. We obtain this ``normal form'' by first solving
 for $\eta$ (see (\ref{eq:etaformal}),(\ref{eq:eta_eqn}) )
 as a functional of $\alpha_0$, $\beta_1$ and the initial data 
 $\eta(0)$
  and then substituting an appropriate expansion (see sections 
 \ref{sec:nfandmaster} and \ref{sec:stab}) 
into  the equations for $\alpha_0$ and $\beta_1$. 

\begin{prop}\label{prop:normalform} {\bf The Normal Form}
There exists a near identity change of variables
\be
 \left(\begin{array}{c}\tilde\alpha_0 \\ \tilde\beta_1\end{array}\right)
\equiv \left(\begin{array}{c}\alpha_0 \\ \beta_1\end{array}\right)\ 
+\ \left(\begin{array}{c}J_\alpha[\alpha_0,\beta_1,t] \\ 
            J_\beta[\alpha_0,\beta_1,t]\end{array}\right)\
\label{eq:ata}\ee
where 
\be J_k[\alpha_0,\beta_1,t]={\cal O}(|\alpha_0|^2+|\beta_1|^2),\ k=\alpha,\beta
 \nn\ee
and bounded uniformly in $t$, and such that
\ba
i\D_t\ta0\ &=&\ (c_{1022}+i\Gamma_{\omega})\ 
 |\tb1|^4\ta0\ +\ 
 F_\alpha[\ta0,\tb1,\eta,t]\nn\\
i\D_t\tb1\ &=&\ (c_{1121}-2i\Gamma_{\omega})
 |\ta0|^2|\tb1|^2\tb1\ 
 +\ F_\beta[\ta0,\tb1,\eta,t].
\label{eq:normalform}\ea
The properties of $F_\alpha$ and $F_\beta$ are briefly discussed
in the remark
 \ref{rmk:approach} following Corollary
 \ref{cor:nlPME} below and described in detail in
 section \ref{sec:stab}.

Furthermore, 
\ba \Gamma\ &=&\ \Gamma_{\omega_*}\ + {\cal O}(|\alpha_0(T)|^2) > 0,
 \ {\rm where},\nn\\
 \Gamma_{\omega_*} &\equiv& \lambda^2
\pi\langle\psi_{0*}\psi_{1*}^2,\delta(H-\omega_*)\psi_{0*}\psi_{1*}^2\rangle
 > 0\label{eq:Gamma}\nn\\
\omega_*\ &=&\ 2E_{1*}-E_{0*}\label{eq:omegastar}
\ea
The coefficients $c_{klmn}=\cO_0^{(0,1)}$
 are real constants multiplying monomials
of the form $\alpha_0^k\bar\alpha_0^l\beta_1^m\bar\beta_1^n$.
\end{prop}

\begin{rmk}
Due to our choice of the phase correction, $\tTheta(t)$ 
 (see (\ref{eq:tTheta}), a term of the
form $c_{1011}\cO_0^{(0,1)}|\tb1|^2\ta0$,
  is absent from the differential equation for
$\ta0$ in (\ref{eq:normalform}).
\end{rmk}

Now  let
\be P_0\equiv |\tilde\alpha_0|^2 \ {\rm and}\
    P_1\equiv |\tilde\beta_1|^2
\label{eq:power_def}\ee
 denote the (renormalized) ground state and excited state powers.
Then, by (\ref{eq:normalform}) we have the {\it Nonlinear Master Equation}:
\begin{cor}\label{cor:nlPME}
\ba
{dP_0\over dt} &=& 2\Gamma P_0P_1^2\  +\  R_0
 \label{eq:P0eqn}\\
{dP_1\over dt} &=& -4\Gamma P_0P_1^2 \ +\ R_1.
\label{eq:P1eqn}
\ea
where
\ba R_0 &=& R_0[\ta0,\tb1,\eta,t]=  2\Im (\ol{\ta0}F_\alpha),\nn\\
    R_1 &=& R_1[\ta0,\tb1,\eta,t]=  2\Im (\ol{\tb1}F_\beta)
\label{eq:Rjdef}
\ea
\end{cor}
A more precise and revealing variant of Corollary \ref{cor:nlPME} is 
 Proposition \ref{prop:Master}, which is
stated and proved in section \ref{sec:stab}.

\bigskip

\begin{rmk}\label{rmk:approach}
The terms $F_\alpha$ and $F_\beta$ are such that $R_0$ and $R_1$ are 
not small perturbations of the leading order terms in (\ref{eq:P0eqn},
\ref{eq:P1eqn}) for all $t\ge0$. In fact there are three time intervals 
defined in terms of transition times $t_0$ and $t_1$ 
 (see section \ref{sec:stab}), 
  in which we consider the system (\ref{eq:P0eqn}-\ref{eq:P1eqn}): $I_0=[0,t_0]$,
 $I_1=[t_0,t_1]$ and $I_2=[t_1,\infty)$.
 It is only for sufficiently large time, ($t\in I_2$), 
 where $R_0$ and $R_1$ are negligible.
 The behavior on short ($0\le t\le t_0$) and intermediate ($t_0\le t\le t_1$)
time scales can be very different.
 We go into the details
of $R_0$ and $R_1$ in section~\ref{sec:stab} but wish to make some remarks 
at this stage which indicate our approach. 

If we drop the terms $R_j$
then we have a flow, which evolves in the first quadrant of the $P_0-P_1$
plane according to:
\ba
{dp_0\over dt} &=& 2\Gamma p_0p_1^2\  
 \label{eq:p0eqn}\\
{dp_1\over dt} &=& -4\Gamma p_0p_1^2,
\label{eq:p1eqn}
\ea
 where solutions for typical data converge to the $p_0$ axis with a rate
$\la t\ra^{-1}$. In order for the  corrections coming from $R_0$ and $R_1$ to 
be small, {\it intuitively} it is sufficient that
\be
R_j\ \sim \cE_0^\rho\left(\ P_0 P_1^2 + \la t\ra^{-3}\ \right),
\label{eq:intuitI2}
\ee
where $\cE_0$ is small and $\rho>0$.  
This is what we show for $t\ge t_1$. 
  For the intermediate time range,
 $t_0\le t\le t_1$, we show that the behavior is controlled
by the system:
\ba
\frac{dP_0}{dt}\ &=& 2\Gamma P_0P_1^2 + \cO(\la t\ra^{-3})\nn\\
\frac{dP_1}{dt}\ &=& -4\Gamma P_0P_1^2 + \cO\left(\sqrt{P_0}P_1^m\right)\ 
 +\ \cO(\la t\ra^{-3}),
\label{eq:I1nf}\ea
where $m\ge3$.
Therefore, for intermediate times we need to show: 
\ba
R_0\ &\sim& \cE_0^\rho\left(\ P_0 P_1^2 + \la t\ra^{-3}\ \right)\nn\\
R_1\ &\sim& \cE_0^\rho\left(\ P_0 P_1^2 + \sqrt{P_0}P_1^m + \la t\ra^{-3}\ 
 \right).
\label{eq:intuitI1}
\ea
What makes the analysis subtle is the dependence of $R_j$ on $P_0$ and $P_1$
 in a manner which is {\it nonlocal in time}.  That is, 
\be
\eta \sim \int_0^t e^{-i\cH_0(t-s)}P_c(\cH_0)\chi
  \alpha_0^{m_1}\beta_1^{m_2}\ol{\alpha_0}^{m_3}\ol{\beta_1}^{m_4} \ ds 
\nn\ee
Local in time terms are simple to dominate by the leading terms.
 However, nonlocal terms require
careful analysis. Note in particular, that due to the 
 ``history dependence'' of such terms, being expressed as 
time integrals from $0$ up to $t$, an analysis of the effect of such  
terms for $t\ge t_1$ requires use of estimates on other time regimes
 $t\le t_0$ and $t_0\le t\le t_1$ as well. Furthermore, there are no
decay estimates on either $P_0$ or $P_1$ in the intermediate interval
 $t_0\le t\le t_1$ or on the size of this interval.
\end{rmk}

The normal form of Proposition \ref{prop:normalform} is essentially
 the Poincar\'e - Dulac normal form which can be constructed along
the lines explictly implemented in \cite{SW99}; see also \cite{Arnold}.
We now give a detailed outline of the procedure with explicit illustrative
 detail of key points concerning the treatment of resonant and nonresonant
terms.
\medskip

\noindent{\bf Resonant terms and removal nonresonant terms:}
Here we illustrate, by way of a simple example, how non-resonant terms
can be removed by near identity changes of variables. Consider the 
scalar ordinary differential equation
\be A'(t)\ =\ |A(t)|^2e^{i\Omega t}\label{eq:cODE}\ee
where $A(t)$ is a complex valued function. We shall introduce a change of 
variables $A\mapsto \tA = A + q_2(A,{\bar A}, t)$, where 
$q_2(A,{\bar A},t)=\cO(|A|^2)$ and $q_2(A,{\bar A},t+{2\pi\over\Omega})
 = q_2(A,{\bar A},t)$, which is therefore approximately the identity for 
$|A|$ small, and such that 
\be \tA'(t)\ =\ {i\over\Omega}|\tA(t)|^2\tA(t)+ 
 {i\over\Omega} e^{2i\Omega t}|\tA(t)|^2 \bar{\tA(t)} + 
 E_4(\tA(t),\bar{\tA(t)},t),
 \label{eq:cODE1}\ee
Here, $E_4$ is $2\pi/\Omega$ periodic in $t$ and 
\be E_4(A(t),\tA(t),t)\ =\ \cO(|\tA(t)|^4)\label{eq:E4estimate}\ee

The change of variables can be derived by elementary means. Integration 
of (\ref{eq:cODE}) gives:
\ba
A(t)-A(0)\ &=&\ \int_0^t |A(s)|^2e^{i\Omega s}ds\nn\\
           &=&\ \int_0^t |A(s)|^2\frac{1}{i\Omega}\frac{d}{ds}e^{i\Omega s}ds
                                                \nn\\
           &=&\ |A(s)|^2 \frac{1}{i\Omega}e^{i\Omega s}\left.\right|_0^t\ 
 -\ \int_0^t \frac{1}{i\Omega}e^{i\Omega s}\frac{d}{ds} |A(s)|^2 \ ds\nn\\
           &=&\ |A(s)|^2 \frac{1}{i\Omega}e^{i\Omega s}\left.\right|_0^t\
 - \frac{1}{i\Omega}\int_0^t e^{i\Omega s}\bar{A(s)}|A(s)|^2e^{i\Omega s} ds
 - \frac{1}{i\Omega}\int_0^t |A(s)|^2A(s) ds\nn\\
&&\label{eq:manip}
\ea
Define $\tA(t)= A(t)-(i\Omega)^{-1}|A(t)|^2e^{i\Omega t}$. Then,
$\tA$ satisfies the renormalized ODE, in which resonant quadratic terms
have been removed. The process can be repeated; by introducing further 
changes of variables $\tA\mapsto \tA_1=\tA +$\  higher order in $\tA$ and period
in $t$, non-resonant (oscillatory) cubic terms  can be removed to obtain:
\be
\tA_1'(t)\ =\ \frac{i}{\Omega}|\tA_1(t)|^2\tA_1(t)+ ik|\tA_1(t)|^4\tA_1^(t)
 +\ \dots,
\label{eq:tA1eqn}\ee
where $k$ is real. That the coefficients in the first to terms of this 
 {\it normal form} are purely imaginary implies that, to this order, 
 the amplitude $|\tA_1(t)|$ is independent in time. This is the 
 typical situation of the norm form finite dimensional Hamiltonian systems, 
in which resonances occur between isolated discrete frequencies. 

We next examine resonances between discrete and frequencies and the continuum of
 frequencies, associated with the continuous spectral (dispersive) 
 part of $\cH_0$. These can introduce {\it nonconservative} terms into 
 the normal from (via coefficients with real as well as imaginary parts),
which are responsible for energy transfer between discrete modes (bound states)
and radiation. 

\medskip

\noindent{\bf Nonconservative resonant terms and energy transfer:}\  
 We explain how to find the key resonant 
 energy transfer terms, the leading terms in (\ref{eq:normalform}). 
 These are terms
responsible for the exchange of energy among the nonlinear ground and excited
states mediated by interaction with continuums modes. We focus on the $\beta_1$
 equation. Analogous considerations apply to the $\alpha_0$ equation.

Equation (\ref{eq:beta_eqn2}) can be written the following compact form:
\be
i\D_t\beta_1 = \sum_{p,q,r}C_{pqr}\beta_1^p\ol{\beta_1}^qe^{-i\omega_r t}
 \la \chi_{pqr},\eta\ra + \cO(\eta^2) +...,
\label{eq:b1formal}
\ee
where $C_{pqr}$ are of order $1$, $\alpha_0$ or higher order in $\alpha_0$,
 $\omega_r\in\{\pm\lambda_\pm,\pm 2\lambda_\pm, 0\}$, and $\chi_{pqr}$ denote
functions which are exponentially localized in space.
The equation for $\alpha_0$ has a similar structure.
The equation for $\eta = \eta_0 + \eta_1 + \eta_2$, can be formally solved
giving:
\ba
\eta &=& \cO(\eta_0) +  \cO(\eta_0^2) + \cO(\eta^2)\nn\\
 &+& \sum_{p_1, q_1, r_1} D_{p_1q_1r_1}\bG\left(\ \beta_1^{p_1}\ol{\beta_1}^{q_1}
 e^{-i\nu_{r_1} s}\ \tchi_{p_1q_1r_1}\ \right)
\label{eq:etaformal}
\ea
Here, $\bG$ denotes the operator
\be
f\ \mapsto\ \bG f\ \equiv -i\int_0^t e^{-i\cH_0(t-s)}\ P_c\ f(s)\ ds.
\label{eq:G_def}
\ee
We insert the expansion for $\eta$, (\ref{eq:etaformal}) into
the terms involving inner products $\la\chi_{pqr},\eta\ra$
 in (\ref{eq:b1formal}) and in the corresponding equation for
  $\alpha_0$. This yields
a coupled system for $\alpha_0$ and $\beta_1$ which is closed up to 
 higher order. The terms in the resulting equations  are of the form
\be
\sum_{p,q,r}\sum_{p_1,q_1,r_1} C_{pqr} D_{p_1q_1r_1}
 \beta_1^p\ol{\beta_1}^q e^{-i\omega_r t}\left\la\chi_{pqr}\ ,\  
 \bG\beta_1^{p_1}\ol{\beta_1}^{q_1}e^{-i\nu_{r_1} s}\ \tchi_{p_1q_1r_1}\right\ra.
\label{eq:terms}
\ee
We now use the integration by parts lemma:
\begin{lem}\label{lem:ibp}
\ba
\int_0^t e^{iAs} f(s)  ds &=&
 \lim_{\delta\downarrow 0}\int_0^t e^{i(A\pm i\delta)s} f(s) ds\nn\\
&=& -i(A\pm i0)^{-1}e^{iAt}f(t) +i(A\pm i0)^{-1}f(0)\nn\\
&+&i(A\pm i0))^{-1}\int_0^t e^{iAs} f'(s) ds \label{eq:ibp}
\ea
\end{lem}
Applying this lemma, we obtain
\ba
&& e^{-i\omega_r t}\left\la\chi_{pqr}\
 \bG\beta_1^{p_1}\ol{\beta_1}^{q_1}e^{-i\nu_{r_1} s}\ \tchi_{p_1q_1r_1}\right\ra
\nn\\
&=& - e^{-i(\omega_r+\nu_{r_1})t}\beta_1^{p_1}\ol{\beta_1}^{q_1}
 \left\la\chi_{pqr}\ , (\cH_0(T)-\nu_{r_1}\mp i0)^{-1} P_c(T)
 \tchi_{p_1q_1r_1}\right\ra
\nn\\
&+& \beta_1^{p_1}(0)\ol{\beta_1}^{q_1}(0)
 \left\la\chi_{pqr}\ , (\cH_0(T)-\nu_{r_1}\mp i0)^{-1}e^{-i\cH_0(T)t}
 P_c(T) \tchi_{p_1q_1r_1} \right\ra\nn\\
&+& \int_0^t\left\la \chi_{pqr}\ , 
 (\cH_0(T)-\nu_{r_1}\mp i0)^{-1}\ e^{-i\cH_0(T)(t-s)} e^{-i\nu_{r_1}s}
  P_c(T) {d\over ds}
 \left( \beta_1^{p_1}\ol{\beta_1}^{q_1}\tchi_{p_1q_1r_1} \right) ds
 \right\ra
\nn\\
\label{eq:terms_exp}
\ea
We first focus on the first term in the expansion (\ref{eq:terms_exp}).
This first contributes a resonant term, which cannot be transformed by  a
near identity transformation  to higher order if
\be \omega_r + \nu_{r_1} = 0.\label{eq:res1}\ee
Now consider such a resonant term. We find that in the $\beta_1$ equation
 they are of the form:
\be
 -|\alpha_0|^{2a_1}|\beta_1|^{2b_1} \beta_1
 \left\la\vec\chi\ , (\cH_0 - \nu_{r_1}\mp i0)^{-1}P_c \vec\chi
 \right\ra\label{eq:sampleres}
\ee
where $\nu_{r_1} = -\omega_r \in \{ \pm\lambda_\pm, \pm 2\lambda_\pm, 0 \}$.

There are two cases to consider:
(I) $\nu_{r_1}$ {\it not in} the continuous spectrum of $\cH_0$ and
(II) $\nu_{r_1}$ {\it in} the continuous spectrum of $\cH_0$
\footnote{In case  (II)
 we consider the generic case where if $\nu_{r_1}$ lies in the
interior of the continuous spectrum of $\cH_0$.}.
If $\nu_{r_1}$ is not in the continuous spectrum of $\cH_0$ then
the inner produce in  (\ref{eq:sampleres}) does not involve a singular limit
and we get the limit
\be
\left\la\vec\chi\ , (\cH_0-\nu_{r_1})^{-1}P_c \vec\chi
 \right\ra.
\ee
In this case, the coefficient of $|\alpha_0|^{2a_1}|\beta_1|^{2b_1} \beta_1$ 
 is {\it real}.  
Such a term results only in a nonlinear distortion of the phase of $\beta_1$
 and does not effect the amplitude.
If $\nu_{r_1}$  is in the interior of the continuous spectrum of $\cH_0$
then the limit is singular. We choose the plus sign ($+i0$) if we study
the evolution for $t>0$ and the negative sign ($-i0$) for $t<0$. This choice
related to the condition of outgoing radiation explained below; see
also \cite{SW99}, for example. Evaluation of this singular limit
gives:
\ba
&& \left\la\chi\ , (\cH_0-\nu_{r_1}\mp i0)^{-1}P_c \chi
 \right\ra
 = \tLambda + i \tGamma,\ \ {\rm where}\nn\\ 
&& \tGamma = \pi\left\la\chi\ , \delta(\cH_0-\nu_{r_1}) \chi\right\ra, \ \
\tLambda = \left\la\chi\ ,\ {\rm P.V.}\ (\cH_0-\nu_{r_1})^{-1}\chi\right\ra
\nn\ea
Contributions to the  imaginary part are therefore
 responsible for a change in amplitude (here damping of $\beta_1$).

{\it Now, when does a frequency $\nu$ lie in the continuous spectrum of $\cH_0$?}
By  Proposition \ref{prop:Hspectrum}, we must have $\nu>-E_0=|E_0|$
or $\nu< E_0$. By (\ref{eq:lambda_def}), $\lambda_\pm \sim \pm(E_1-E_0)$.
Since $\nu$ varies over the frequencies $0,\pm\lambda_\pm,\pm 2\lambda_\pm$
we find that $\nu_{r_1}=\pm 2\lambda_\pm = -\omega_r$ resonances are in the
 continuous spectrum and therefore are those
giving rise to energy transfer, provided $2E_1-E_0>0$; 
 see (\ref{eq:omegastar}). 
We now embark on the details.

\subsection{Expansion of $\eta$}

We expand  $\eta$ as follows:
\be \eta(t)\ =\ \eta_0(t) +\eta_1(t) + \eta_2(t),\label{eq:eta-exp}\ee
where $\eta_0(t)$ corresponds to the linear homogeneous evolution 
with initial data $\eta(0)=P_c(T)\Phi_2(0)$ and $\eta_1$ solves 
 the inhomogeneous linear equation driven by  $\alpha_0$, 
 $\alpha_1$ and $\eta_0(t)$.

\noindent\underline{Equation for $\eta_0(t)$:}
\ba
i\D_t\eta_0\ &=&\ {\cal H}_0(T)\eta_0\nn\\
\eta_0(0)\ &=&\ P_c(T)\Phi_2(0)\label{eq:eta0ivp}\ea
Thus,
\be \eta_0(t)\ =\ e^{-i{\cal H}_0(T) t}P_c(T)\Phi_2(0)\label{eq:eta0}\ee

\noindent\underline{Equation for $\eta_1(t)$:}

\ba
i\D_t\eta_1\ &=&\ {\cal H}_0(T)\eta_1\ +\ 
 P_c(T)\left[{\cal H}_0(t)-{\cal H}_0(T)\right]\Phi_2[\eta_0] 
+\ E_{10}(t)P_c(T)\sigma_3
 \left(\begin{array}{c} \Psi_1\\ c.c.\end{array}\right) \nn\\
&+&\  \lambda P_c(T)\sigma_3
\left(\begin{array}{c} 2|\Psi_0|^2\Psi_1 + \Psi_0^2\ol{\Psi_1}
      \\ c.c. \end{array} \right)\nn\\
&+&\ \lambda P_c(T)\sigma_3
\left(\begin{array}{c}
 2\Psi_0|\Psi_1 +\pi_1\Phi_2[\eta_0]|^2 + 
 \ol{\Psi_0}(\Psi_1+\pi_1\Phi_2[\eta_0])^2\\ c.c.
\end{array}\right)\nn\\
&+&\ \lambda P_c(T)\sigma_3
\left(\begin{array}{c}
|\Psi_1 +\pi_1\Phi_2[\eta_0]|^2 (\Psi_1 +\pi_1\Phi_2[\eta_0])
 - |\Psi_1|^2\Psi_1\\ c.c.
\end{array}\right)\nn\\
&-&\ i P_c \left(\begin{array}{c}
 \vec\nabla_1\Psi_1(t)\cdot\D_t\vec\alpha_1(t) +
 \vec\nabla_0\Psi_0(t)\cdot\D_t\vec\alpha_0(t)\\
 c.c. \end{array}\right)
\label{eq:eta1_eqn}\ea
The initial data for $\eta_1$ is $\eta_1(0)=0$.

\begin{rmk}
A direct computation using the biorthogonal decomposition
of the discrete subspaces of  
 $\cH_0$ and $\cH_0^*$  of section \ref{sec:linearization}
yields that the second term in (\ref{eq:eta_eqn}) is of a 
higher order than is explicit:
\be
E_{10}(t)P_c(T)\sigma_3
 \left(\begin{array}{c} \Psi_1\\ c.c.\end{array}\right)
 = \cO\left( |\alpha_0|^2|\alpha_1| +|\alpha_1|^3 \right)\  
 \left(\chi_0^{(0)} + \chi_0^{(1)} + \chi_0^{(0,1)} \right)
\label{eq:direct}\ee
for $|\alpha_0|<<1$ and $|\alpha_1|\downarrow 0$.
\end{rmk}
\noindent\underline{Equation for $\eta_2(t)$:}

\ba
i\D_t\eta_2\ &=&\ {\cal H}_0(T)\eta_2\ +\
P_c(T)\left[{\cal H}_0(t)-{\cal H}_0(T)\right]\Phi_2[\eta_1+\eta_2]\nn\\
&+&\ \lambda P_c(T)\sigma_3
\left(\begin{array}{c}
2\Psi_0\left[(\Psi_1+\pi_1\Phi_2[\eta_0])
\overline{\pi_1\Phi_2[\eta_1+\eta_2]} + c.c.\right]\\ c.c.\end{array}\right)  
\nn\\ &+&\ \lambda P_c(T)\sigma_3
\left(\begin{array}{c}
2\Psi_0|\pi_1\Phi_2[\eta_1+\eta_2]|^2\ +\ 
 2\ol{\Psi_0}
 (\Psi_1+\pi_1\Phi_2[\eta_0]) \pi_1\Phi_2[\eta_1+\eta_2]
\\ c.c. \end{array} \right)\nn\\
&+&\  \lambda P_c(T)\sigma_3\left(\begin{array}{c}
\ol{\Psi_0}(\pi_1\Phi_2[\eta_1+\eta_2])^2 \\ c.c.
\end{array} \right)\nn\\
&+&\ \lambda P_c(T)\sigma_3\left(\begin{array}{c}
2|\Psi_1+\pi_1\Phi_2[\eta_0]|^2\pi_2\Phi[\eta_1+\eta_2]\\
c.c. \end{array} \right)\nn\\
&+&\ \lambda P_c(T)\sigma_3\left(\begin{array}{c}
(\pi_1\Phi_2[\eta_1+\eta_2])^2(\ol{\Psi_1}+
 \overline{\pi_1\Phi_2[\eta_0]} )\ +\ 
\overline{\pi_1\Phi_2[\eta_1+\eta_2]} 
 (\Psi_1+\pi_1\Phi_2[\eta_0])^2\\ c.c.
\end{array} \right)\nn\\
&+&\ \lambda P_c(T)\sigma_3\left(\begin{array}{c}
2(\Psi_1+\pi_1\Phi_2[\eta_0]) |\pi_1\Phi_2[\eta_1+\eta_2]|^2\
 +\ |\pi_1\Phi_2[\eta_1+\eta_2]|^2 \pi_1\Phi_2[\eta_1+\eta_2]\\ c.c.
\end{array} \right)
\nn\\ \label{eq:eta2_eqn}
\ea

We expect that $\eta_2={\cal O}(\langle t\rangle^{-1})$. 
Let
\be \eta_2\ =\ \eta_{2a} + \eta_{2b}.\label{eq:eta2_def}\ee
By construction we will show that 
   $\eta_{2a}={\cal O}(\langle t\rangle^{-1})$
and $\eta_{2b}={\cal O}(\langle t\rangle^{-{3\over2}})$.

\ba
i\D_t\eta_{2a}\ &=&\ {\cal H}_0(T)\eta_{2a}\ +\ 
 P_c(T)[{\cal H}_0(t)-{\cal H}_0(T)]\Phi_2[\eta_1]\nn\\
&+&\ 2\lambda P_c(T)\sigma_3\left(\begin{array}{c}
\Psi_0(\Psi_1\pi_1\ol{\Phi_2[\eta_1]} +
 \ol{\Psi_1}\pi_2\Phi_2[\eta_1]) + \Psi_0|\pi_1\Phi_2[\eta_1]|^2\\ c.c.
\end{array}\right)\nn\\
&+&\ \lambda P_c(T)\sigma_3\left( \begin{array}{c} 
2\ol{\Psi_0}\pi_1\Phi_2[\eta_1](\Psi_1 + \pi_1\Phi_2[\eta_0])
  + \ol{\Psi_0}(\pi_1\Phi_2[\eta_1])^2 \\ c.c.
\end{array}\right)\nn\\
&+& \lambda P_c(T)\sigma_3\left( \begin{array}{c}
2|\Psi_1+\pi_1\Phi_2[\eta_0]|^2\pi_1\Phi_2[\eta_1] 
+(\Psi_1+\pi_1\Phi_2[\eta_0])^2\ol{\pi_1\Phi_2[\eta_1]}\\ c.c
\end{array}\right)\nn\\
&+& \lambda P_c(T)\sigma_3\left( \begin{array}{c}
2(\Psi_1+\pi_1\Phi_2[\eta_0])\ |\pi_1\Phi_2[\eta_1]|^2 
+\ol{(\Psi_1+\pi_1\Phi_2[\eta_0])}\ (\pi_1\Phi_2[\eta_1])^2\\ c.c
\end{array}\right)\nn\\
&+& \lambda P_c(T)\sigma_3\left( \begin{array}{c}
|\pi_1\Phi_2[\eta_1]|^2 \pi_1\Phi_2[\eta_1]\\ c.c.
\end{array}\right)
 \label{eq:eta2a_eqn}\ea
\medskip


\subsection{ Normal form and master equations}

Using (\ref{eq:direct})
 and explicitly inserting in (\ref{eq:eta1_eqn}) the representation
\be \alpha_1\ =\ \left( X(t)\beta(t)\right)_1
 = e^{-i\lambda_+(T)t}\beta_1 + 
\cO_0^{(0,1)}\alpha_0^2(T)E_{10}^{-1}(T) e^{-i\lambda_-(T)t}\ol{\beta_1}
\label{eq:a1b1}\ee
gives the following equation for $\eta_1$: 
\ba
i\D_t\eta_1\ &=&\ {\cal H}_0(T)\eta_1\ +\
 P_c(T)\left( {\cal H}_0(t)-{\cal H}_0(T) \right)\Phi_2[\eta_0]\nn\\
&+& E_{10}(t)P_c(T)\sigma_3 (\ |\ta0|^2\chi_0^{(0)} +\ |\tb1|^2\chi_1^{(1)}\ )
 \nn\\
&& \times
 \left(\begin{array}{c} e^{-i\lambda_+(T)t}\beta_1 +
     \cO_0^{(0,1)}\alpha_0^2(T)E_{10}^{-1}(T)e^{-i\lambda_-t}\ol{\beta_1}
         \\ c.c.\end{array}\right)\nn\\
&+&  \lambda P_c(T)\sigma_3\psi_{0*}^2\psi_{1*}
\left(\begin{array}{c}
2|\alpha_0|^2\beta_1 e^{-i\lambda_+(T)t} + \alpha_0^2\ol{\beta_1}
 e^{i\lambda_+(T)t}\\ c.c. \end{array}\right)\nn\\
&+& \lambda P_c(T)\sigma_3\psi_{0*}^2\psi_{1*}
\left(\begin{array}{c}
2\alpha_0|\beta_1|^2 + \ol{\alpha_0}\beta_1^2e^{-2i\lambda_+(T)t}\\ c.c.
 \end{array}\right) + \vec{\cal R}_{\eta_1}, 
\label{eq:eta1_eqn2}\ea
with initial condition $\eta_1(0)=0$.
\bigskip

Substitution of $\eta_1$ into the equations (\ref{eq:a0eqn})
 and (\ref{eq:beta_eqn2}) for $\alpha_0$ and $\beta_1$ 
 gives rise to terms which make explicit the resonant exchange of energy
between the ground state and excited state. We next isolate the key terms 
 in the expansion of $\eta_1$ relating to this energy exchange. 

Let's begin with the $\beta_1$ equation, (\ref{eq:beta_eqn2}). Written 
out in greater detail we have:
\ba
i\D_t\beta_1 &=& 2\lambda\la\psi_{0*},\psi_{1*}^3\ra|\beta_1|^2\alpha_0         e^{i\lambda_+(T) t} + 
 2\lambda\la\psi_{0*}\psi_{1*}^2,\pi_2\Phi_2\ra\beta_1\alpha_0\nn\\
&+& \lambda\la\psi_{0*},\psi_{1*}^3\ra\beta_1^2\ol{\alpha_0}
  e^{-i\lambda_+(T)t}
+ {\bf 2\lambda\la\psi_{0*}\psi_{1*}^2,\pi_1\Phi_2\ra} ({\bf \ol{\beta_1}\alpha_0 
 e^{2i\lambda_+(T) t}} + \beta_1\ol{\alpha_0})\nn\\
&+& 
2\lambda\la\psi_{0*}\psi_{1*},|\pi_1\Phi_2|^2\ra\alpha_0 e^{i\lambda_+(T)t}
+ 
 \lambda\la\psi_{0*}\psi_{1*},(\pi_1\Phi_2)^2\ra\ol{\alpha_0}e^{i\lambda_+(T)t}
\nn\\
&+& \lambda\la\psi_{1*}^3,\pi_1\Phi_2\ra |\beta_1|^2e^{i\lambda_+(T)t} +
 \lambda \la\psi_{1*}^2,(\pi_1\Phi_2)^2\ra\ol{\beta_1}e^{2i\lambda_+(T)t}\nn\\
&+&\lambda\la\psi_{1*}^3,\pi_2\Phi_2\ra\beta_1^2 e^{-i\lambda_+(T)t}
 + 2\lambda\la\psi_{1*}^2,|\pi_1\Phi_2|^2\ra\beta_1 e^{-i\lambda_+(T)t} \nn\\
&+& \lambda\la\psi_{1*},|\pi_1\Phi_2|^2\pi_1\Phi_2\ra e^{i\lambda_+(T)t}
 + {\cal R}_\beta,\label{eq:beta_eqn3}\ea
where ${\cal R}_\beta$ is defined in (\ref{eq:Rbeta_def}).

We claim that the key term in (\ref{eq:beta_eqn3}) responsible for energy 
transfer is the term  
\be {\bf  2\lambda\la \psi_{0*}\psi_{1*}^2,\pi_1\Phi_2\ra\ol{\beta_1}\alpha_0 
 e^{2i\lambda_+(T)t} }\label{eq:betarespo} \ee
on the second line of (\ref{eq:beta_eqn3}).
To see this we decompose $\eta_1$ into ``resonant'' and ``nonresonant'' parts
\be \eta_1 = \eta_{1R} + \eta_{1NR}\nn\ee
The part $\eta_{1NR}$ gives rise to the key resonant energy transfer 
 term in the equation for $\beta_1$ and is the solution to the initial value 
  problem
\ba
i\D_t\eta_{1R} &=& {\cal H}_0(T)\eta_{1R} + \lambda P_c(T)\sigma_3
 \psi_{1*}^2\psi_{0*}\ol{\alpha_0}\beta_1^2 e^{-2i\lambda_+(T) t}
 \left(\begin{array}{c}1\\ 0\end{array}\right)\nn\\
\eta_{1R}(0) &=& 0.
\label{eq:eta1R_eqn}\ea
Solving (\ref{eq:eta1R_eqn}) using DuHamel's principle we have 
\ba
\eta_{1R}(t) &=& -i\lambda\int_0^t e^{-i{\cal H}_0(T)(t-s)}P_c(T)\sigma_3
  \psi_{1*}^2\psi_{0*}\ol{\alpha_0(s)}\beta_1^2(s)e^{-2i\lambda_+(T)s}
 \left(\begin{array}{c} 1 \\ 0\end{array}\right) ds
\nn\\
&=& -i\lambda e^{-i{\cal H}_0(T)t} \int_0^t e^{i[{\cal H}_0(T)-2\lambda_+(T)]s}
 P_c(T)\sigma_3 \psi_{1*}^2\psi_{0*}\ol{\alpha_0(s)}\beta_1^2(s) 
\left(\begin{array}{c} 1 \\ 0\end{array}\right) ds\nn\\
\label{eq:eta1R_solved}\ea
Recall that the continuous spectrum  of ${\cal H}_0(T)$ is given by points 
 $\omega$  such that $|\omega|\ge|E_0|$. By Proposition \ref{prop:xdef}, 
for $|\alpha_0|^2/E_{10}$ sufficiently small
\be \lambda_+(T) = E_{10}(T) + \cO_0^{(0,1)}|\alpha_0(T)|^2.\nn\ee
By the hypothesis (\ref{eq:omegastar}), $2E_{1*}-E_{0*}>0$, if $|\alpha_0|$ 
 is sufficiently small then
 $2\lambda_+(T)$ lies in the continuous spectrum of ${\cal H}_0(T)$. Therefore,
 (\ref{eq:eta1R_eqn}) is a resonantly forced system. We expand the solution
as follows. Let $\delta>0$ and set
\be
\eta_{1R}^\delta(t)\ =\ -i\lambda e^{-i{\cal H}_0(T)t} 
 \int_0^t e^{i[{\cal H}_0(T)-2\lambda_+(T)-i\delta]s}
 P_c(T)\sigma_3 \psi_{1*}^2\psi_{0*}\ol{\alpha_0(s)}
 \beta_1^2(s) \left(\begin{array}{c} 1 \\ 0\end{array}\right) ds.
\label{eq:eta1Rdelta_def}\ee
Then, $\eta_{1R}=\lim_{\delta\to0}\eta_{1R}^\delta$.

We now apply Lemma \ref{lem:ibp} to 
$\eta_{1R}$ with $A={\cal H}_0(T)-2\lambda_+(T)$.
The result is
\begin{prop}\label{prop:eta1R_expand}
 The limit $\lim_{\delta\to0}\eta_{1R}^\delta=\eta_{1R}$ exists in ${\cal S}'$
 and  
\ba
\eta_{1R}(t) &=& -\lambda e^{-i\lambda_+(T) t} \ol{\alpha_0(t)}\beta_1^2(t) 
 \left({\cal H}_0(T)-2\lambda_+(T)-i0\right)^{-1}P_c(T)
 \left(\begin{array}{c} 1 \\ 0\end{array}\right) \psi_{1*}^2\psi_{0*}\nn\\
&+& \lambda\ol{\alpha_0(0)}\beta_1^2(0) e^{-i{\cal H}_0(T)t}
  \left({\cal H}_0(T)-2\lambda_+(T)-i0\right)^{-1}P_c(T)
\left(\begin{array}{c} 1 \\ 0\end{array}\right) \psi_{1*}^2\psi_{0*}\nn\\
&+& \lambda e^{-i{\cal H}_0(T)t}
 \left({\cal H}_0(T)-2\lambda_+(T)-i0\right)^{-1}\cdot\nn\\
&&\ \ \  \int_0^t e^{i[{\cal H}_0(T)-2\lambda_+(T)]s}P_c(T)
 \left(\begin{array}{c} 1 \\ 0\end{array}\right) \psi_{1*}^2\psi_{0*}
{d\over ds}\left(\ol{\alpha_0(s)}\beta_1^2(s)\right) ds\nn\\
&=& \eta_{1Ra} + \eta_{1Rb} + \eta_{1Rc}
\label{eq:eta1R_expanded}
\ea
\end{prop}
We substitute (\ref{eq:eta1R_expanded}) into the key term 
 (\ref{eq:betarespo}) in (\ref{eq:beta_eqn3}).
\be
2\lambda\la \psi_{0*}\psi_{1*}^2,\pi_1\Phi_2[\eta]\ra\ol{\beta_1}\alpha_0
 e^{2i\lambda_+(T)t} = 
2\lambda\la \psi_{0*}\psi_{1*}^2,\pi_1\eta_{1Ra}\ra\ol{\beta_1}\alpha_0
 e^{2i\lambda_+(T)t} + \bR.
\label{eq:resp1}\ee
Here, $\bR$ denotes rapidly dispersively decaying terms plus higher order
terms in $|\alpha_0\beta_1|$. 
By Proposition \ref{prop:eta1R_expand} and the Plemelj formula
(\ref{eq:Plemelj})
\ba
&& 2\lambda\la \psi_{0*}\psi_{1*}^2,\pi_1\eta_{1Ra}\ra\ol{\beta_1}\alpha_0
 e^{2i\lambda_+(T)t}\nn\\
&=& -2\lambda^2\left\la \psi_{0*}\psi_{1*}^2,
 \pi_1({\cal H}_0(T)-2\lambda_+(T)-i0)^{-1}P_c(T)
 \left(\begin{array}{c}1 \\ 0\end{array}\right)\psi_{0*}\psi_{1*}^2\right\ra
 |\alpha_0|^2|\beta_1|^2\beta_1\nn\\
&=& -2(\Lambda+i\Gamma) |\alpha_0|^2|\beta_1|^2\beta_1,
\label{eq:resp2}\ea
where (using that $E_{0}-2E_{10}=E_0-2(E_1-E_0)= 
 \omega_*+{\cal O}(|\alpha_0(T)|^2)$)
we have 
\ba
\Lambda &=& \Lambda_{\omega_*}\ \cO_0^{(0)} \nn\\
&=& \lambda^2\la \psi_{0*}\psi_{1*}^2,
 {\rm P.V.} ( H-\omega_*)^{-1} 
 \psi_{0*}\psi_{1*}^2\ra\ \cO_0^{(0)}\label{eq:Lambda_def}\\
\Gamma &=& \Gamma_{\omega_*}\ \cO_0^{(0)}\nn\\ 
&=& \lambda^2\pi \la \psi_{0*}\psi_{1*}^2,
  \delta(H-\omega_*) 
 \psi_{0*}\psi_{1*}^2\ra\ \cO_0^{(0)} 
\label{eq:Gamma_def}
\ea
Recall that $\cO_0^{(0)}$ denotes a term of the form $1+\cO(|\alpha_0|^2)$.
Returning to the equation for $\beta_1$ we have:

\ba i\D_t \beta_1\ &=&\ (\Lambda_{\omega_*} -i\Gamma_{\omega_*})
 |\alpha_0|^2|\beta_1|^2\beta_1+....
\label{eq:beta_eqn4}
\ea

We now seek the key terms in the $\alpha_0$ equation, (\ref{eq:a0eqn}).
Using that $\phi_1=\Psi_1+\phi_2$ 
 and the representation (\ref{eq:a1b1}),
 we have 
\ba
2i\D_t \alpha_0\ &=& 
  \lambda\la\psi_{0*}^2,\psi_{1*}^2\ra e^{-2i\lambda_+(T)t}
     \beta_1^2\ol{\alpha_0}
\nn\\
&+& 2\lambda\la\psi_{0*}^2\psi_{1*},\phi_2\ra e^{-i\lambda_+(T)t}
 \ol{\alpha_0}\beta_1 + 
 \lambda\la\psi_{0*}^2,\phi_2^2\ra \ol{\alpha_0}
\nn\\
&+& 
 + 2\lambda\la\psi_{0*}\psi_{1*}^2, \phi_2\ra|\beta_1|^2\nn\\
&+& \lambda\la\psi_{0*}\psi_{1*}, \phi_2^2\ra\ol{\beta_1}e^{i\lambda_+(T)t}
 + {\bf 
 \lambda\la\psi_{0*}\psi_{1*}^2, \ol{\phi_2}\ra\beta_1^2e^{-2i\lambda_+(T)t}
    }
\nn\\
&+& 2\lambda\la\psi_{0*}\psi_{1*},|\phi_2|^2\ra\beta_1e^{-i\lambda_+(T)t}
 + \lambda\la\psi_{0*},|\phi_2|^2\phi_2\ra\nn\\
&+& {\cal R}_{\alpha_0}
\label{eq:alpha0_eqn2}
\ea

We first focus on the key resonant term in (\ref{eq:alpha0_eqn2}) which is 
 responsible for the system settling onto the nonlinear ground state.
We claim this term is:
\be
{\bf \lambda\la\psi_{0*}\psi_{1*}^2, \ol{\phi_2}\ra\beta_1^2
 e^{-2i\lambda_+(T)t} }
\label{eq:alpharespo}\ee
In analogy with the previous calculation, we have
\be
\lambda\la\psi_{0*}\psi_{1*}^2, \ol{\phi_2}\ra\beta_1^2e^{-2i\lambda_+(T)t}
 = \lambda\la\psi_{0*}\psi_{1*}^2, \pi_2\eta_{1Ra}\ra\beta_1^2e^{-2i\lambda_+(T)t}
 + \bR,\label{eq:resp3}
\ee
where $\bR$ is as above.
Therefore, applying Proposition \ref{prop:eta1R_expand} we get
\be
\lambda\la\psi_{0*}\psi_{1*}^2, \pi_2\Phi_2\ra\beta_1^2e^{-2i\lambda_+(T)t}
 = (-\Lambda_{\omega_*} +i\Gamma_{\omega_*})|\beta_1|^4\alpha_0
 + \bR 
\label{eq:keyterm}
\ee
The expression $\bR$ denotes terms which are higher order, oscillatory type
and dispersively decaying. See also Remark \ref{rmk:approach}. 

In summary, we have the following system for $\alpha_0$ and $\beta_1$:
\begin{prop}\label{eq:prenormalform}
\ba
i\D_t \alpha_0 &=& (-\Lambda_{\omega_*} +i\Gamma_{\omega_*})|\beta_1|^4\alpha_0
 + \bR \nn\\
i\D_t \beta_1 &=&  2(\Lambda_{\omega_*} -i\Gamma_{\omega_*})
 |\alpha_0|^2|\beta_1|^2\beta_1 + \bR
\label{eq:pre_nf}\ea
\end{prop}

The proof of Proposition \ref{prop:normalform} 
 follows by constructing an appropriate near-identity change of 
variables, transforming  (\ref{eq:pre_nf}) to 
 (\ref{eq:normalform}). This is implemented as in \cite{SW99}. 
\bigskip
\section{Stability analysis on different time scales\ -\ 
 overview}\label{sec:stab}

In Corollary \ref{cor:nlPME}  we obtained  coupled power equations
 or nonlinear master equations governing the (renormalized) ground state and
excited state square amplitudes: 
\be
P_0=|\talpha_0|^2 \ \ {\rm and}\ \  
P_1=|\tbeta_1|^2
\nn\ee
If we neglect the correction terms $R_0$ and $R_1$, 
 in (\ref{eq:P0eqn}-\ref{eq:P1eqn}) we obtain the simpler autonomous system
 of differential equations: 
\ba
{dp_0\over dt} &=& 2\Gamma p_0p_1^2\label{eq:modelP0_eqn}\\
{dp_1\over dt} &=& -4\Gamma p_0p_1^2\label{eq:modelP1_eqn}
\ea
Note that this  system is exactly solvable.  
Addition of twice (\ref{eq:modelP0_eqn}) to (\ref{eq:modelP1_eqn}) yields that 
 along any solution trajectory:
\be 2p_0(t)+p_1(t)\ =\ 2p_0(0) + p_1(0).\label{eq:conserv}\ee 
This relation can be used to eliminate $p_1$ from (\ref{eq:modelP0_eqn}) or 
 $p_0$ from (\ref{eq:modelP1_eqn}). $p_0(t)$ and $p_1(t)$ are thus obtained by quadrature.
 The dynamics of this 
  finite dimensional reduced system
 anticipates that an initial 
state, arbitrarily close to but not exactly on the excited state branch, 
 with energy distributed among the ground state and excited state, will evolve
to a state with an increased ground state energy and no energy in the 
excited state. While not strictly correct, since there are nongeneric 
 data giving rise to solutions which converge to the excited 
  state \cite{TY01b},  
  this captures the generic very large time dynamics. 
 The correction terms $R_0$ and $R_1$ 
 in (\ref{eq:P0eqn}-\ref{eq:P1eqn}) lead to different
transient behavior which may be quite different from that suggested
 by the system (\ref{eq:modelP0_eqn}-\ref{eq:modelP1_eqn}).
However, we show that eventually ($t\ge t_1$), this system dominates.
 Moreover,  a  
large class of data, for which  the system
(\ref{eq:modelP0_eqn}-\ref{eq:modelP1_eqn}) controls the behavior is 
that for which $P_0(0)>P_1(0)$ and sufficiently small initial 
 dispersive part.

Before embarking on the details we give a brief overview of the strategy.
Using the change of variables $(\alpha,\beta)\mapsto(\talpha,\tbeta)$
 of Proposition \ref{prop:normalform}
we have transformed away all {\it local in time} nonresonant terms. This
 introduces contributions to $F_\alpha$ and $F_\beta$, and therefore
contributions to $R_0$ and $R_1$ in equations (\ref{eq:P0eqn}-\ref{eq:P1eqn}),
which are of two types: 

\nit (i) {\it local in time} terms depending on $\talpha_0$ and $\tbeta_1$,
 which can be absorbed by the
 leading terms in (\ref{eq:P0eqn}-\ref{eq:P1eqn}), with a small correction
to the coefficient to the coefficient $\Gamma$ and are of order $b_0\Jt^{-2}$;
 see Proposition \ref{prop:propA} below.

\nit {\it nonlocal in time} functions of $\talpha_0$
 and $\tbeta_1$
defined in terms of $\eta=\eta[\eta_0,\eta_1,\eta_2]$ in $F_\alpha$ and
 $F_\beta$. These contribute
terms to (\ref{eq:normalform}) with the same (anticipated) time-decay
rate as the leading order terms in (\ref{eq:normalform}). Correspondingly,
there are {\it nonlocal in time} functions of $\talpha_0$ and $\tbeta_1$
 which contribute to $R_j$ in equations (\ref{eq:P0eqn}-\ref{eq:P1eqn}) which
are of the same (anticipated) decay rate as the leading order terms
 in (\ref{eq:P0eqn}-\ref{eq:P1eqn}).
 The goal is to control these nonlocal terms,
 to the extent possible,
  by the leading order terms. However, due to the
 different behavior
of $\talpha_0$ and $\tbeta_1$ on different time scales the argument is somewhat
tricky and we now explain our strategy.

 Let
\be
t_0 + 1\ \equiv
 \sup_{\tau\ge0}\left\{
 \tau:\ 0\le\tau'\le\tau,\ P_0(\tau')\le\frac{5}{2}
 \frac{ [\eta_0]_X+\cE_0 }{ {\la \tau-1\ra} {\la\tau'\ra^2} }
          \right\}\label{eq:t0def}
\ee
Propositions \ref{prop:biR0i}-\ref{prop:ODE} and \ref{prop:Riteta}
 will justify this choice, by implying the inequalities (\ref{eq:below}).  
If $t_0<\infty$, then we have  the bound:
\be
P_0(t) \le \frac{5}{2}
 \frac{[\eta_0]_X+\cE_0}{ {\la t_0\ra} {\la t\ra^2} },\ 0\le t\le t_0.
\nn\ee

Consider the system for $P_0(t)$ and $P_1(t)$,
  (\ref{eq:P0eqn}-{\ref{eq:P1eqn}).
Decomposing $R_j$ into local and nonlocal in time  parts we have:
\ba R_j &=&
 R_j^{l}(P_0,P_1,\eta_0,\eta) + \int_0^tK(t,s)r_j^{nl}(s) ds
\nn\\
 &=& R_j^{l}(P_0,P_1,\eta_0,\eta) + \int_0^{t_0}K(t,s)r_j^{nl}(s) ds
\nn\\
\ \ &+&\ \int_{t_0}^t K(t,s)r_j^{nl}(s) ds
\nn\ea
We have that
\ba
R_0^{l}(P_0,P_1,\eta_0,\eta) &\le& \cO(\cE_0^{\rho_0})\left(\ P_0P_1^2 +
                               \frac{b_0}{\Jt^2}\right)
                                \nn\\
R_1^{l}(P_0,P_1,\eta_0,\eta) &\le& \cO(\cE_0^{\rho_1})\left(\ P_0P_1^2 +
                               \frac{b_1}{\Jt^2} +
                                \cO\left(\sqrt{P_0}P_1^m\right)\ \right),
\label{eq:below}\ea
where $\cE_0$ is the initial total energy and $\rho_j>0$. 
 Therefore, for $\cE_0$ small
\ba
\frac{dP_0}{dt} &\ge& 2\Gamma' P_0P_1^2 - \frac{b_0}{\Jt^2}
         + \int_{t_0}^t K_0(t,s)r_0^{nl}(s) ds\nn\\
\frac{dP_1}{dt} &\le& -4\Gamma' P_0P_1^2 + \frac{b_1}{\Jt^2} +
           \int_{t_0}^t K_1(t,s)r_1^{nl}(s) ds
 + \cO\left(\sqrt{P_0}P_1^m\right),
\nn\ea
with $\Gamma'\sim\Gamma$. The reverse inequalities hold with $\Gamma'$
replaced by $\Gamma''=\Gamma + o(\Gamma)$.

Next (Proposition \ref{prop:propB}) we introduce the auxiliary quantities
\be Q_0 = P_0-k_0\Jt^{-1}\ {\rm and}\ Q_1 = P_1 + k_1\Jt^{-1},\nn\ee
where $k_0(b_0,b_1,\cE_0)$ and $k_1(b_0,b_1,\cE_0)$ are chosen appropriately
and derive equations of the form
\ba
\frac{dQ_0}{dt} &\ge& 2\Gamma'' Q_0Q_1^2
         + \int_{t_0}^t K_0(t,s)r_0^{nl}(s) ds\nn\\
\frac{dQ_1}{dt} &\le& -4\Gamma'' Q_0Q_1^2  +
           \int_{t_0}^t K_1(t,s)r_1^{nl}(s) ds
 + \cO\left(\sqrt{Q_0}Q_1^m\right),
\label{eq:Q0Q1dp}\ea
where $\Gamma''\sim\Gamma'\sim\Gamma$.

We then proceed with the following continuity argument. At $t=t_0$,
\be
\frac{dQ_0(t_0)}{dt} \ge 2\Gamma'' Q_0(t_0)Q_1^2(t_0),\ \
\frac{dQ_1(t_0)}{dt} \le -4\Gamma'' Q_0(t_0)Q_1^2(t_0)  +
  \cO\left(\sqrt{Q_0(t_0)}Q_1^m(t_0)\right).
\label{eq:t0est}\ee
Therefore, by continuity, the following inequalities hold for some time interval
 $t_0\le t\le t_{0,1}$, with $\Gamma''$ replaced by ${\Gamma''\over2}$:
\be
\frac{dQ_0}{dt} \ge \Gamma'' Q_0Q_1^2,\ \
\frac{dQ_1}{dt} \le -2\Gamma'' Q_0Q_1^2  +
  \cO\left(\sqrt{Q_0}Q_1^m\right).
\label{eq:t0t01}\ee
Let
$t^* \ \equiv\
 \sup\{t\ge t_0:\ $ inequalities (\ref{eq:t0t01}) hold $\}$.
We show that
\be
 \int_{t_0}^t K_1(t,s)r_1^{nl}(s) ds \le
\cO(\cE_0^\rho)\left(\ Q_0Q_1^2 + \frac{b_j}{\Jt^2}\ \right)
\label{eq:nlbound}\ee
and therefore, up to renormalization of $Q_j$ (adding higher order terms
 of order $\cE_0^\rho k_j\la t\ra^{-1}$ to the definition of $Q_j$)).
 Use of this estimate
in (\ref{eq:Q0Q1dp}) implies (\ref{eq:t0t01}), for $\cE_0$ sufficiently small.
 The argument can be repeated   
and therefore, $t^*=T$.

\section{ Finite dimensional reduction and its analysis 
 on different time scales}\label{sec:finite-dim-red}

We now begin our study of the generic case, where $t_0<\infty$ and
the solution converges to the nonlinear ground state family as $t\to\infty$. 
The following three propositions concern the various time-scales which
enter the analysis. The first is a basic result, a normal form, which  
is the point of departure for our analysis 
 on all time scales.
 
\begin{prop}\label{prop:propA}
\nit Let $m\ge4$. Let
\ba
b_0\ &=&\ \la t_0\ra^{-1}\left(\ [\eta_0]_X\ +\ c_*\cE_0^2\ \right),
\label{eq:b0}\\
\ b_1\ &=&\ \la t_0\ra^{-{1\over2}}
 \left(\ [\eta_0]_X^{2\over3}\ +\ d_*\cE_0^2\ \right), \label{eq:b1}
\ea
for some order one constants $c_*$ and $d_*$.

 If for some $t_0$, positive and finite,
\be P_0(t_0)\ge \frac{ 3b_0\left(t_0,[\eta_0]_X\right)}{\la t_0\ra},
\label{eq:thresh}\ee
 then
for $t\ge t_0$
\ba
{dP_0\over dt} &\ge& 2(1-\delta_1)\Gamma P_0P_1^2\ -\ {b_0\over \Jt^2}\ +\ J_0
\label{eq:P0P1aa}\\
{dP_1\over dt} &\le& -4(1-\delta_1)\Gamma P_0P_1^2\ +\ \cO(\sqrt{P_0}P_1^m)
\ +\ {b_1\over \Jt^2}\ +\ J_1,
\label{eq:P0P1ab}
\ea
where $J_0$ and $J_1$ are nonlocal in time terms, which have the form:
\be
J_j \ =\ \int_{t_0}^t\ K(t,s)\ r_j^{nl}(s)\ ds. \nn 
\ee
The terms encompassed in $J_j$ are derived and estimated in the coming sections. 
\end{prop}
\begin{rmk}
\medskip

\nit The reverse inequalities of (\ref{eq:P0P1aa}- \ref{eq:P0P1ab}) 
hold as well with a different constant $\delta_2\sim\delta_1$.
\end{rmk}

 The proof of Proposition \ref{prop:propA} will be given
 following the estimates on the remainder terms, $R_i$ (Proposition
 \ref{prop:Master}).
\bigskip

\begin{prop}\label{prop:propB}
Assume that $t_0$ is positive and finite as in Proposition \ref{prop:propA}.
 Then,  there exist $k_0=k_0(b_0,b_1,\cE_0)$ 
  and $k_1=k_1(b_0,b_1,\cE_0)$, such that 
 for $t\ge t_0$ the auxiliary functions
\be Q_0(t)\ \equiv \ P_0(t) - { k_0\over \Jt},\ \ \  
 Q_1(t)\ \equiv \ P_1(t) + { k_1\over \Jt}
\label{eq:Q0Q1def}
\ee
satisfy
\ba
{dQ_0\over dt} &\ge& 2\Gamma' Q_0Q_1^2\ +\ J_0\ +\
           \frac{\cE_0^2c_*}{\la t_0\ra\ \la t\ra^2}  \nn\\
{dQ_1\over dt} &\le& -4\Gamma' Q_0Q_1^2  + \cO(\sqrt{Q_0}Q_1^m)\ 
 +\ J_1\ - \ \frac{\cE_0^2d_*}{\la t_0\ra^{1\over2}\ \la t\ra^2},
\label{eq:Q0Q1b}
\ea
for some positive free constants $c_*$ and $d_*$.
In particular, $Q_0(t)$ is monotonically increasing for $t\ge t_0$.
\end{prop}
The next result shows that $c_*$ and $d_*$ can be chosen to control the terms 
 $J_j$.

\begin{prop}\label{prop:propB.5}(Monotonicity of $Q_0$ for $\t\ge t_0$)
There exist $c_*$ and $d_*$ of order one, such that for $t\ge t_0$.
\ba
{dQ_0\over dt} &\ge& 2\Gamma' Q_0Q_1^2\nn\\
{dQ_1\over dt} &\le& -4\Gamma' Q_0Q_1^2  + \cO(\sqrt{Q_0}Q_1^m)\
\label{eq:Q0Q1b.5}
\ea
\end{prop}
The above three propositions are established in the next two sections. We complete 
the current section by working out the consequences of the 
 finite dimensional reduction (\ref{eq:Q0Q1b.5}).
\bigskip

The next proposition, shows that even if $Q_0$ very small at some stage,
 it will eventually become large relative to $Q_1$ and the 
$\cO(\sqrt{Q_0}Q_1^m)$ term in (\ref{eq:Q0Q1b}) will become negligible.
\begin{prop}\label{prop:propC}
Assume
$t\ge t_0$ and suppose for some $r>0$ that 
\be
\frac{Q_0}{Q_1} \leq \cE_0^r.\label{eq:smallQ0}
\ee
Then,
\be \frac{Q_0(t)}{Q_1(t)}\ {\rm is\ increasing\ for\ } t\ge t_0 \ee
and there exists $t_1$ such that
\be \frac{Q_0(t_1)}{Q_1(t_1)}=\cE_0^r.\nn\ee
Furthermore, for $t\ge t_1$:
\ba
&&\frac{d Q_0}{dt}\geq 2\Gamma' Q_0 Q_1^2\nn\\
&&\frac{d Q_1}{dt} \leq - 4\Gamma' Q_0 Q_1^2
\label{eq:Q0Q1c}\ea
\nit Finally, for $t\ge t_1$:
\be
Q_1(t) \le {Q_1(t_1)\over 1+4\Gamma'' Q_1(t_1)(\inf_{[t_1,T]}|Q_0|)\cdot (t-t_1)}
\label{eq:Qbound}
\ee
\end{prop}
\nit{\bf Proof of Proposition  \ref{prop:propB}:}
Define
\be Q_0 \equiv P_0-k_0\la t\ra^{-1},\ \ 
 Q_1 \equiv P_1+k_1\la t\ra^{-1},\label{eq:Q01_def}\ee
where $k_0,k_1>0$ are to be appropriately chosen.  Note that 
\be Q_0(t)\le P_0(t)\ {\rm and}\ Q_1(t)\ge P_1(t).\nn\ee 
 Using (\ref{eq:P0P1aa}-\ref{eq:P0P1ab})
 and some estimation we deduce a simplified system for $Q_0$ and $Q_1$.
We calculate, omitting the terms $J_0$ and $J_1$, which
are carried along passively.
 
We begin with $Q_0$.
\ba
{dQ_0\over dt} &=& {dP_0\over dt} + {k_0\over\la t\ra^2}\nn\\
 &\ge& 2\Gamma P_0P_1^2 - \frac{b_0}{\la t\ra^2} + {k_0\over\la t\ra^2}\nn\\
 &=& 2\Gamma \left(Q_0+{k_0\over\la t\ra}\right)
        \left(Q_1-{k_1\over\la t\ra}\right)^2
       -\frac{b_0}{\la t\ra^2} + {k_0\over\la t\ra^2}\nn\\
 &\ge& 2\Gamma Q_0 Q_1^2 -4\Gamma {k_1\over\la t\ra} Q_0 Q_1
       + 2\Gamma {k_1^2\over \la t\ra^2}\nn\\
   &+& {k_0-b_0\over\la t\ra^2}.
\label{eq:Q0a}
\ea
We estimate the second term on the right hand 
side as follows. For any $s>0$,
\be
-4\Gamma {k_1\over\la t\ra} Q_0 Q_1 \ge -2s\Gamma Q_0Q_1^2
 -2\Gamma Q_0 {k_1^2\over s\la t\ra^2}.\nn\ee
Therefore,
\be
{dQ_0\over dt} \ge 2\Gamma(1-s)Q_0Q_1^2 + \left(
  2\Gamma k_1^2 Q_0\left(1-{1\over s}\right)
+ k_0-b_0\right){1\over\la t\ra^2}
\label{eq:Q0b}
\ee
Now set $s={1\over10}$ and assume 
\be k_1\ge b_1.\label{eq:k1con}\ee
Then,  using that $k_1=10b_1$, we have
\be
{dQ_0\over dt} \ge
2\cdot{9\over10}\Gamma Q_0Q_1^2 + \left(k_0 - b_0 
 -18\Gamma b_1^2 Q_0\right){1\over\la
 t\ra^2}
\label{eq:Q0c}
\ee
If
\be k_0 \equiv  b_0 +18\Gamma b_1^2 \sup_{t_0\le t\le T} Q_0\ +\ 
 \frac{\cE_0^2c_*}{\la t_0\ra},
\label{eq:k0_def}
\ee
we have  by (\ref{eq:Q0c}) and  (\ref{eq:Q1c})
\be
{dQ_0\over dt} \ge 2\cdot{9\over10}\Gamma Q_0Q_1^2\ +\ 
 \frac{\cE_0^2c_*}{\la t_0\ra\ \la t\ra^2}
\label{eq:Q0d}\ee
In particular, $Q_0$ is increasing for $t\ge t_0$.
\medskip

We now turn to $Q_1$, we have
\ba
{dQ_1\over dt} &=& {dP_1\over dt} - {k_1\over\la t\ra^2}\nn\\
               &\le& -4\Gamma P_0P_1^2 + 
 {b_1\over\la t\ra^2}- {k_1\over\la t\ra^2} 
 + \cO(\sqrt{P_0}P_1^m)\nn\\
               &=& -4\Gamma \left(Q_0+{k_0\over\la t\ra}\right)
                \left(Q_1-{k_1\over\la t\ra}\right)^2
  + {b_1-k_1\over \la t\ra^2} + \cO(\sqrt{P_0}P_1^m)\nn\\
               &\le & -4\Gamma Q_0\left(Q_1-{k_1\over\la t\ra}\right)^2 + 
 {b_1-k_1\over \la t\ra^2} + \cO(\sqrt{P_0}P_1^m)\nn\\
               &=& -4\Gamma Q_0Q_1^2 + 8\Gamma {k_1\over\la t\ra}Q_0Q_1
                   -4\Gamma Q_0{k_1^2\over\la t\ra^2} + 
  {b_1-k_1\over \la t\ra^2} + \cO(\sqrt{P_0}P_1^m)\nn\\
&& \label{eq:Q1a} 
\ea
The second term on the right hand side is estimated as follows. For any $r>0$
we have, since $2ab\le ra^2 + r^{-1}b^2$, that
\be
8\Gamma {k_1\over\la t\ra}Q_0Q_1 \le {4\Gamma k_1 r Q_0\over\la t\ra^2}
 +{4\Gamma k_1 Q_0 Q_1^2\over r}.\nn\ee
Therefore,
\be
{dQ_1\over dt} \le -4\Gamma(1-{k_1\over r}) Q_0Q_1^2 + 
              4\Gamma k_1 Q_0 {r-k_1\over\la t\ra^2}
                + {b_1-k_1\over\la t\ra^2}
 + \cO(\sqrt{P_0}P_1^m)
\label{eq:Q1b}
\ee
Let
\be  r\equiv 20k_1\ {\rm and}\ b_1\ \equiv\ k_1(1-76\Gamma k_1Q_0)\ -\ 
 \frac{\cE_0^2d_*}{\la t_0\ra^{1\over2}},
 \label{eq:k1b1}\ee
which is consistent with the constraint (\ref{eq:k1con}).
Then, 
\ba
{dQ_1\over dt} &\le& -4\cdot{19\over20}\Gamma Q_0Q_1^2 
              \  -\  \frac{\cE_0^2d_*}{\la t_0\ra^{1\over2}\ \la t\ra^2} + 
                   \cO(\sqrt{P_0}P_1^m).
\label{eq:Q1c}
\ea
By (\ref{eq:Q0Q1b}), $Q_0$ is increasing
 for $t\ge t_0$. Therefore, 
\ba Q_0(t_0)&\ge& {1\over 100}\frac{k_0}{\la t_0\ra}\ \Longrightarrow\nn\\ 
  P_0(t)&\le& Q_0(t) + \frac{k_0}{\la t\ra}
 =  Q_0(t) + 100 \frac{k_0}{100\la t\ra}\nn\\
 &\le& Q_0(t) + 100 Q_0(t_0)\le 101 Q_0(t) 
\nn\ea
Also, by definition $P_1\le Q_1$ and so
\ba
{dQ_1\over dt} &\le& -4\cdot{19\over20}\Gamma Q_0Q_1^2 +\cO(\sqrt{Q_0}Q_1^m) 
\label{eq:Q1d}
\ea
This completes the proof of Proposition \ref{prop:propB}.

\nit{\bf Proof of Proposition \ref{prop:propC}: }
\ba
&&\frac{d}{dt}\left(\frac{Q_0}{Q_1}\right) = \frac{\dot Q_0 Q_1 - \dot
Q_1 Q_0}{Q_1^2} \geq \frac{1}{Q^2_1} \left\{ \Gamma  Q_0Q_1^2Q_1 +
4\Gamma Q_0^2 Q_1^2 - C Q_0^{3/2} Q_1^m \right\}\nn\\
&&
\geq \frac{1}{Q^2_1} \left\{ 2\Gamma Q_0 \left[Q^3_1 - C Q_0^{1/2}
Q_1^{-1/2} Q_1^{m + 1/2} \right] + 4\Gamma Q^2_0 Q_1^2\right\}\nn
\ea
$\frac{Q_0}{Q_1}\leq \cE_0^r$   then implies that  
\ba
\frac{d}{dt}\left(\frac{Q_0}{Q_1}\right)
 &&\geq \frac{1}{Q_1^2}\left\{ 2\Gamma Q_0\left[Q_1^3 - \cE_0^{r/2} 
 Q_1^{m+1/2}\right]+4\Gamma
Q_0^2 Q_1^2\right\}\nn\\
&&\geq \Gamma \left(\frac{Q_0}{Q_1}\right) Q_1^2 + 4\Gamma
Q_1^2\left(\frac{Q_0}{Q_1}\right)^2 
\nn\\
&&\ge \Gamma \left(\frac{Q_0}{Q_1}\right) Q_1^2,\nn\\
\ea
for $m + \frac{1}{2} > 3$ and $|Q_1| \leq \cE_0 << 1$.
Hence
$$
\frac{Q_0}{Q_1}\bigg|_t\geq \frac{Q_0}{Q_1}\bigg|_{t_0} 
\exp(\Gamma\int^{t}_{t_0} Q^2_1(s)\ ds).
$$
Since $Q_1>0$, either $Q_1\downarrow 0$ or $\frac{Q_0}{Q_1}$ grows
 exponentially with $t\ge t_0$. In either case, there exists
 $t_1\ge t_0$, such that for $t=t_1$,  
$ \frac{Q_0}{Q_1}\bigg|_{t_1} = \cE_0^r$. 
Now whenever, 
\be \frac{Q_0}{Q_1}\ge \cE_0^r,\label{eq:Q0overQ1}\ee
 we have
\ba
\frac{dQ_1}{dt} &\leq& - 4\Gamma Q_0Q_1^2 + 
+ \cO(\sqrt{Q_0}Q_1^m)\nn\\
&\leq& - 4\Gamma Q_0Q_1^2 + Q_0Q_1^{m-1/2}
\cE_0^{-r/2}\nn\\
&\leq& - 4\Gamma Q_0\left[Q_1^2 - C\cE_0^{-r/2} \cE_0^{m -5/2}
Q_1^2\right]\nn\\
&\leq& - 4\Gamma Q_0 Q_1^2\label{eq:ddtQ1bd}\ea
for $m > \frac{5}{2} + \frac{r}{2}$.
Therefore, by (\ref{eq:Q0overQ1}) we have
 $\frac{dQ_1}{dt}\bigg|_{t_1}<0$. 
Since $Q_0$ is increasing for $t\ge t_1\ge t_0$,
 the inequality  $\frac{Q_0}{Q_1}\ge \cE_0^r$ persists 
 and (\ref{eq:ddtQ1bd}) holds for all $t\ge t_1$.
Hence
\be
\frac{dQ_1}{dt} \leq - 4\Gamma' Q_0 Q_1^2 \ {\rm and}\ 
    \frac{dQ_0}{dt} \geq 2\Gamma'
Q_0 Q_1^2.
\label{eq:Q0Q1f}
\ee

Finally, for $t\ge t_1$
\be {dQ_1\over dt} \le -4{19\over20}\Gamma (\inf_{[t_0,T]}Q_0)Q_1^2.
\label{eq:Q1e}
\ee
Solving this scalar inequality 
\be
Q_1(t)\le {Q_1(t_1) \over 1+4\Gamma'' Q_1(t_1)
 |Q_0(t_1)|(t-t_1)}
\label{eq:Q1bound}
\ee
Since $P_1\le Q_1$, $P_1(t)$ decays to zero like $\la t-t_1\ra^{-1}$.
This completes the proof of Proposition \ref{prop:propC}.

\section{Decomposition and estimation of the dispersion} 
In this section, we revisit the decomposition of the dispersive part, $\eta$,
which satisfies equation (\ref{eq:eta_eqn}). Here, we decompose $\eta$ 
in a manner suitable for consideration of the solution on the various time scales.

\begin{prop}\label{prop:eta-decomp}

\be \eta(t)\ =\ \eta_0(t)\ +\ e_0(t)\ +\ \tetab(t).\label{eq:eta-decomp}
\ee
The three terms can be described as follows:\ 

\nit (i)
  $\eta_0(t)$ is a dispersive wave generated by the data $\eta(0)=\eta_{\rm in}$:
\be i\D_t\ \eta_0\ =\ \cH_0\ \eta_0,\ \eta(0)=\eta_{\rm in},
\label{eq:eta0-eqn}\ee

\nit (ii) $e_0(t)$ is driven principally by $\eta_0(t)$:
\be
i\D_t\ e_0\ =\ \cH_0\ e_0\ + P_cS^{(0)}[e_0,\tetab;\eta_0],\ \ e_0(0)=0,
 \label{eq:e0-eqn}
\ee
and

\nit (iii)
 $\eta_b(t)$, which is driven by bound state dynamics: 
\be
i\D_t\ \tetab\ =\ \cH_0\ \tetab\ + P_cS^{(b)}[e_0,\tetab;\eta_0],\ 
 \ \tetab(0)=0.\label{eq:etab-eqn} 
\ee

We display expressions for $S^{(0)}$ and $S^{(b)}$ 
 with the detail required
in our analysis. We let $\chi$ denote a generic 
 exponentially localized function of
 position, $x$.
\ba
S^{(0)}\ 
&\equiv&\ S^{(0)}_0+S^{(0)}_1e_0+S^{(0)}_2e_0^2+e_0^3\nn\\
&\sim& \left(\alpha_0^2(t)-\alpha_0^2(T)\right)\chi\eta_0
 + \eta_0^3 + \alpha_0\alpha_1\chi\eta_0 +\alpha_0\chi\eta_0^2  
\nn\\
&+&\ \left( \alpha_0\alpha_1+\alpha_1^2 \right)\chi e_0  +
\left(\alpha_0+\alpha_1\right)\chi (\eta_0 +\tetab) e_0 + \eta_0\tetab e_0
\nn\\
&+&\ \left(\alpha_0^2(t)-\alpha_0^2(T)\right)\chi e_0  
+\ (\alpha_0+\alpha_1)\chi e_0^2 + (\eta_0+\tetab)e_0^2\nn\\ 
&+&\ e_0^3 
\label{eq:S0-def}\\
S^{(b)}\ &\equiv&\ S^{(b)}_0+S^{(b)}_1\tetab+S^{(b)}_2\tetab^2+\tetab^3\nn\\
&\sim& \left(\alpha_0^2(t)-\alpha_0^2(T)\right)\chi \tetab\nn\\
&+&  
 i P_c(T)\left(\begin{array}{c}
 \vec\nabla_1\Psi_1(t)\cdot\D_t\vec\alpha_1(t) +
 \vec\nabla_0\Psi_0(t)\cdot\D_t\vec\alpha_0(t)\\
 c.c. \end{array}\right)\nn\\ 
&+& \left( E_1(t)-E_0(t) \right)P_c(T)\sigma_3
             \left(\begin{array}{c}\Psi_1(t)\\ c.c.\end{array}\right) 
\nn\\
&+&\ \left(\ |\alpha_0|^2\ |\alpha_1|\ +\ |\alpha_0|\ |\alpha_1|^2\ \right)  \chi 
\nn\\
&+&\ \alpha_0\alpha_1\chi\tetab +\left(\alpha_0+\alpha_1\right)\chi\tetab^2 + 
 \left(\eta_0+e_0\right)\tetab^2\nn\\
&+&\ |\alpha_0|\chi\eta_0\tetab\ +\ \left(\eta_0^2+e_0^2\right)\tetab + \tetab^3 
\label{eq:Sb-def}
\ea
\end{prop}

By  (\ref{eq:phi2H1bound}) we have the following $H^1$ bounds on $e_0$ 
and $\tetab$ in terms of one another:
\ba
\| e_0(t)\|_{H^1}\ &\le&\ \cE_0 + \|\eta_0\|_{H^1} + \|\tetab(t)\|_{H^1}
 \label{eq:e0H1}\\
\| \tetab(t)\|_{H^1}\ &\le&\ \cE_0 + \|\eta_0\|_{H^1} + \|e_0(t)\|_{H^1}.
 \label{eq:etab0H1}
\ea

Using DuHamel's formula, both the $e_0$ and $\tetab$ equations can be written as
equivalent integral equations: 
\ba
e_0(t)\ &=&\ -i\int_0^t\ e^{-i\cH_0(t-s)}\ P_c\ S^{(0)}(s)\ ds\label{eq:e0-int}\\
\tetab(t)\ &=&\ -i\int_0^t\ e^{-i\cH_0(t-s)}\ P_c\ S^{(b)}(s)\ ds
\label{eq:etab-int}
\ea
Therefore, in both cases we must estimate an expression of the form:
\be
w(t)\ =\ \int_0^t\ e^{-i\cH_0(t-s)}\ P_c\ S(s)\ ds.
\label{eq:w-int}
\ee
For estimation, we shall require the following class of dispersive estimates:
\begin{prop}\label{prop:regdecay}
\nit (1)\ Let $2\le p<\infty,\ p',q\ge2,q'$ and $s$ be related by: 
\be
3\left({1\over q}-{1\over p}\right) = s,\
\  p^{-1} + (p')^{-1} = 1,\
  q^{-1} + (q')^{-1} = 1.
\nn\ee
Then,
\be
\| e^{-i\cH_0t}P_cf\|_p\ \le\
|t|^{-3({1\over2}-{1\over q})}
\Jt^{-3({1\over q}-{1\over p})}
\left(\ \| f\|_{p'}\ +\ \|\D^s f\|_{q'}\ \right)
\label{eq:ge0}
\ee

\nit (2) Assume $q\ge2$ and $s>3/q$. Then,
\be
\| e^{-i\cH_0t}P_cf\|_\infty\ \le\
|t|^{-3({1\over2}-{1\over q})}
\Jt^{-{3\over q}}
\left(\ \| f\|_1\ +\ \|\D^s f\|_{q'}\ \right)
\label{eq:ge0A}
\ee
\end{prop}
{\bf Proof of Proposition \ref{prop:regdecay}:}
We use the classical Sobolev inequality for functions
defined on $\R^3$:
\be
\|f\|_p\le C\ \|\D^sf\|_q,\ \ \  3(q^{-1}-p^{-1}) = s
\label{eq:sobolev}
\ee
and the $L^p$ -- $L^{p'}$ estimate (Theorem \ref{theo:lin_asymp_stab})
\be \| e^{-i{\cal H}_0t } P_c f\|_p\ \le\ C_{1,p}\ t^{-{3\over2}+{3\over p}}
 \|f\|_{p'},\ p^{-1}+(p')^{-1}=1.\label{eq:ge1}\ee

For $|t|\ge1$, we use (\ref{eq:ge1}). For $|t|\le1$, we have
by (\ref{eq:sobolev}) that
\ba
\| e^{-i{\cal H}_0t}P_c f\|_p\ &\le&\ \|\D^s e^{-i{\cal H}_0t}P_c f\|_q\nn\\
 &\le&\ C\ \| \cH_0^{s\over2}e^{-i{\cal H}_0t}P_c f\|_q\nn\\
 &\le&\ C\ |t|^{-3({1\over2}-{1\over q})}\ \| D^s f\|_{q'},\label{eq:le1}\\
 && q^{-1} + (q')^{-1}=1\nn
\ea
From estimates (\ref{eq:ge1}) and (\ref{eq:le1})
 we obtain (\ref{eq:ge0}). Estimate (\ref{eq:ge0A})
 is obtained similarly.
 This completes the proof of Proposition
 \ref{prop:regdecay}).
\medskip

We now apply Proposition \ref{prop:regdecay} with $q=4$ and $s=1>3/4$
 to the integral 
 equation (\ref{eq:w-int}) and obtain the bound: 
\be
\|w(t)\|_\infty\ \le\ \int_0^t\ \frac{1}{|t-s|^{3\over4}}
 \frac{1}{\la t-s\ra^{3\over4}}\ 
\left(\ \|S(s)\|_1+\|\D S(s)\|_{4\over3}\ \right)\ 
  ds
\label{eq:w-est}
\ee
We shall use this bound as the first step in estimating $\|e_0(t)\|_\infty$
and $\|\tetab(t)\|_\infty$.
\medskip

\nit More specifically, our estimation strategy seeks a closed system of 
inequalities for the following

\nit norms of $e_0$:\ 
 {\bf 1.}\ $[e_0]_{\infty,\3half},\ \ 
 {\bf 2.}\ [\D e_0]_{\L2loc;\3half},\ \ {\bf 3.}\ [e_0]_{H^1;0}$

\nit norms of $\tetab$:\ 
$\ {\bf 4.}\ [\eta_b]_{\infty;0}, \ \ {\bf 5.}\  [\eta_b]_{H^1;0}$,  

\nit in terms of norms of the initial data 
$[\phi(0)]_X$.
\medskip

\centerline{\bf 1.\ Estimation of $\|e_0\|_\infty$}

By (\ref{eq:w-est}), we have
\be
\|e_0(t)\|_\infty\ \le\ \int_0^t\ \frac{1}{|t-s|^{3\over4}}
 \frac{1}{\la t-s\ra^{3\over4}}\
\left(\ \|S^{(0)}(s)\|_1+\|\D S^{(0)}(s)\|_{4\over3}\ \right)\
  ds,
\label{eq:e0-est}
\ee
so it sufficies to bound $\|S^{(0)}\|_1$ and 
 $\|\D S^{(0)}\|_{4\over3}$. 
\medskip

\nit\underline{\bf $\|S^{(0)}\|_1$:}
 For any $t\ge0$, 
\ba
\|S_0^{(0)}\|_1\ &\le&\ C\la s\ra^{-{3\over2}}
 \left(\ \cE_0[\eta_0(0)]_X + \sqrt{\cE_0} [\eta_0(0)]_X^2\ \right)
 \le C(\cE_0)D(\eta(0))\ \la s\ra^{-{3\over2}}
\nn\\
\|S_1^{(0)}e_0\|_1\ &\le&\ C\la s\ra^{-{3\over2}}\ [e_0]_{\infty;{3\over2}}
 \left(\ \cE_0 + \cE_0^{1\over2}[\tetab]_{\infty;0} + 
 \cE_0^{1\over2}[\eta_0]_{\infty;0} + [\eta_0]_{\infty;{3\over2}}^2\ \right)\nn\\ 
 &+& C\la s\ra^{-{3\over2}}\ [e_0]_{\infty;{3\over2}}
  [\eta_0]_{\infty;{3\over2}} [\tetab]_{\infty;0}\ \le\
 C\la s\ra^{-{3\over2}}
\nn\\
\|S_2^{(0)}e_0^2\|_1\ &\le&\ C\la s\ra^{-3}\ [e_0]_{\infty;{3\over2}}^2
 \left(\ C\sqrt{\cE_0} + [\eta_0]_{\infty;{3\over2}} + [\tetab]_{\infty;0}
\ \right)\nn\\
\|e_0^3\|_1\ &\le&\ C\ \la s\ra^{-{3\over2}}[e_0]_{\infty;{3\over2}}
                                      [e_0]_{H^1;0}^2
\label{eq:S00L1}
\ea

\nit\underline{\bf $\|\D S^{(0)}\|_{4\over3}$:}
 For any $t\ge0$,

\ba
\|\D S_0^{(0)}\|_{4\over3}\ &\le&\ C\la s\ra^{-{3\over2}}\ \left(\
\cE_0 [\D\eta_0]_{\infty;{3\over2}}\ +\ 
\cE_0^{1\over2} [\eta_0]_{\infty;{3\over2}} [\eta_0]_{H^1;0}\ 
+\ \cE_0^{1\over4} [\eta_0]_{\infty,{3\over2}}^{3\over2}\ 
 \right)\nn\\
\|\D S_1^{(0)}e_0\|_{4\over3}\ &\le&\ C\la s\ra^{-{3\over2}}\left(\
\cE_0 [\eta_0]_{\infty;{3\over2}}\ +\ \cE_0 [e_0]_{\L2loc;{3\over2}}
\right) \nn\\ 
&+& C\la s\ra^{-{3\over2}}\left(\
 \cE_0^{1\over2} [\tetab]_{\infty;0} [\eta_0]_{\infty;{3\over2}}\ 
+\ \cE_0^{1\over2} [e_0]_{\infty,{3\over2}}  [\eta_b]_{H^1;0}
 + \cE_0^{1\over2} [\eta_b]_{\infty;0} [\D e_0]_{\L2loc;{3\over2}}
 \right)\nn\\
\|\D S_2^{(0)}e_0^2\|_{4\over3}\ &\le&\ 
C \la s\ra^{-{3\over2}}  [e_0]_{\infty;{3\over2}} 
\left(\
\cE_0^{1\over2}[e_0]_{\infty;{3\over2}}+\cE_0^{1\over2}[e_0]_{H^1;0}\right)\nn\\ 
 &+& C \la s\ra^{-{3\over2}} [e_0]_{\infty;{3\over2}}
 \left( [\eta_0]_{H^1;0} + [\tetab]_{H^1;0} \right)\ [e_0]_{H^1;0} 
 \nn\\
 &+&\ [e_0]_{\infty;{3\over2}}^{3\over2}\ [e_0]_{L^2;0}^{1\over2}
  \left( [\eta_0]_{H^1;0}+[\tetab]_{H^1;0}\right) 
\nn\\
\|\D e_0^3\|_{4\over3}\ &\le&\ C\la s\ra^{-{3\over2}}\ 
 [e_0]_{\infty;{3\over2}}\ [e_0]_{H^1;0}^2
\label{eq:DrS00L43}
\ea

Use of the bounds (\ref{eq:S00L1}) and (\ref{eq:DrS00L43}) 
 in (\ref{eq:e0-est}) implies
\begin{prop}\label{prop:e0infty-est}
\ba \|e_0(t)\|_\infty\ &\le&\ \la t\ra^{-\3half}\left(\
 \cC(\cE_0,[\eta_0(0)]_X)\ +\ \cE_0^\half[\tetab(t)]_{\infty;0}
\ +\ [\eta_0(0)]_X\ [\tetab(t)]_{\infty;0}\ [e_0]_{\infty;\3half}\ \right)\nn\\
&+&\ \la t\ra^{-\3half}
 \left(\ \cE_0^\half+[\eta_0(0)]_X\ +\ [e_0(t)]_{H^1;0}^2\ +\ 
 [\tetab(t)]_{\infty;\3half}\ \right) 
 \left(\ [e_0]_{\infty;\3half}+[e_0]_{\infty;\3half}^2\ \right)\nn\\
&+&\ \la t\ra^{-\3half}\left(\ 
 [e_0]_{\infty;\3half}\ [e_0]_{H^1;0}^2\ +\ \cE_0 [e_0]_{\L2loc;\3half}
\ +\ \cE_0^{1\over2} [e_0]_{\infty;\3half}^2\ \right)
\label{eq:e0-inftyest}
\ea
\end{prop}

\medskip

\centerline{\bf 2.\ Estimation of $[\D e_0(t)]_{\L2loc;{3\over2}}$}

\ba
\|\D e_0(t)\|_{\L2loc}\ &\sim& \|\chi\D e_0(t)\|_2\ \ \ \left(\ 
 \chi\ {\rm localized}\ \right)\nn\\ 
 &\sim&\ \| \chi \la\cH_0\ra^{1\over2} e_0(t)\|_2
\ \ \left(\ \D\ \la\cH_0\ra^{-{1\over2}}\in \cB(L^2)\ \right)\nn\\
 &\le&\ \int_0^t\ \|\chi \la\cH_0\ra^{1\over2} e^{-i\cH_0(t-s)}P_cS^{(0)}(s)\|_2\ ds
\ \ \ \left(\ {\rm by}\ (\ref{eq:e0-int})\ \right)\nn\\
&& \label{eq:ldest-start}
\ea
For the purpose of continuing the estimation, we regard $S^{(0)}$ as consisting
of terms of two types: (i)\ terms having spatially localized (exponentially)
 functions of $x$ as a factor, coming from $\Psi_j,\ j=0,1$, and (ii)\ terms like 
 $e_0^3$ or $\tetab \eta_0 e_0$ and others which are not of this type.
 It is convenient to refer to such terms as $S^{(0)}_{\rm LOC}$ and 
 $S^{(0)}_{\rm NLOC}$ below. From 
(\ref{eq:ldest-start})  we have:
\ba
\|\D e_0(t)\|_{\L2loc}\ &\le&\ 
 \int_0^t \|\chi \la\cH_0\ra^{1\over2} e^{-i\cH_0(t-s)}
 P_cS^{(0)}_{\rm LOC}(s)\|_2\ ds\nn\\
 &+&\ \left(\int_0^{t-1}+\int_{t-1}^t\right)
  \|\chi \la\cH_0\ra^{1\over2} e^{-i\cH_0(t-s)}
 P_cS^{(0)}_{\rm NLOC}(s)\|_2\ ds \nn\\
&\le&
  \int_0^t \frac{C}{\la t-s\ra^{3\over2}}\ \|\la x\ra^\sigma S^{(0)}_{\rm LOC}\|\ ds
\nn\\
&+&\ \int_0^{t-1} \| e^{-i\cH_0(t-s)}
 P_c\la\cH_0\ra^{1\over2}S^{(0)}_{\rm NLOC}(s)\|_\infty\ ds\ +\ 
 \int_{t-1}^t \|\la\cH_0\ra^{1\over2} e^{-i\cH_0(t-s)}
 P_cS^{(0)}_{\rm NLOC}(s)\|_2\ ds\nn\\
&\le& \int_0^t 
 \frac{C}{\la t-s\ra^{3\over2}}\ \|\la x\ra^\sigma S^{(0)}_{\rm LOC}\|\ ds
\nn\\
&+&\ C\ \int_0^{t-1} \frac{1}{|t-s|^{3\over2} }\ \|
 \ \D\ S^{(0)}_{\rm NLOC}(s)\|_\infty\ ds\ +\ 
 \int_{t-1}^t \|
 \D S^{(0)}_{\rm NLOC}(s)\|_2\ ds\nn\\ 
&=&\ {\bf A}\ +\ {\bf B}\ +\ {\bf C}
\label{eq:ABCdef}
\ea

We estimate the integrands of ${\bf B}$ and ${\bf C}$; that of ${\bf A}$ 
 is similar.

\be
\left|\ {\rm integrand\ of\  {\bf B}}\ \right|\ \le\ C\ \la t\ra^{-{3\over2}}\ 
 \left(\ [e_0]_{\infty;{3\over2} }+[\eta_0]_{\infty;{3\over2} }\ \right)
\ \left(\ 
 [e_0]_{H^1;0}^2\ +\ [\tetab]_{H^1;0}^2\ +\ [\eta_0]_{H^1;0}^2\ \right)  
\label{eq:Best}
\ee

\ba
\left|\ {\rm integrand\ of\  {\bf C}}\ \right|\ 
 &\le&\ \la t\ra^{-{3\over2}}\left(\     
  [e_0]_{\infty;{3\over2}}^2\left(\ [e_0]_{H^1;0}+[\tetab ]_{H^1;0}\ \right)
\right.
\nn\\
&+&\ \left. [e_0]_{\infty;{3\over2}} [\eta_0]_{\infty;{3\over2}}[\tetab]_{H^1;0}
 \right. \nn\\
&+&\ \left. [\eta_0]_{W^{1,\infty};{3\over2}}\left(
 [\tetab]_{\infty;0} [e_0]_{H^1;0}+[\eta_0]_{\infty;{3\over2}}[\tetab]_{H^1;0}
 + [e_0]_{\infty;{3\over2}}[e_0]_{H^1;0} \right) \right.\nn\\
&+& \left. [\eta_0]_{W^{1,\infty};{3\over2}}^2[e_0]_{H^1;0}\ \right)
\nn\ea
\medskip
 
Substitution of these bounds into (\ref{eq:ABCdef}), we have:

\begin{prop}\label{prop:De0localdecay}
\be  \|\D e_0(t)\|_{\L2loc;{3\over2}}\ \le\  
 C\la t\ra^{-\3half}\left(\ [e_0]_{\infty;\3half}\ +\ [\eta_0(0)]_X\ \right)
 \ \left(\ [e_0]_{\infty;\3half}\ +\ [e_0]_{H^1;0}\ \right)
\label{eq:De0loc-est}\ee
\end{prop}

\centerline{\bf 3.\ Estimation of $[e_0(t)]_{H^1;0}$}

\be
\|e_0(t)\|_{H^1}\ \sim 
 \|\cH_0^{1\over2}e_0(t)\|_2\ \le\ \int_0^t \|\D S^{(0)}(s)\|_2
 \ ds\label{eq:e0H1start}\ee

\ba
\|\D S_0^{(0)}\|_2\ &\le& C\la s\ra^{-{3\over2}}\ 
\left(\ 
 \cE_0 [\eta_0]_{W^{1,\infty};{3\over2}} + 
 \cE_0^{1\over2}[\eta_0]_{\infty;{3\over2}}^2 \right.\nn\\
&+&\ \left. [\eta_0]_{H^1;0}\ [\eta_0]_{\infty;{3\over2}}^2\ +\ 
 \cE_0\ [\eta_0]_{W^{1,\infty};{3\over2}}\ \right)
\nn\\
\|\D S_0^{(1)}e_0\|_2\ &\le& C\la s\ra^{-{3\over2}}\
\dots\nn\\
\|\D S_0^{(2)}e_0^2\|_2\ &\le& C\la s\ra^{-{3\over2}}\
 \left(\ \cE_0^{1\over2}[e_0]_{\infty;{3\over2}}^2 +
\cE_0^{1\over2}[e_0]_{\infty;{3\over2}}[\eta_0]_{H^1;0}\ \right)\nn\\
&+&\ C\la s\ra^{-{3\over2}}\ [e_0]_{\infty;{3\over2}}^2
 \left( [\eta_0]_{H^1;0}+[\tetab]_{H^1;0} \right)\nn\\
&+&\ C\la s\ra^{-{3\over2}}\ 
 [e_0]_{\infty;{3\over2}} [\eta_0]_{W^{1,\infty};{3\over2}}
 \left( [\eta_0]_{2;0}+[\tetab]_{2;0} \right)
\nn\\
\|\D e_0^3\|_2\ &\le&
 C\la s\ra^{-3}\ [e_0]_{\infty;{3\over2}}^2\ [e_0]_{H^1;0}
\nn\ea

\begin{prop}\label{prop:e0H1} For $t\ge0$,
\be
\|e_0(t)\|_{H^1}\ \le\ \cC(\cE_0,[\eta_0(0)]_X)\ +\ 
 [e_0]_{\infty;\3half}\left(\ [\eta_0(0)]_X\ +\ [e_0]_{H^1;0}\ \right) 
\nn\ee
\end{prop}
\medskip

We now turn to the estimation of $\|\tetab(t)\|_\infty$ and 
 $\|\tetab(t)\|_{H^1}$. By (\ref{eq:w-est}) we have
\be
\|\tetab(t)\|_\infty\ \le\ \int_0^t\ \frac{1}{|t-s|^{3\over4}}
 \frac{1}{\la t-s\ra^{3\over4}}\
\left(\ \|S^{(b)}(s)\|_1+\|\D S^{(b)}(s)\|_{4\over3}\ \right)\
  ds,
\label{eq:etab-est}
\ee
where $S^{(b)}$ is given by (\ref{eq:Sb-def}).

\begin{prop}\label{eq:SbQ}
In terms of $Q_j,\ j=0,1$ we have
\be
S^{(b)}\ =\ S^{(b)}_0 + S_1^{(b)}\tetab + S_2^{(b)}\tetab^2 + \tetab^3 
\nn\ee
where
\ba
S_0^{(b)}\ &\sim&\ 
 Q_0^{1\over2}Q_1 \chi\ +\ Q_0Q_1^{1\over2}\chi\nn\\
S_1^{(b)}\tetab\ &\sim&\ Q_0^{1\over2}Q_1^{1\over2}\chi\tetab
 + (\eta_0^2+e_0^2)\tetab + Q_0^{1\over2}\chi\eta_0\tetab\nn\\
S_2^{(b)}\tetab^2\ &\sim&\ \left( Q_0^{1\over2}+Q_1^{1\over2}\right)\chi\tetab^2
 + \left(\eta_0+e_0\right)\tetab^2
\nn\ea
\end{prop}

We now proceed with estimates of $S^{(b)}$ in $L^1$ and in $W^{1,{4\over3}}$.

Beginning with $\|S^{(b)}\|_1$ we have:
\ba
&&\|S^{(b)}\|_1\nn\\
&\le& C\left(\  
    Q_0^{1\over2}Q_1+Q_0Q_1^{1\over2}\  
 +\ \ Q_0^{1\over2}Q_1^{1\over2}\|\tetab\|_\infty\ +\  
  \|\tetab\|_\infty \left( \|\eta_0\|_2^2+\|e_0\|_2^2\right)
\ \right)\nn\\
 &+&\ C\left(\ 
  Q_0^{1\over2}\|\eta_0\|_\infty\|\tetab\|_2\ 
 +\  \ \left( Q_0^{1\over2}+Q_1^{1\over2}\right) \|\tetab\|_\infty^2\ 
+\ \|\tetab\|_\infty\left(\ \|\eta_0\|\ \|\tetab\|_2\ +\ 
 \|e_0\|_2\ \|\tetab\|_2\ \right)\
  \right)\nn\\
&+&\ C\|\tetab\|_\infty\|\tetab\|_2^2
\label{eqw:L1Sb}\ea

We now turn to $\|\D S^{(b)}\|_{4\over3}$.
\ba
&&\|\D S^{(b)}\|_{4\over3}\nn\\
&\le&\ C\left(\ 
Q_0^{1\over2}Q_1 + Q_0Q_1^{1\over2}+
Q_0^{1\over2}Q_1^{1\over2}\ \|\tetab\|_{H^1}
\ \right)\nn\\
&+&\ C\left(\ 
\| (\eta_0^2+e_0^2)\D\tetab\|_{4\over3} + 
 \|\eta_0\D\eta_0\tetab\|_{4\over3} + 
 Q_0^{1\over2}\|\tetab\|_2\left(\|\eta_0\|_\infty+\|\D\eta_0\|_\infty
 \right)\ 
\right)
\nn\\
&+&\ C\left(\ 
Q_0^{1\over2}\|\eta_0\|_\infty\|\D\tetab\|_2 +
 \|(\eta_0+e_0)\tetab\D\tetab\|_{4\over3}+ 
\|(\D\eta_0+\D e_0)\tetab^2\|_{4\over3}\ +\ \|\tetab^2\D\tetab\|_{4\over3}
\ \right)\nn\\
\label{eq:L43Sb}
\ea
To further estimate $\|\D S^{(b)}\|_{4\over3}$ we use the following estimates
of individual terms:
\ba 
\|\eta_0^2\D\tetab\|_{4\over3} &\le& \|\eta_0\|_\infty\ \|\eta_0\|_2^{1\over2}\
 \|\tetab\|_{H^1}\nn\\
\|e_0^2\D\tetab\|_{4\over3} &\le& \|\eta_0\|_\infty^{3\over2}\ 
 \|e_0\|_2^{1\over2}\ \|\tetab\|_{H^1}\nn\\
\|\eta_0\D\eta_0\tetab\|_{4\over3} &\le& \|\D\eta_0\|_\infty\ 
 \|\eta_0\|_\infty^{1\over2}\ \|\eta_0\|_2^{1\over2}\ \|\tetab\|_2\nn\\
\|\eta_0\tetab\D\tetab\|_{4\over3} &\le& \|\tetab\|_\infty\ 
 \|\eta_0\|_\infty^{1\over2}\ \|\eta_0\|_2^{1\over2}\ \|\tetab\|_{H^1}\nn\\
\|e_0\tetab\D\tetab\|_{4\over3} &\le& \|\tetab\|_\infty\ 
 \|e_0\|_\infty^{1\over2}\ \|e_0\|_2^{1\over2}\ \|\tetab\|_{H^1}\nn\\
\|\D\eta_0\tetab^2\|_{4\over3} &\le& \|\tetab\|_\infty\ 
 \|\D\eta_0\|_\infty^{1\over2}\ \|\D\eta_0\|_4\ \|\tetab\|_2\nn\\
\|\D e_0\tetab^2\|_{4\over3} &\le& \|e_0\|_{H^1}\ 
 \|\tetab\|_\infty^{3\over2}\ \|\tetab\|_2^{1\over2}\nn\\
\|\tetab^2\D\tetab\|_{4\over3} &\le& \|\tetab\|_\infty\ 
 \|\D\tetab\|_2\ \|\tetab\|_2 
\label{eq:forDSb}
\ea
\bigskip

Recall that $\|\tetab(t)\|_{H^1}$ can be estimated in terms of
  $\|e_0(t)\|_{H^1}$; see (\ref{eq:etab0H1}).


Since $\tetab$ is driven by the bound state amplitudes ($Q_0$ and $Q_1$),
 which have different behavior on the intervals $I_j$, we 
  now estimate $\tetab(t)$ separately on 
$I_0~=~[0,t_0],\ I_1~=~[t_0,t_1]$ and $I_2~=~[t_1,\infty)$. 

We now introduce appropriate norms on different time scales. Define
\ba
M_{I_0}(t)\ &\equiv&\ \sup_{0\le t\wedge t_0}\ 
                    \la t\ra^\3half\ \|e_0(t)\|_\infty 
\ +\ \sup_{0\le t\wedge t_0}\ \la t\ra\ \|\tetab(t)\|_\infty\ +\ 
 \sup_{0\le t\wedge t_0}\ \|e_0(t)\|_{H^1}\nn\\
         &+&\ \sup_{0\le t\wedge t_0}\ 
 \la t\ra^\3half\ \|\D e_0(t)\|_{\L2loc} 
\ +\ \sup_{0\le t\wedge t_0} \la t\ra^2\ |Q_0(t)|\ +\ 
 \sup_{0\le t\wedge t_0}\ |Q_1(t)| 
 \label{eq:MI0def}\\
M_{I_1}(t)\ &\equiv&\ \sup_{t_0\le t\wedge t_1}\ 
              \la t\ra^\3half\ \|e_0(t)\|_\infty 
\ +\ \sup_{t_0\le t\wedge t_1}\ \|\tetab(t)\|_\infty\ +\  
 \sup_{t_0\le t\wedge t_1}\ \|e_0(t)\|_{H^1}\nn\\
      &+&\ 
 \sup_{t_0\le t\wedge t_1}\ \la t\ra^\3half\ \|\D e_0(t)\|_{\L2loc}
 \ +\ \sup_{t_0\le t\wedge t_1}\ |Q_0(t)|\ +\ 
 \sup_{t_0\le t\wedge t_1}\ |Q_1(t)| \label{eq:MI1def}\\
M_{I_2}(t)\ &\equiv&\ \sup_{t_1\le t\wedge T}\ 
 \la t\ra^\3half\ \|e_0(t)\|_\infty 
\ +\ \sup_{t_1\le t\wedge T}\ \la t-t_1\ra^\half\ \|\tetab(t)\|_\infty\ +\  
 \sup_{t_1\le t\wedge T}\ \|e_0(t)\|_{H^1}\nn\\
      &+&\
 \sup_{t_1\le t\wedge T}\ \la t\ra^\3half\ \|\D e_0(t)\|_{\L2loc}
 \ +\ \sup_{t_1\le t\wedge T}\ |Q_0(t)|\nn\\
& +&\ 
 \sup_{t_1\le t\wedge T}\ \la t-t_1\ra
\ \Gamma'|Q_0(t_1)|\ |Q_1(t_1)|\  |Q_1(t)| \label{eq:MI2def} 
\ea
\medskip
\begin{rmk}\label{rmk:e0control}
By Propositions \ref{prop:e0infty-est}, \ref{prop:De0localdecay} and 
 \ref{prop:e0H1}, the $e_0$ contributions to the norms $M_{I_k}$ are controlled
in terms of the initial conditions. Therefore, 
 to control $M_{I_k}$, it suffices to 
bound $Q_j$ and $\tetab$. 
\end{rmk}
 
The above estimates can be used together with the bounds on $Q_j$ of section
 \ref{sec:finite-dim-red} to obtain the following three propositions, which
give bounds for $\tetab$ on the intervals $I_0,\ I_1$ and $I_2$.

\begin{prop}\label{prop:etabonI0} ($\tetab(t)$ for  $t\in I_0$)
Assume $t\in I_0=[0,t_0]$, {\it i.e.}\ 
$Q_0(t)\le C\left([\eta_0]_X+\cE_0\right)\la t\ra^{-2}$ for $t\in [0,t_0]$. 
Then, for $\|\phi(0)\|_X$ sufficiently small
\be
\|\tetab(t)\|_\infty\ \le\ \cC\left(\ \|\phi(0)\|_X,M_{I_0}(t_0)\ \right)
  \ \la t\ra^{-1}\label{eq:etabI0}
\ee
\end{prop}


\begin{prop}\label{prop:etabonI1} ($\tetab(t)$ for  $t\in I_1$)
Assume $t\in I_1=[t_0,t_1]$.
Then, for $\|\phi(0)\|_X$ sufficiently small
\be
\|\tetab(t)\|_\infty\ \le\ \cC\left(\ \|\phi(0)\|_X,M_{I_0}(t_0),
            M_{I_1}(t_1)\ \right)
\label{eq:etabI1}
\ee
\end{prop}

\begin{prop}\label{prop:etabonI2} ($\tetab(t)$ for  $t\in I_2$)
Assume $t\in I_2=[t_1,T]$.
Then, for $\|\phi(0)\|_X$ sufficiently small
\be
\|\tetab(t)\|_\infty\ \le\ \cC\left(\ \|\phi(0)\|_X,M_{I_0}(t_0),
            M_{I_1}(t_1), M_{I_2}(T)\ \right)\ \la t-t_1\ra^{-\half}
\label{eq:etabI2}
\ee
\end{prop}




In our estimates of section \ref{sec:BeginProof}, we shall use the following
 result to estimate the size of correction terms in the system for $Q_0$ and 
 $Q_1$ for $t\ge t_0$, where $Q_0$ is monotonically increasing.

\begin{prop}\label{prop:disip}
Let $\zeta= \zeta(x,t),\ x\in \R^3,\ t\ge t_0 $, with $\zeta(x,t_0) = 0$ 
 satisfy the following dispersive equation 
\be
i\partial_t \zeta = 
 \cH_0\zeta + P_c\left(\    
 S_1(t) + S_2(t)\zeta + \chi\zeta^2 + \zeta^3\ \right) ,\label{eq:zetaeqn}
\ee
where for all $k\ge0$ and $j=1,2$:
\ba
\|S_1(t)\|_{H^k} &=& \cO\left(\cC([\phi(0)]_X\ Q_0^{1\over2}(t)\right)\nn\\
 \D_tQ_0&\ge& 0\  {\rm for }\ t\ge t_0,\ {\rm and }\  Q_0 \leq\cE_0.
\label{eq:Sbound}\ea

Suppose
 $\|\zeta(t)\|_{H^1} \leq \cC([\phi(0)]_X, M_{I_1}(t_1),M_{I_2}(T)\ )$ 
 for all
$t_0\le t\le T$, where  $[\phi(0)]_X$ is sufficiently small.
Then,
$
\|\zeta(t)\|_\infty\ \le\ C\left(\ 
 [\phi(0)]_X, M_{I_1(t_1)},M_{I_2}(T)\ \right)\ Q_0(t)^{1/2} $. 
\end{prop}

\begin{cor}\label{cor:Corollary 1}

Let $p>6$.  $\cE_0^{\bar m}=\sup_{t'\leq t} \epsilon_1 (t'),$ then 
$$
\|\zeta\|_p \leq c Q_0^{1/2}(t)
 \int^t_0\frac{\epsilon_1(s)}{\la t-s\ra^{3/2 - 3/p}}\ ds.
$$
\end{cor}

\begin{cor}\label{cor:Corollary 2} Let $p>6$. 
$$
|\la\chi,\zeta\ra| \leq cQ_0^{1/2}(t)
 \int^t_0\frac{\epsilon_1(s)}
{\la t-s\ra^{3/2 - 3/p}}\ ds.
$$
which follows by using
$$
\|\chi e^{-iHt} P_c\bar \chi g\|_2\leq c \langle t\rangle^{-3/2}
 \| g\|_2.
$$
\end{cor}

\begin{cor}\label{cor:Corollary 3}
$$
\| \D^k\zeta\|_\infty\leq CQ_0^{1/2}(t)\int^t_0 \frac{\epsilon_1(s)}{\langle
t-s\rangle^{3/2}}\ ds.
$$
\end{cor}
\nit{\bf Proof:} This result follows from applying $\D^k$ to the equation
for $\zeta$ and estimating, as above, in $L^p$ for any $\alpha$ and 
$p>6$. By the Sobolev inequality this implies control derivatives
  in $L^\infty$.   
\bigskip

\begin{rmk}
In our applications of Proposition \ref{prop:disip} and its corollaries,
 $\epsilon_1(s)$ will be given by the source terms depending on $Q_0$
and $Q_1^{1\over2}$. For $t>t_1$, ~$Q_1=\cO(\la t-t_1\ra^{-1})$.  
 Therefore, since
 the lowest order term in $Q_1$ contributing to $\epsilon_1(s)$ is 
 $\cO(Q_1)$ (\ see equation (\ref{eq:eta1_eqn})\ ), it follows
that for $t>t_1$
\be
\|\tetab\|_{W^{k,\infty}} = 
 \cO(\la t-t_1\ra^{-{1\over2}}) + \cO(\la t-t_1\ra^{-1}) 
 +\cO(\la t-t_1\ra^{-{3\over2}}). 
\nn\ee

Since $\eta = \cO(\tetab) + \cO(\eta_0)$, the conclusion of the main 
theorem for large $t$, $t>t_1$, follows. Namely, for 
\be
\|\eta\|_{W^{k,\infty}} = \cO(\la t-t_1\ra^{-{1\over2}}) + 
 e^{-i\cH_0t}P_c[\eta(0) + \cE_0^2],
\nn\ee 
where the non-free wave part is coming from spatially localized source terms.
\end{rmk}


\section{Beginning of proof of Proposition \ref{prop:propA} }
\label{sec:BeginProof}

The key to Proposition  \ref{prop:propA} is the following more detailed
version of Corollary \ref{cor:nlPME}:

\begin{prop}\label{prop:Master}
Let $t\ge 0$.
The equations for $P_0, P_1$ can be written in the following form
\ba
\frac{d P_0}{dt} &=& 2\Gamma P_0 P_1^2 + R_0^{(0)}[\eta_0] + R_1^{(0)}
[\tetab] + R_2^{(0)}[P_0, P_1]\nn\\
\frac{d P_1}{dt} &=& -4\Gamma P_0 P_1^2 + R_0^{(1)}[\eta_0] + R_1^{(1)}
[\tetab] + R_2^{(1)}[P_0, P_1].
\label{eq:P0P1}
\ea
where

\nit (i) $R_0^{(i)}[\eta_0]$ are $\eta_0$-dependent terms only both local and
nonlocal in time, and may also depend on $\talpha_0, \tbeta_1$.

\nit (ii) $\tetab$  is the bound state driven part of the dispersion;
 see section \ref{sec:dec-mod}. 
 
\nit (iii) $R_2^{(0)}$ depends only on $P_0, P_1, t$, but not on $\eta$;
  It is
formally linear in $P_0$, of high order in $P_1$,
and contains both local and nonlocal in time terms.

\nit  (iv)
 $R_2^{(1)}$ depends only on $P_0, P_1, t$, but not on $\eta$. It is of
high order in $P_1$, local and nonlocal terms included, but has terms
which are linear in $\tilde \alpha_0 \sim \sqrt{P_0}$.
\end{prop}

The proof uses repeated application of near-identity 
transformations of the variables $\tilde \alpha_0, \tilde \beta_1$, 
 derivable by integration by parts 
( see the 
discussion of resonant and nonresonant terms in section
 \ref{sec:nfandmaster} )  and  the
decomposition of $\eta$: $\eta=\eta_0+e_0+\tetab$; see equation 
 (\ref{eq:eta-decomp}).

The proof is long so we break it up into three parts, which are presented
in three different sections. The following is an overview.

\nit {\bf Part 1:} 
  The terms $R_0^{(0)}[\eta_0]$ and $R_0^{(1)}[\eta_0]$ are forcing terms
 in the ODE dynamics,
 which are driven by the
 dispersive part of the initial conditions. They are studied and  estimated 
  in this section;
 Proposition \ref{prop:biR0i}

\nit {\bf Part 2:}  The terms $R_1^{(0)}[\eta_b]$ and $R_1^{(1)}[\eta_b]$
are studied and estimated 
 in section \ref{sec:Rj1teta}; Proposition \ref{prop:Riteta}.

\nit {\bf Part 3:} The terms $R_2^{(0)}[P_0,P_1]$ and $R_2^{(1)}(P_0,P_1)$
are studied and estimated 
  in section \ref{sec:R120P0P1}; Proposition \ref{prop:ODE}

In Parts 1, 2 and 3, we require estimates for all $t\ge0$.
On $I_0=\{t:0\le t\le t_0\}$ we use  
  the {\it \'a priori} bound on $P_0(t)$, implied by the definition 
of $t_0$; see eqn. (\ref{eq:t0def}). For $t>t_0$, we use the  

\nit{\bf Monotonicity Property Q}

\be 
Q_0\ {\rm and}\ \frac{Q_0}{Q_1}\ {\rm are\ montonically\  increasing},
\label{eq:PropertyQ}\ee
where $Q_0$ and $Q_1$
are  the {\it modified} bound state energies related to $P_0$
and $P_1$ (\ see (\ref{eq:Q0Q1def}), (\ref{eq:b0}), (\ref{eq:b1})\ ). 
 This monotonicity 
property shown to hold at $t=t_0$ and is then 
 shown to continue for all time, $t$,
 by a continuity argument; see section \ref{sec:bootstrap}.
Since there are many terms, we focus on those which are most problematic,
namely, those which are {\it nonlocal} in time and of 
 slowest time-decay rate. This calculations are very lengthy 
and before embarking on them we present a calculation, related to the 
normal form discussion in section \ref{sec:nfandmaster}, and which is
repeatedly in order to exploit rapid oscillations in time.
\bigskip

\noindent{\bf Expansion of oscillatory integrals, resonances  and improved 
 time decay} 

In deriving and estimating the terms $R_j^{(i)}$ in (\ref{eq:P0P1}), 
 we must frequently
expand and/or estimate terms of the form:
\be
 \left\la \chi_1\ ,\int^t_0 e^{-i\cH_0(t-s)} P_c\  
 \chi_2(s)\ e^{i\Omega_1 s}\ ds
 \right\ra
\label{eq:osc-int}
\ee
Here,  $\chi_1$ and $\chi_2(s)$ are localized functions of $x$, $\chi_1$ is
 independent of $s$ and $\chi_2(s)$ depends on $s$ through its dependence on
 $\alpha_0(s),\beta_1(s)$ or $\eta_0(s)$. 

Recall that $\cH_0$, define in (\ref{eq:H_def}) is of the form
\be
\cH_0\ =\ {\tilde\cH_0}\ +\ E_0\sigma_3
\nn\ee 
and ${\tilde\cH_0}=\sigma_3(-\Delta)$ plus a matrix potential which 
decays to zero rapidly as $|x|$ tends to infinity. In 
(\ref{eq:osc-int}) we would like to integrate by parts, exploiting the 
oscillation of frequency $E_0$. However, ``peeling off'' these oscillations
is a little tricky because $\sigma_3$ does not commute with ${\tilde\cH_0}$.
We handle this as follows.

In \cite{Cucc1,Cucc2}, it is shown that 
\ba
Z\cH_0 W\ &=&\ \sigma_3\left(-\Delta-E_0\right), 
\label{eq:ZH0W}\nn\\
Ze^{-i\cH_0tP_c} W\ &=&\ e^{-i\sigma_3\left(-\Delta-E_0\right)t}
\label{eq:ZexpH0tW}
\ea
where $W$ denotes the wave operator
\ba W\ &=&\ \lim_{t\to\infty} e^{-i\cH_0t}e^{-i\sigma_3(-\Delta-E_0)t},\nn\\
    Z\ &=&\ W^{-1}
\label{eq:Waveop}
\ea
By (\ref{eq:ZexpH0tW}), we can rewrite (\ref{eq:osc-int}) as
\ba
&& \left\la \chi_1\ ,e^{-i\cH_0t}\int^t_0 
 We^{i\sigma_3\left(-\Delta-E_0\right)s} Z P_c\
  e^{i\Omega_1 s}\ \chi_2(s)\ ds\right\ra\nn\\
\ \ &=&\ 
 \left\la \chi_1\ ,e^{-i\cH_0t}\int^t_0 We^{-i\sigma_3\left(-\Delta-E_0\right)s} Z P_c\
  e^{i\Omega_1 s}\ \chi_2(s)\ ds\right\ra\nn\\
\ \ &=&\ 
  \left\la \chi_1\ ,e^{-i\cH_0t}\int^t_0 We^{i\sigma_3\Delta s}Z\ \cdot\
  We^{i\sigma_3E_0s} 
 Z P_c\
 e^{i\Omega_1 s}\ \chi_2(s)\ ds\right\ra\nn\\
\ \ &=&\
  \left\la \chi_1\ ,e^{-i\cH_0t}\int^t_0 We^{i\sigma_3\Delta s}Z\ \cdot\
  We^{i(\sigma_3E_0+I\Omega_1)s}
 Z P_c\
\chi_2(s)\ ds\right\ra\nn\\
\ \ &=&\ \left\la \chi_1\ ,e^{-i\cH_0t}\int^t_0 We^{i\sigma_3\Delta s}Z\ \cdot\
  (-i)\frac{d}{ds}W(\sigma_3E_0+I\Omega_1)^{-1} e^{i(\sigma_3E_0+I\Omega_1)s}
 Z P_c\
\chi_2(s)\ ds\right\ra\nn\\
\ \ &\sim&\  -\ 
 \left\la \chi_1\ ,e^{-i\cH_0t}\int^t_0 W\sigma_3\Delta e^{i\sigma_3\Delta s}Z\ \cdot\
  W(\sigma_3E_0+I\Omega_1)^{-1} e^{i(\sigma_3E_0+I\Omega_1)s}
 Z P_c\
\chi_2(s)\ ds\right\ra\nn\\
\ \ &\sim&\  -\ 
 \left\la \chi_1\ ,\int^t_0 e^{i\cH_0 (t-s)}{\tilde \cH}_0
  (\sigma_3E_0+I\Omega_1)^{-1} e^{i\Omega_1s} 
  P_c\
\chi_2(s)\ ds\right\ra
\label{eq:int-by-pts}
\ea
In the previous string of equations we have used the notation $f\sim g$ to
mean equality up to terms which are local in time. 
 Note that $\sigma_3E_0+I\Omega_1$ is  invertible since its determinant
is $\Omega_1^2+E_0^2$. We can therefore carry out this procedure any 
finite number of times to arrange, up to local in time terms, an expression
involves an operator of the form $\chi\exp(i\cH_0 (t-s)){\tilde\cH_0^k}\chi, 
\  k\ge1$,
where $\chi$ is spatially localized. Therefore, the enhanced 
 local decay estimate 
(\ref{eq:H0kldest}) of Theorem \ref{theo:local_decay} applies. We shall use
these observations, together with the detailed dependence of $\chi_2(s)$ on  
on $\alpha_0,\beta_1$ etc. to control certain terms in $R_j^{(i)}$. 
\bigskip

\noindent{\bf Part 1: Estimation of 
$\left|R_0^{(0)}[\eta_0]\right|$ and $\left|R_0^{(1)}[\eta_0]\right|$}

\begin{prop}\label{prop:biR0i}
Assume either $t\le t_0$ or {\bf Montonicity Property  Q} on 
 $[t_0,t_0+\delta_*]$. Then, 
\ba
\left|R_0^{(0)}[\eta_0]\right|,\ \left|R_0^{(1)}[\eta_0] \right| 
 &\leq& b_i\left(t_0, [\eta_0]_X\right) \langle t
\rangle^{-2} + \cO(\cE_0) 2\Gamma P_0P_1^2
\label{eq:R0i}\\
b_0\left(t_0,[\eta_0]_X\right) &=& \cO( \la t_0\ra^{-1}[\eta_0]_X )\nn\\
b_1\left(t_0,[\eta_0]_X\right) &=& \cO( \la t_0\ra^{-{1\over2}} 
 [\eta_0]_X^{2/3} )
\nn\ea
\end{prop}
\medskip

\nit {\bf Proof of Proposition \ref{prop:biR0i}:}
\medskip

\nit \underline{\it Proof of esimate (\ref{eq:R0i}) for $R_0^{(0)}[\eta_0]$:}\ 
 The key terms  are those in the  $P_0$ equation, which are decaying most slowly 
 with $t$. These are linear in $\eta_0$, since 
$\|\eta_0\|_\infty = \cO(t^{-3/2})$.  We focus on the most difficult terms.
These are {\it nonlocal} in $\eta_0(t)$.
Recall equation (\ref{eq:alpha0_eqn2}) for $\alpha_0$ and  that
the equation for $P_0$, (\ref{eq:P0eqn}), 
is derived from the $\talpha_0$ equation
 (related to (\ref{eq:alpha0_eqn2}) by a near-identity change of variables)
by multiplication by $\talpha_0$ and taking the imaginary part of the
$\partial_t\alpha_0$ equation. 

We consider the following representative ``most problematic'' terms in
 $R_0^{(0)}[\eta_0]$, 
 whose estimation
introduces the necessary methods for treating them all:
\ba
&&(T1)\ 
 \cO(\talpha_0) \left\la\chi,\int^t_0 e^{-i\cH_0(t-s)} P_c\Psi_0(s) \Psi_1(s)
{\ol \eta_0}(s) ds \right\ra
\cO\left(\ol{\talpha_0}\tbeta_1e^{-i\lambda_+t}+|\tbeta_1 |^2\right),
\nn\\
&&\label{eq:bi1}\\
&& (\ {\rm also\ with}\ \Psi_0\Psi_1{\ol\eta_0}\ {\rm replaced\ by}\
  \Psi_0{\ol\Psi_1}\eta_0,
\  {\ol\Psi_0}\Psi_1\eta_0\ )\nn\\
&&(T2)\ 
\cO(\talpha_0) \cO(\talpha_0\tbeta_1) \left\la
 \chi, \int^t_0 e^{-i\cH_0(t-s)}
P_c\tetab^2\eta_0 \right\ra ds.
\label{eq:T2}
\ea

%
%

\noindent\underline{\it Estimation of T1:}
 Since the time-integral is bounded by 
$\cO(\cE_0) \la t\ra^{-3/2}\ [\eta_0]_X$, 
by the Cauchy-Schwarz inequality the second term in
 (\ref{eq:bi1}) is bounded as follows:
\be
\bigg|\ \cO(\talpha_0) \cO(|\tbeta_1|^2)\left\la\chi,\int^t_0\cdots\right\ra
\bigg|
\le \cO(\cE_0) P_0 P_1^2 + \cO(\cE_0) \la t\ra^{-3} \|\eta_0\|^2_X.
\nn\ee

We now  control the first term in (\ref{eq:bi1}). 
 $\cO(\bar{\talpha_0}\tbeta_1)\cO(\talpha_0)$.
We argue that the key contribution from this term which must be bounded 
is of the form:
\be
\cO(\talpha_0) \bar{\talpha_0}\tbeta_1 e^{-i\lambda_+t}
 \frac{\partial}{\partial t}
\left\la \chi, \int^t_0 e^{-i\cH_0(t-s)}P_c \Psi_0(s)\Psi_1(s)
 {\ol\eta_0(s)}ds\right\ra
\nn\ee
To see this, consider the term in the $\ta0$ equation which corresponds
to the first term in (\ref{eq:bi1}):
\be
 \left\la \chi,\int^t_0 e^{-i\cH_0(t-s)} P_c\Psi_0(s) \Psi_1(s)
{\ol\eta_0(s)} ds \right\ra
\cO\left(\ol{\talpha_0}\tbeta_1e^{-i\lambda_+t}\right),
\label{eq:a0-bi1}\ee
 Next we integrate with respect to $t$,
and integrate by parts, making using the oscillatory exponential factor. 
 The result
is a boundary term, which can be subsumed in the definition of $\ta0$, by a 
near identity transformation, followed by a time-integral $0$ to $t$.
 The latter contributes  terms to the 
  $P_0$ equation (which has been modified 
  due to the slight redefinition of $\ta0$) of the following type:  
\ba
&&\cO(\talpha_0) \cO(\tb1\D_t\ta0+\ta0\D_t\tb1)\ 
\left\la \chi, \int^t_0 e^{-i\cH_0(t-s)}P_c \Psi_0\Psi_1{\ol\eta_0} ds
\right\ra\label{eq:fst}\\ 
&&\cO(\talpha_0) \cO(\bar{\talpha_0}\tbeta_1)\frac{\partial}{\partial t}
\left\la\chi, \int^t_0 e^{-i\cH_0(t-s)}P_c \Psi_0\Psi_1{\ol\eta_0} ds\right\ra
\label{eq:snd}\ea
Since $\D_t\ta0 = \cO(|\ta0|^2|\tb1|)$ and  $\D_t\tb1=\cO(|\ta0|\ |\tb1|^2)$
(\ref{eq:fst}) is bounded by
\be
\cO(\cE_0)\ \cO\left(\ta0^3\tb1^2\la t\ra^{-\frac{3}{2}}\ [\eta_0]_X\ \right),
\nn\ee
 where we have used
\be\|\eta_0\|_\infty \leq C[\eta_0]_X\ \langle t \rangle^{-3/2}.\nn\ee
By the Cauchy-Schwarz inequality this is bounded by
 $\cO(\cE_0) (2\Gamma P_0
P_1^2+ [\eta_0]_X^2 \langle t\rangle^{-3})$. Therefore, the contribution 
from this first term satisfies the estimate (\ref{eq:R0i}).

Obtaining a bound on (\ref{eq:snd}) is more involved. 
We use the local decay estimate of Theorem \ref{theo:local_decay}:
\be \|\chi e^{-i\cH_0t} P_c{\tilde\cH_0}f\|_2\le\la t\ra^{-5/2}
 \|\D^k\la x\ra^\sigma f\|_2.
\nn\ee
The expression (\ref{eq:snd}) bounded by: 
\ba
&=&\ \cO(\talpha_0) \cO(\bar{\talpha_0}\tbeta_1) \left( \la \chi,
\Psi_0\Psi_1\eta_0\ra -i \la \chi, \int^t_0 e^{-i\cH_0(t-s)}
P_c{\tilde\cH_0}\Psi_0\Psi_1{\ol\eta_0} ds \ra\right)\nn\\
&\le&\ \cO(\cE_0)\left(\ \Gamma Q_0Q_1^2\ +\ \la t\ra^{-3}\ \right)\nn\\ 
&&\ \ +\ \cO(\talpha_0^2 \tbeta_1)\ [\eta_0]_X\  
\int^t_0\frac{ds}{\langle t-s\rangle^{5/2}} \frac{1}{\la s\ra^{3\over2}}
|\talpha_0| |\tbeta_1 | .
\label{eq:bi2}
\ea
 The latter integral requires detailed
estimation on different time scales. 

To estimate the last integral, we split the range of integration
into three regions:
\ba I_0 \equiv \{s: 0\le s\le t_0\}\nn\\
    I_1 \equiv \{s: t_0 <s\le t_1\}\nn\\ 
    I_2 \equiv \{s: t_1 <s\le  T\}\label{eq:intervals}
\ea
\nit{\bf Estimate on $I_0$:}
 Assume $t\le t_0$. Recall that by (\ref{eq:t0def}),  $|\talpha_0(s)|\le
 \cO(\cE_0)\la s\ra^{-1}$.
Using this we have
\ba
&&\cO(\talpha_0^2\tbeta_1)\ [\eta_0]_X\  
\int_0^t \frac{1}{\langle t-s\rangle^{5/2}} \frac{1}{\la s\ra^{3\over2}}
|\talpha_0|\ |\tbeta_1|\ ds\nn\\
&=&\ \cO(|\ta0|^{3\over2}) \left(4\Gamma\talpha_0^{1\over2}\tbeta_1\right)
\ \left(\frac{[\eta_0]_X}{4\Gamma}\ 
 \int_0^t\ \frac{1}{\langle t-s\rangle^{5/2}} \frac{1}{\la s\ra^{3\over2}}
 |\talpha_0|\tbeta_1|\ ds\right)
 \nn\\
&\le& \cO(\cE_0^{3\over4}) \left[ 2\Gamma P_0 P_1^2 + [\eta_0]_X^{4/3}
\la t\ra^{-\frac{5}{2} \cdot \frac{4}{3}}\right],
\nn\ea
where we have used that $ab\le a^4/4 + 3b^{4\over3}/4$.
\bigskip

\nit{\bf Estimate on $I_1$:}
Let $t$ be such that $t_0\le t\le t_1$.
We break the integral into an integral over $[0,t_0]$ plus an integral over
 $[t_0,t]$.
 Recall the definitions of $Q_j(t),\ j=0,1$ in terms of $|{\tilde\alpha}_j(t)$
 displayed in (\ref{eq:Q0Q1def}).
 Using that
\begin{itemize}
\item for $s\in I_0$, $|\ta0(s)|\le
 \cO(\cE_0)\la s\ra^{-1}$ 
and for
\item $s\in I_1 $,
 $Q_0$ is increasing
 and $Q_0(s)\le \cE_0^r Q_1(s)$, 
\end{itemize}
we have 
\ba
&&\cO(|\ta0(t)|^2|\tb1(t)|) [\eta_0]_X\
\left(\ \int_0^{t_0}+\int_{t_0}^t\ \right)
\frac{1}{\la t-s\ra^{5/2}\la s\ra^{3/2}}
 |\tb1(s)|\ |\ta0(s)|\ ds\nn\\
&\le&\
I_0\ {\rm type\ bound}\ +\ \cO(|\ta0|^2\ |\tb1|)\ [\eta_0]_X\ 
 \int_{t_0}^t\ 
\frac{|\tb1(s)|\ |\ta0(s)|}{\la t-s\ra^{5/2}\la s\ra^{3/2}}
\ ds\nn\\
&\le&\ I_0\ {\rm type\ bound}\ +\ \cO(|\ta0|^3\ |\tb1|\ \|\tb1\|_\infty)\ 
 [\eta_0]_X\ 
 \int_{t_0}^t\ \frac{1}{\la t-s\ra^{5/2}\la s\ra^{3/2}}
 \ ds\nn\\ 
&=&\ I_0\ {\rm type\ bound}\ +\ \cO\left(\cE_0^{1\over2}|\ta0|\cdot 
 |\ta0|\cdot |\ta0| |\tb1|\right)\ [\eta_0]_X\  \la t\ra^{-{3\over2}} 
 \ ds\nn\\
&=&\ I_0\ {\rm type\ bound}\ +\ \cO\left(\cE_0\cdot 
 Q_0^{1\over2}\cdot\cE_0^{r\over2}Q_1\right)\ [\eta_0]_X\ 
 \la t\ra^{-{3\over2}} 
 \nn\\
&=&\ I_0\ {\rm type\ bound}\ +\
 \cO\left(\cE_0^{{r+2\over2}}\right)\
  Q_0^{1\over2}Q_1\ \la t\ra^{-{3\over2}}\nn\\
&=&\ I_0\ {\rm type\ bound}\ +\
 \cO\left(\cE_0^{{r+2\over2}}\right)\left(\ Q_0Q_1^2\ +\ \la t\ra^{-3}\ \right),
\ea
which is a bound of the type in (\ref{eq:R0i}).

\nit{\bf Estimate on $I_2$:}
 
\ba
&& \cO(\talpha_0^2 \tbeta_1(t))
 [\eta_0]_X \int^t_0
\frac{|\talpha_0| | \tbeta_1 |}{\la  t-s\ra^{5/2} \la s\ra^{3/2}}\ ds
\nn\\
&=& \cO(\talpha_0^2\tbeta_1 (t) [\eta_0]_X) \left(\int^{t_0}_0 +
\int^{t_1}_{t_0} + \int^t_{t_1}\right)\ 
\frac{|\talpha_0| | \tbeta_1 |}{\la
t-s\ra^{5/2} \la s\ra^{3/2}}\ ds\nn\\
&=& I_0\ \&\ I_1\ {\rm type\ bounds}\ +\ 
\cO(\talpha_0^2 \tbeta_1(t))
 [\eta_0]_X\ \int^t_{t_1}
 \frac{|\talpha_0| | \tbeta_1 |}{\la
t-s\ra^{5/2} \la s\ra^{3/2}}\ ds\nn\\
&&\label{eq:soit}\ea

The latter integral must be treated differently from the previous terms.
For $t\in I_1$, we used that $Q_0$ monotonically increasing and 
bounded by a small constant times
 $Q_1$ to treat terms perturbatively. On $I_2$, $Q_0$ dominates $Q_1$
 (which decays) and we must use a different argument. We return to 
expression from which the last term in (\ref{eq:soit}) is derived: 
\be
\cO(\talpha_0) \cO(\bar{\talpha_0}\tbeta_1) \left( \la \chi,
\Psi_0\Psi_1\eta_0\ra -i \la \chi, \int^t_0 e^{-i\cH_0(t-s)}
P_c{\tilde\cH_0}\Psi_0\Psi_1{\ol\eta_0} ds \ra\right).
\label{eq:soit1}
\ee
We need to expand and estimate time integral:
\be
\int_{t_1}^t e^{-i\cH_0(t-s)} P_c{\tilde\cH_0}
\ta0(s)\tb1(s)e^{-i\lambda_+s}{\ol\eta_0(s)} ds\
 =\ e^{-i\cH_0t}\ \int^t_{t_1} e^{i(\cH_0-\lambda_+)s} P_c\cH_0 
\ta0(s)\tb1(s){\ol\eta_0(s)} ds\nn\ee
which we do using integration by parts. We carry this out, 
 then take the inner 
 product of the result with a localized function, $\chi$, and 
then finally multiply by 
$\cO(|\ta0|^2\ |\tb1|)$. The result is of order
\ba
&& |\ta0|^3\ |\tb1|^2\ \|\eta_0(t)\|_\infty\ +\ 
 |\ta0|^3\ |\tb1|\ \left|\left\la\chi,e^{-i\cH_0t}
 {\tilde\cH_0}\chi\right\ra\right|
\nn\\
&&+\ \cO(|\ta0|^2\ |\tb1|)\ \int_{t_1}^t e^{-i\cH_0(t-s)}\ P_c  
 (\cH_0-\lambda_+)^{-1}{\tilde\cH_0}\chi e^{-i\lambda_+s}
\frac{d}{ds}\left( \ta0(s)\tb1(s){\ol\eta_0(s)}\right)\ ds
\nn\\
\label{eq:soit2}
\ea
The first two terms in (\ref{eq:soit2}) are bounded as follows:
\ba
&&|\ta0|^3\ |\tb1|^2\ \|\eta_0(t)\|_\infty\ = \cO(\cE_0)\ (|\ta0|\ |\tb1|^2)\ 
\|\eta_0(t)\|_\infty\nn\\
&& \le\ \cO(\cE_0)\left(\ Q_0Q_1^2\ +\ \|\eta_0(t)\|_\infty^2\
 \right)\nn\\
&& \le\ \cO(\cE_0)\left(\ Q_0Q_1^2\ +\ \la t\ra^{-3}\ \right)
\nn\ea
\ba
&& |\ta0|^3\ |\tb1|\ \left|\left\la\chi,e^{-i\cH_0t}{\tilde\cH_0}
 \chi\right\ra\right|
\le\ C|\ta0|^{5/2}\ (|\ta0|^{1/2}\ |\tb1|)\ \la t\ra^{-5/2}\nn\\ 
&&\le \cO(\cE_0^{5/2})\left( Q_0Q_1^2\ +\ \la t\ra^{-{5\over2}\cdot{4\over3}}\
  \right)
\ \le\ \cO(\cE_0^{5/2})\left( Q_0Q_1^2\ +\ \la t\ra^{-3}\ \right).
\ea

We now turn to the nonlocal term, 
  in (\ref{eq:soit2}),  which we denote $\cI_1$:
\ba
\cI_1\ \sim\  \cO(|\ta0|^2\ |\tb1|)\  &&\int_{t_1}^t e^{-i\cH_0(t-s)}\ P_c
 (\cH_0-\lambda_+)^{-1}{\tilde\cH_0}\chi e^{-i\lambda_+s}\nn\\
   && \left(\ \D_s\ta0(s)\tb1(s){\ol\eta_0(s)}+
                   \ta0(s)\D_s\tb1(s){\ol\eta_0(s)}+
                  \ta0(s)\tb1(s)\D_s{\ol\eta_0(s)}\ \right).\nn\\
&&\label{eq:int1120}
\ea
Using that
\be \D_t\ta0\sim\cO(\tb1^2\ta0),\ \D_t\tb1\sim\cO(\ta0\tb1^2),\
 {\rm and}\ \D_t\eta_0\ =\ -i\cH_0\eta_0,\label{eq:Dt-ests}
\ee
we have
\ba
\cI_1\ \sim\  \cO(|\ta0|^2\ |\tb1|)\ &&\int_{t_1}^t e^{-i\cH_0(t-s)}\ P_c
 (\cH_0-\lambda_+)^{-1}{\tilde\cH_0}\chi e^{-i\lambda_+s}\nn\\
   &&\ \ \left(\ \ta0\tb1^3{\ol\eta_0}+
                   \ta0^2\tb1^2{\ol\eta_0(s)}+
                  \ta0(s)\tb1(s){\tilde\cH_0}{\ol\eta_0(s)}\ \right)\nn\\
&=&\  \cI_{1a}\ +\ \cI_{1b}\ +\ \cI_{1c}
\label{eq:I1abc}
\ea
Each of the three terms $\cI_{1j},\ j=a,b,c$ satisfies a bound of the form:
\be
\left|\cI_{1j}\right|\ \le\ C\ \cE_0^2\ [\eta_0(0)]_X\
                        \frac{1}{\la t_0\ra}\frac{1}{\la t-t_1\ra^2}
\label{eq:Ij-bound}
\ee

We illustrate this by estimating $\cI_{1a}$; the other two terms
are estimated similarly.

Using that $|\ta0|,\ |\tb1|\ =\ \cO(\cE_0^{1\over2})$ we have  
\ba
\left|\cI_{1a}\right|\ &\le&\ |\ta0|^2\ |\tb1|\
  \int_{t_1}^t \frac{1}{\la t-s\ra^{5\over2}}\
               |\tb1|^3\ |\ta0|\ |\eta_0|\ ds\nn\\
&\le&\ \cE_0^2\ [\eta_0(0)]_X\ 
 \left(\sup_{t\ge t_1} \la t-t_1\ra^{1\over2}|\tb1(t)|\right)^3
\int_{t_1}^t \frac{1}{\la t-s\ra^{5\over2}}
 \frac{1}{\la s-t_1\ra^{3\over2}}\frac{1}{\la s\ra^{3\over2}}
\nn
\ea
Separate estimation of the contributions from the intervals
 $[t_1,{1\over2}(t+t_1)]$ and $[{1\over2}(t+t_1),t]$ yields the bound:
\be
\left|\cI_{1a}\right|\ \le\ C\ \cE_0^2\ [\eta_0(0)]_X\ 
   \left(\sup_{t\ge t_1} \la t-t_1\ra^{1\over2}|\tb1(t)|\right)^3                     \frac{1}{\la t_0\ra}\frac{1}{\la t-t_1\ra^2}
\label{eq:I1a-bound}
\ee


\noindent\underline{\it Estimation of T2:}
Consider the term
\be
{\rm T2}\ =\ \cO(\talpha_0(t)) \cO(\talpha_0(t)\tbeta_1(t)) \left\la 
 \chi, \int^t_0 e^{-i\cH_0(t-s)}
P_c\tetab^2(s)\eta_0(s) \right\ra\ ds.
\label{eq:T2-def}
\ee
\medskip

\nit \underline{\it $t\in I_1\ \equiv\ [t_0,t_1] $}:\ 
For $t\in I_1$,  
\be
|\talpha_0|^2\ |\tbeta_1| \leq\ c\cE_0^{r\over2} |\talpha_0|\ |\tbeta_1|^2;\
 \ {\rm  Proposition\  \ref{prop:propC}}.
\nn\ee
Therefore,
\ba
|{\rm T2}|\ &\le&\ c\cE_0^{r\over2}\ |\talpha_0|\ |\tbeta_1|^2 
 \left|\left\la\ \chi, 
 \int^t_0 e^{-i\cH_0(t-s)}P_c \tetab^2 \eta_0 ds\ \right\ra\right|\nn\\
&\le& 
 c\cE_0^{r\over2}\ |\talpha_0|\ |\tbeta_1|^2 
 \la t\ra^{-3/2} \left(\sup_{0\le s\leq t}\la s\ra^{3\over2}
  \| \tetab^2 (s) \eta_0(s) \|_{W^{k,2}\cap L^1} \right)\nn\\
&\le& c\cE_0^{r\over2}\ |\talpha_0|\ |\tbeta_1|^2\ \la t\ra^{-3/2}
  \|\tetab\|^2_{W^{k,2}} [\eta_0]_X \nn\\
&\le&\ c
\cE_0^\rho [\eta_0]_X\ |\talpha_0|\ |\tbeta_1|^2\  \la t \ra^{-3/2}\nn\\
&\le&\ c\cE_0^\rho [\eta_0]_X\left(\ P_0P_1^2\ +\ \la t \ra^{-3}\ \right)
\nn\ea
\bigskip

\nit \underline{$t\in I_2\equiv [t_1,\infty)$:}

\nit To see the relevant terms for $t> t_1$, we integrate by parts
and obtain, besides easily estimable local terms and
 terms with faster time decay, 
\be
O(\ta0^2\tb1) \left\la\ \chi, \int^t_0 e^{-i\cH_0(t-s)}
  {\tilde\cH_0}P_c\tetab^2\eta_0\ ds\ \right\ra  \quad t> t_1.
\label{eq:T2term}
\ee
Consider the contribution to the integral 
in (\ref{eq:T2term}) coming from  
$s\in[0,t_0]$: 
\be
\cO\left(\talpha_0^2(t)\beta_1(t)\right) 
\left\la\chi, \int_0^{t_0} e^{-i\cH_0(t-s)}
  {\tilde\cH_0}P_c\tetab^2\eta_0\ ds\ \right\ra \quad t> t_1>t_0
\label{eq:T2terma}
\ee
Consider, the inner produce
\be
R(t)\ =\ \left\la \chi,\int_0^{t_0} e^{-i\cH_0(t-s)}
  {\tilde\cH_0}P_c\tetab^2\eta_0\ ds\  \right\ra 
\label{eq:Rdef}
\ee 
We have 
\be \left|{\rm (\ref{eq:T2terma})}\right|\ \le\ 
 C\ |\ta0(t)|^2 |\tb1(t)|\ |R(t)|\  \sim\ CQ_0(t) Q_1^{1\over2}(t) |R(t)|
\label{eq:T2termz}
\ee
and therefore it suffices to prove that 
\be |R(t)|\ \le\ CQ_1^{3\over2}(t)\label{eq:Rbound}\ee  
To prove (\ref{eq:Rbound}), recall that for $t\ge t_1$ y 
\ba
\frac{dQ_0}{dt}\ &\asymp&\ \Gamma Q_0Q_1^2\nn\\
\frac{dQ_1}{dt}\ &\asymp&\ -2\Gamma Q_0Q_1^2,\
\label{eq:T2c}
\ea
and
\be \cE_0^rQ_1(t_1)\ =\ Q_0(t_1).
\label{eq:T2d}
\ee
Therefore,
\be
Q_1(t)\ \asymp\  \frac{2Q_1(t_1)}{1\ +\ \Gamma\int_{t_1}^tQ_0(s)ds}
\label{eq:Q1asymp}
\ee
Therefore, to establish (\ref{eq:Rbound}) we need:
\be
|R(t)|\ \le\ C\  
\left(\ \frac{2Q_1(t_1)}{1\ +\ \Gamma\int_{t_1}^tQ_0(s)ds}\ 
     \right)^{3\over2}
\label{eq:Rbound1}
\ee
We consider two cases

\nit Case 1: $\Gamma\ \sup_{s\in[t_1,t]}|Q_0(s)|\ |t-t_1|\ \le\ 1$,
and

\nit Case 2: $\Gamma\ \sup_{s\in[t_1,t]}|Q_0(s)|\ |t-t_1|\ \ge\ 1$,
where it suffices to prove
\be
|R(t)|\ \le\ \left(\ \frac{3Q_1(t_1)}{ \Gamma\int_{t_1}^tQ_0(s)ds}\ 
     \right)^{3\over2} 
\label{eq:case2bound}
\ee
In case 2 we prove the bound on $R(t)$, (\ref{eq:Rbound}), while 
in case 1 we prove that the expression (\ref{eq:T2term}) of order
$\la t-t_1\ra^{-{3\over2}}\la t-t_0\ra^{-{3\over2}}$.
\nit We first handle case 2, the bound (\ref{eq:case2bound}).
  From (\ref{eq:T2c}) we have 
\ba
\frac{d}{dt}\left(\ Q_0\ +\ 2Q_1\ \right)\ &\asymp&\ 0,\ t\ge t_1
\nn\\
Q_0(t)\ +\ 2Q_1(t)\ &\asymp&\ Q_0(t_1)\ +\ 2Q_1(t_1)\nn\\
                    &=&\ (2+\cE_0^r)\ Q_1(t_1)
\label{eq:T2e}
\ea
Therefore, since
\be
Q_1(t)\ \asymp\ \frac{2Q_1(t_1)}{1\ +\ \Gamma\int_{t_1}^tQ_0(s)ds}
\label{eq:Q1asym}
\ee
we have, as $t\to\infty$
\ba
Q_0(t)\ &\asymp&\ (2+\cE_0^r)Q_1(t_1)\ -\ 2Q_1(t)\nn\\
        &\longrightarrow&\ (2+\cE_0^r)Q_1(t_1)
\label{eq:T2f}
\ea
Therefore, for $t>t_1$ ($t-t_1$ large enough)
\be
(1+\cE_0^r)Q_1(t_1)\ \le\ Q_0(t)\ \le\ (2+\cE_0^r)Q_1(t_1)
\nn\ee
and therefore
\ba
 \frac{Q_1(t_1)}{1\ +\ \Gamma\int_{t_1}^tQ_0(s)ds}
    &\ge&\ \frac{Q_1(t_1)}{1\ +\ \Gamma(2+\cE_0^r)\ |t-t_1|\ Q_1(t_1)}\nn\\
    &\ge&\ \frac{1}{3}\frac{1}{\Gamma\ |t-t_1|}\ \ge 
 \frac{1}{3}\frac{1}{\Gamma\ |t-t_0|}
\label{eq:Q1lba}
\ea
On the other hand,
\ba
\left| R(t) \right|\ &\le&\
 C\ 
  \left\la \chi,\int_0^{t_0} \frac{1}{\la t-s\ra^{3\over2}}
\| {\tilde\cH_0}\tetab^2(s)\eta_0(s)\|_{L^1}\ ds\ \right\ra \nn\\
 &\le&\ C\ \sup_{\tau\in [0,t_0]}\|\tetab(\tau)\|_{H^2}^2\ 
                                 [\eta_0]_X\ 
      \int_0^{t_0} 
 \frac{1}{\la t-s\ra^{3\over2}\la s\ra^{3\over2}} \ ds\nn\\
&\le&\ C\ \cE_0^2\ [\eta_0]_X\  
       \frac{1}{\la t-t_0\ra^{3\over2}}
\label{eq:T2termb}
\ea 
The bounds (\ref{eq:T2termb}) and (\ref{eq:Q1lba}) imply (\ref{eq:case2bound}).
\medskip

We now turn to case 1. In this case, 
 $\Gamma\sup_{s\in [t_1,t]}Q_0(s)\ |t-t_1|\le 1$, 
\be Q_0(t)\ \le\ {1\over\Gamma}\frac{1}{|t-t_1|}
\label{eq:Q0ub}
\ee
and therefore by (\ref{eq:T2term}) and (\ref{eq:Q0ub})
\be
\left| {\rm (\ref{eq:T2term})}\right| 
\le
\cE_0^2[\eta_0]_X\ \sup_{s\in[t_1,t]}\left(\la s-t_1\ra^{1\over2}Q_1(s)\right)
  \frac{1}{|t-t_1|}\frac{1}{\la t-t_1\ra^{1\over2}} 
  \frac{1}{\la t-t_0\ra^{3\over2}}
\label{eq:T2termc}
\ee

We begin by noting that by Proposition \ref{prop:disip}
and its corollaries, 
 for $p>6$
\ba \|\tetab\|_{W^{k,p}}&\le&\cO(\cE_0) \la t\ra^{-1},\ \  t\leq t_0\nn\\
 \|\tetab \|_{W^{k,p}} &\le& \cO(\cE_0^\rho) Q_0 (t)^{1/2},\ \    
  t_0<t\le t_1\nn\\
\|\tetab\|_\infty &=& \cO(\la t-t_1\ra^{-{1\over2}}), t\ge t_1
\label{eq:tetabWkp}\ea

For $t< t_1$, the previous arguments with the known estimates on
$\tetab$, and the facts that $|\talpha_0(t)| \leq \cO(\cE_0)/\la t\ra$
on $I_0$ and 
 $Q_0(t) \leq \cE_0^rQ_1$  on $I_1$
imply the necessary bounds.
%
%
Collecting all these, we have 
$$
\cO (\talpha^2_0\tbeta_1) \bigg|\langle \chi, \int^t_0 e^{i\cH_0(t-s)}P_c
{\tilde\cH_0}\tetab^2 \eta_0 ds \rangle \bigg| \leq \cO(\cE_0) \left[2\Gamma
Q_0 Q_1^2 + \langle t_0\rangle^{-1} \langle t \rangle^{-2}\right].
$$  
\medskip

\nit\underline{Proof of estimate (\ref{eq:R0i}) for  $R_0^{(1)}$:}
The key terms to consider in the  $P_1$ equation are the slowest
decaying nonlocal terms; see (\ref{eq:beta_eqn3}) and
 (\ref{eq:eta1_eqn}).
Since $b_1$ enters as $b_1^2$ in 
 (\ref{eq:k0_def}), we
need to bound $R_0^{(1)}$ by $\cO(\cE_0)\left[\langle
t_0\rangle^{-1/2} \langle t \rangle^{-2} + 2\Gamma P_0 P_1^2\right]$
for $t> t_0$.

The slowest decaying nonlocal term in the $\beta_1$ equation arises from
the balance:
\be
i\D_t\beta_1\ \sim\ 
\lambda\la\psi_{1*}^3,\pi_1\Phi_2\ra |\beta_1|^2e^{i\lambda_+(T)t} 
\nn\ee
Since the $P_1$ equation is obtain by multiplying by $\ol{\beta_1}$ and 
taking the imaginary part, we must estimate:
\be
\lambda\la\psi_{1*}^3,\pi_1\Phi_2\ra\ |\beta_1|^2\ol{\beta_1}\
  e^{i\lambda_+(T)t},
\nn\ee
the leading order nonlocal part of which is
\ba
&\sim& |\tbeta_1|^2 \ol{\tbeta_1} \left\la \chi, \int^t_0 e^{-i\cH_0(t-s)}
P_c\Psi_0\Psi_1 \eta_0 ds \right\ra e^{i\lambda_+ t}\nn\\
&\sim& |\tbeta_1|^2 \ol{\tbeta_1} \left\la \chi, \int^t_0 e^{-i\cH_0(t-s)} P_c
\talpha_0 \tbeta_1 e^{-i\lambda_+s} \chi \eta_0 (s) ds\right\ra
e^{i\lambda_+ t}
\nn\ea
Integrating by parts (using the oscillatory factor $e^{i\lambda_+t}$) and 
 removing nonresonant local in terms  by near identity transformations the 
key term is
\be 
 |\tbeta_1|^2 \ol{\beta_1} 
\left\la \chi, \int^t_0 e^{-i\cH_0(t-s)}{\tilde\cH_0}
P_c \talpha_0\tbeta_1 \chi e^{i\lambda_+ s} \eta_0 (s) ds\right\ra
e^{i\lambda_+ t}.\nn
\ee

Suppose $t> t_1$.  Since we have no factor of $\talpha_0$ outside the
integral (local in time $\alpha_0(t)$ factor), 
 the estimate for $I_0=\{0 < s < t_0\}$ is
the critical part and requires the $\cO(\la t\ra^{-1})$ bound on $I_0$ 
 and Theorem \ref{theo:local_decay}:
\ba
&&\left|\left\la \chi, \int^t_0 e^{-i\cH_0(t-s)}{\tilde\cH_0} 
 P_c \talpha_0 \tbeta_1 \chi
\eta_0 e^{-i\lambda_+ s} ds \right\ra \right|\nn\\
&\leq& C[\eta_0]_X 
 \int^{t_0}_0 \frac{ds}{\la t-s\ra^{5/2}\la s \ra^{3/2}}|\talpha_0(s) |
|\tbeta_1(s)|\  + \left(\ \int^{t_1}_{t_0} +
\int^t_{t_1}\ \right)\cdots\ ds\nn\\
&\leq& \cO(\cE_0) [\eta_0]_X
 \int^{t_0}_0 
  \frac{ds}{\la t-s\ra^{5/2}\la s \ra \la s\ra^{3/2}} 
\ + [\eta_0]_X\ \int^{t_1}_{t_0}
 \frac{ds}{\la t- s\ra^{5/2}\la s \ra^{3/2} }
|\talpha_0(t_1)|\ |\tbeta_1(s)|  
 + \int_{t_1}^t\cdots\ ds\nn\\
&\leq& \cO(\cE_0) \left[\la t\ra^{-5/2} [\eta_0]_X +
 \cO(\cE_0)[\eta_0]_X  |\talpha_0 (t) | \la t\ra^{-3/2} +
[\eta_0]_X |\talpha_0(t)|\  
 \la t-t_1\ra^{-{1\over2}}\la t\ra^{-{3\over2}} \right].
\nn\ea
Note that we can extract the factor $|\talpha(t)|$ from the integral for 
 $s\ge t_0$ since in this range $Q_0(t)$ is monotonically increasing and 
 $Q_0(t)\sim P_0(t)\sim |\talpha_0(t)|$ with correction terms which are 
 which are rapidly decaying in time and which are 
 therefore dominated by the
first term. Multiplication by $\cO(|\tbeta_1|^3)$ prefactor gives the 
bound
\be
C(\cE_0,[\eta_0]_X)\left[\
 \left(\sup_{s\ge t_1}\la s-t_1\ra^{1\over2}|\tbeta_1(s)|\right)^3
    \la t\ra^{-5/2}\la t-t_1\ra^{-3/2}\ +\ Q_0Q_1^2\ \right]
\nn\ee
This completes the proof of Proposition \ref{prop:biR0i}.
\bigskip

\section{Local and nonlocal ODE terms: $R_2^{(j)}(P_0,P_1)$ of Proposition
 \ref{prop:Master}}\label{sec:R120P0P1}

In this section we prove estimates on the terms $R_2^{(0)}(P_0,P_1)$ 
and $R_2^{(1)}(P_0,P_1)$ of Proposition \ref{prop:Master}. 

\begin{prop}\label{prop:ODE}
Assume either $t\le t_0$ or Monotonicity Property {\bf Q} on 
 $[t_0,t_0+\delta_*]$. Then,
For $m\ge4$,
\ba
&&(1)\ |R_2^{(0)}(P_0,P_1)|\le \cO(\cE_0^\rho)\Gamma P_0P_1^2 + 
 \cO(\la t\ra^{-3})
\  (P_0, P_1 << 1 )\nn\\
&&(2)\ |R_2^{(1)}(P_0,P_1)|\le \cO(\cE_0^\rho)\Gamma P_0 P_1^2 + 
\cO(\alpha_0)P_1^{2m} + \cO(\la t\ra^{-3}) , \ \   |\alpha_0|\sim\sqrt{P_0}. 
\nn\ea
\end{prop}

\subsection{Proof of part (1) of Proposition \ref{prop:ODE}:}

The most problematic terms are nonlocal, slowest
decaying.
The terms which are linear in $\eta$ in the $\alpha_0$
equation contribute the slowest terms; these are nonlocal in time, $t$.

We have to consider terms arising in equation (\ref{eq:alpha0_eqn2})
 of the type:
\be
\la\chi, \eta\ra e^{-i\lambda_+ t} \ol{\alpha_0} \beta_1,\   
\la \chi, \eta\ra |\beta_1|^2,\  
\la \chi, \eta\ra \beta_1^2 e^{-2i\lambda_+ t},
\label{eq:typ1}
\ee
where $\phi_2\sim\eta$ and $\eta(t)=\eta_0(t)+e_0(t)+\eta_b(t)$;
see (\ref{eq:eta-decomp}).

As calculated earlier, the 
 last term is resonant and its contribution is 
$+2\Gamma P_0 P_1^2$ to the $P_0$ equation; see (\ref{eq:keyterm}).

The leading ODE terms in $\eta$ are the source terms of the type
in (\ref{eq:eta1_eqn}):
\be
|\Psi_0|^2\Psi_1, \Psi_0^2 \ol{\Psi_1}, \Psi_0 |\Psi_1|^2 , 
\ol{\Psi_0} \Psi_1^2;
\label{eq:typ2}
\ee
Recall that the $P_0$ equation is obtained from the $\tilde\alpha_0$
equation by multiplication by $\ol{\tilde\alpha_0}$ and taking the  
 imaginary part. 

Solving for $\tetab$ and plugging the source term contributions
into the $\langle \chi, \eta\rangle$ terms in the $P_0$ equation
gives, apart from the resonant term, terms of the type
\be
\ol{\alpha_0} \cO (\bar \alpha_0 \beta_1 e^{i\lambda_+ t} + |\beta_1|^2)
\langle \chi, \int^t_0 e^{-i\cH_0(t-s)} P_c \left(\ |\Psi_0|^2 \Psi_1 +
\Psi_0^2 \bar \Psi_1 + \Psi_0|\Psi_1|^2 + \bar \Psi_0 \Psi_1^2\ \right)
ds
\label{eq:typ3}
\ee
For $t>t_1$ the slowest terms are $\cO(|\Psi_0|^2 \Psi_1 + \Psi_0^2
\bar\Psi_1)$ source terms. For $t_0<t<t_1$ the problematic terms are
$\cO(\Psi_0 |\Psi_1|^2 + \bar \Psi_0 \Psi^2_1)$  source terms, since
on this interval $\beta_1$ does not necessarily decay.
Since $\Psi_1 \sim e^{-i\lambda_+ t} \beta_1 + \ {\rm  higher\ order
 \ terms},$
we can integrate by parts the $\cO(|\Psi_0|^2 \Psi_1 + \Psi_0^2 \bar
\Psi_1)$ terms in (\ref{eq:typ3})
 to get arbitrary order $\beta_1^m$ terms, which are nonlocal.
 The remaining terms are local and its lowest order term is of the order
\be
\bar \alpha_0 \cO(\bar\alpha_0 e^{i\lambda_+t}\beta_1 + 
 |\beta_1|^2) \left\la \chi,  \cO
(|\alpha_0|^2 \beta_1 e^{i\lambda_+t} + \alpha_0^2 
 \bar\beta_1 e^{-i\lambda_+t})\right\ra.
\nn\ee
The nonresonant local 
 terms can be transformed to higher order by a near identity
change of variables giving
$$
\cO(|\alpha_0|^3\beta_1^{2m} e^{i\Omega t}),\ \Omega>0\ {\rm and}\ 
\  m\ {\rm arbitrary\  large},
$$
while resonant terms are of the type already derived but of higher order
 (bounded by $\cO(\cE_0)Q_0Q_1^2)$).
We are therefore left to consider the following nonlocal terms 
\be
\bar \alpha_0
  \cO\left( \bar\alpha_0\beta_1 e^{-i\lambda_+ t} + |\beta_1|^2\right) 
\left\la \chi,
\int^t_0 e^{-i\cH_0(t-s)}P_c \alpha_0 \beta_1^{2m} e^{i\Omega_m s}\
  \tilde \chi ds\right\ra
\label{eq:nlc}
\ee

Consider the first term in (\ref{eq:nlc}). 
 Due to the oscillatory factor $e^{-i\lambda_+ t}$ in the 
 $\cO(\bar\alpha_0
\beta_1 e^{-i\lambda_+ t})$ term we can, 
 by a near-identity transformation,  
in the $\alpha_0$ equation, remove this term in exchange for one 
 higher order
\ba
 && \cO(|\alpha_0|^2 \beta_1^3) \left\la \chi, 
 \int^t_0\cdots \right\ra + 
 \cO(\bar\alpha_0^2 \beta_1 e^{i\Omega s}) \partial_t \left\la \chi, \int^t_0 e^{-i\cH_0(t-s)}
P_c\ \cdots\ \right\ra\nn\\
&& +\ {\rm higher\ order}\  + \ {\rm local\  terms}, 
\nn\ea
and by further near identity transformations, we are left 
 with the following slowest term
\be
\cO(|\alpha_0|^2\beta_1) \partial^m_t
\left\la \chi, \int^t_0 e^{-i\cH_0(t-s)}
\ P_c\cdots\ ds\ \right\ra.
\label{eq:slow}
\ee
All other terms are of order $o(\cE_0)\cO(P_0 P_1^2)$, 
 $o(\cE_0)\la t\ra^{-3}$  
 or higher order, for $\cE_0$ small.

\ba
&&\partial_t \langle \chi, \int^t_0 e^{-i\cH_0(t-s)}
P_c\ \alpha_0\beta_1^{2m}\tilde \chi ds\rangle\nn\\
&=& \left\la\chi, \alpha_0(t) \beta_1^{2m}(t) \tilde\chi \right\ra -
\left\la \chi, i \int^t_0 e^{-i\cH_0(t-s)} {\tilde\cH_0}
  P_c \tilde\chi\alpha_0 \beta_1^{2m} e^{i\Omega s}
ds\right\ra.
\nn\ea 
The first term is local and will contribute $o(1) \cO (P_0 P_1^2)$.
It remains to estimate the  nonlocal term.
 Note that 
$$
\int^t_0 e^{-i\cH_0(t-s)} e^{i\omega s} \tilde \chi \alpha_0\beta_1^{2m}ds 
 = e^{-i\bar \omega t}
\int^t_0 e^{-i(\cH_0-\bar\omega)(t-s)} e^{i(\omega+\bar\omega)s}
 \tilde\chi \alpha_0\beta_1^{2m} ds
$$

Consider now the second term in  (\ref{eq:nlc}).
Using the oscillatory factor $e^{-i\bar \omega t}$, one can transform  
 the $\cO(\bar\alpha_0|\beta_1|^2) \langle \chi,
\int\cdots \rangle$ to higher order.

We now turn to the second term in (\ref{eq:nlc}). First, let's
consider the case where $t\ge t_1$.  
\ba
&&\cO(\bar\alpha_0 |\beta_1|^2) \partial^m_t \left\la\ \chi,
\int_0^t e^{-i\cH_0(t-s)} \tilde \chi \alpha_0\beta_1^{2m}\ ds\ 
 \right\ra\label{eq:typ4}
\ea
We consider (\ref{eq:typ4})
  for $t\ge t_1$. For this we require the following

\begin{lem}\label{lem:Q1lbound}  
For $t> t_1,$
\be
Q_1(t) \geq Q_1(t_1) \left[ 1+4(\Gamma' + \delta) Q_1 (t_1) Q_0 (t_1)
(t-t_1) \right]^{-1}
\label{eq:Q1lb}
\ee
\end{lem}
{\bf Proof of Lemma \ref{lem:Q1lbound}:}  For $t\ge t_1$ 
\be
\frac{dQ_1}{dt} \geq - 4 (\Gamma' + \delta) Q_0 Q_1^2
\nn\ee
Therefore, 
\be
-\frac{dQ_1^{-1}}{dt} \geq - 4 (\Gamma' + \delta)Q_0(t)
 \ge - 4 (\Gamma' + \delta)Q_0 (t_1)\ {\rm or}\ \
  \frac{dQ_1^{-1}}{dt} \leq 4 (\Gamma' + \delta) Q_0 (t_1)
\nn\ee
since $Q_0 (t_1) \leq Q_0 (t)$ ($Q_0\uparrow$ for $t>t_0$).
Integrating from $t$ to $t_1$, we get
\be
Q_1(t)^{-1} - Q_1(t_1)^{-1} \leq 4 (\Gamma' + \delta) Q_0 (t_1) (t-t_1),
\nn\ee
which is equivalent (\ref{eq:Q1lb}). This completes the proof of Lemma 
\ref{lem:Q1lbound}. 

\nit{\bf Estimation of (\ref{eq:typ4}) for $t\ge t_1$} 
Carrying out the differentiation in (\ref{eq:typ4}) we find that it suffices
to bound terms of the type:
\be 
\cO\left(\alpha_0 |\beta_1|^2\right) 
\left\la\chi, P_c\chi \right\ra\ \alpha_0\beta_1^{2m}
 \ =\ \cO(\cE_0)Q_0Q_1^2\label{eq:1st}\\
\cO\left(\alpha_0 |\beta_1|^2\right) 
 \left\la\chi,
 \int_0^te^{-i\cH_0(t-s)}P_c{\tilde\cH_0}^m \tilde\chi \alpha_0\beta_1^{2m}\ ds
\right\ra
\label{eq:2nd}\ee

We now consider (\ref{eq:2nd}), which we  break into the sum of 
three integrals:   
\be \cO(\bar\alpha_0 |\beta_1|^2) \left\langle\ \chi,
\left(\ \int_0^{t_0}+\int_{t_0}^{t_1}+\int_{t_1}^t\right)
 e^{-i\cH_0(t-s)}\ P_c\ {\tilde\cH_0}^k \tilde \chi \alpha_0\beta_1^{2m}\ 
 ds\ \right\rangle\
\label{eq:typ5}
\ee
\bigskip

\nit {\bf $\int_{t_1}^t:$}
By the local decay estimate for $e^{-i\cH_0 t}\cH_0^k P_c$ of
   Theorem \ref{theo:local_decay}, 
the integral in (\ref{eq:typ5}) is bounded above by:
\be
\int^t_{t_1}\frac{ds}{\la t-s\ra^k} |\alpha_0 \beta_1^{2m} |
\nn\ee
This in turn is bounded above by 
\be
Q_0(t)^{1/2} \int^t_{t_1} \frac{ds}{\langle t-s\rangle^k} Q_1^m(s),
\label{eq:typ6}\ee
since $Q_0$ is increasing for $t\ge t_0$.

We now aim to  further bound (\ref{eq:typ6}) by ``extracting'' powers
of $Q_1(t)$ from under the integral.  Recall that for $s\ge t_1$  
\ba
Q_1(s) &\le& \frac{Q_1 (t_1)}{1+2\Gamma Q_0(t_1)Q_1(t_1)|s-t_1|}\nn\\
 &=& \frac{1}{2\Gamma Q_0(t_1)} \left[\frac{Q_1 (t_1) Q_0(t_1)
2\Gamma}{1+2\Gamma Q_0(t_1) Q_1 (t_1) |s-t_1|}\right]\nn\\
&\le& \min \left\{
Q_1(t_1), 2\Gamma^{-1} Q_0^{-1} (t_1) \langle s-t_1\rangle^{-1}\right\}
\label{eq:Q1bound1}
\ea
and 
\be
Q_0^{-1} (t_1) \equiv Q_1(t_1)^{-1} \cE_0^{-r}, \ {\rm  and }\
  Q_1 \leq \cE_0.
\label{eq:Q0toQ1}\ee
We write 
\ba Q_1^m(s) &=& Q_1^{m-\bar k}(s) Q_1^{\bar k}(s)\ \le 
 Q_1^{m-\bar k}(s)\frac{1}{\la s-t_1\ra^{\bar k}}
                  \frac{1}{(2\Gamma Q_0(t_1))^{\bar k}}\nn\\
             &=& Q_1^{m-\bar k}(s)\frac{1}{\la s-t_1\ra^{\bar k}} 
\frac{\cE_0^{-r\bar k}}{(2\Gamma Q_1(t_1))^{\bar k}}\nn\\
             &\le& Q_1^{m-2\bar k}(s) \cE_0^{-r\bar k} 
               \la s-t_1\ra^{-\bar k},
\label{eq:Q1mbound}\ea
where we have used that $Q_1(t_1)\ge Q_1(s)$.

We consider separately the cases $t\ge 2t_1$ and $t_1\le t\le 2t_1$.
 
\nit{\bf $t\geq 2t_1$:}\  Take ${\bar k}=2$ and $m>2(r+2)$. Then, 
the integral in (\ref{eq:typ6}) is bounded by 
\be
\cE_0^\rho\ Q_0(t)^{1\over2} \langle t \rangle^{-\min(k, \bar k)}, \ \rho>0.
\nn\ee
This implies, for $k$  sufficiently large, 
\be
\bigg|\cO(\alpha_0 |\beta_1|^2) \langle \chi, \int_0^t\cdots\rangle \bigg|
  \leq
\cO(\cE_0^\rho)
 \left(\ Q_0 Q_1^2 + \langle t \rangle^{-4}\ \right).\  \rho>0
\nn\ee
\bigskip

\nit{\bf $t_1 \leq t < 2 t_1$:}
 Let $t = t_1 + 2M,  M = (t-t_1) /2$ and rewrite the integral as
\be
\int^t_{t_1}\frac{ds}{\la t-s\ra^k}|\alpha_0\beta_1^{2m}| = 
 \left(\int^{t_1 + M}_{t_1} + \int^{t_1 + 2 M}_{t_1 + M})\right)
 \frac{ds}{\la t-s\ra^k}|\alpha_0\beta_1^{2m}|.
\label{eq:split}
\ee
In a manner similar to the previous estimate, using that $Q_1$ is decreasing
 we have, by (\ref{eq:Q1mbound}):
\ba
\int^{t_1 + M}_{t_1} \frac{ds}{\langle t-s\rangle^k} Q_1 (s)^{m/2}
\langle s -t_1\rangle^{-\bar k}
&\le& c \langle t - (t_1 + M)
\rangle^{-k} Q_1(t_1)^{m/2}\nn\\
&\le& c \langle t-t_1\rangle^{-k}Q_1(t_1) Q_1(t_1)^{{m\over2}-1}\nn\\
&\le& c \langle t-t_1\rangle^{-k} \left[\ 1+4 (\Gamma' + \delta)
Q_1(t_1) Q_0(t_1)|t - t_1|\ \right] Q_1 (t) Q_1 (t_1)^{m/2-1}\nn\\
&\le& \cO(\cE_0)Q_1(t)
\nn\ea
for $k\ge1$.
Hence
\be
\cO(\alpha_0|\beta_1|^2) Q_0^{1/2} \int^{t_1 + M}_{t_1} \cdots \leq \cO
 (\cE_0) Q_0 Q_1^2.
\nn\ee
Furthermore, by (\ref{eq:Q1mbound}) the integral over $[t_1+M, t_1+2M]$ 
is bounded by
\be
\int^{t_1+2M}_{t_1+M} 
 \frac{ds}{\la t-s\ra^k \la s-t_1\ra^{\bar k}} Q_1(s)^{m/2}
 \leq c Q_1(t_1)^{m/2} \langle
M\rangle^{-\bar k}\leq c Q_1(t_1)^{m/2} \langle t-t_1\rangle^{-\bar k}
\nn\ee
and, as above, using the upper bound (\ref{eq:Q1lb}) for $Q_1(t_1)$
 in terms of $Q_1(t)$ we have 
\be
\int^{t_1 + 2 M}_{t_1+M} \cdots \leq\cO (\cE_0) Q_1 (t) 
\nn\ee
for $\bar k \geq 1$.
Therefore, for all $t > t_1$ the nonlocal (and local) ODE terms in the
$Q_0$ equation are bounded by 
\be
\cO(\cE_0) \left[Q_0 Q_1^2 + \langle t \rangle^{-3}\right],
\nn\ee
 provided we
control the integral $\int^{t_1}_0 \cdots\ ds$. 
\bigskip

\nit{\bf $\int_0^{t_1}$:}
Consider
\be
I(t) \equiv Q_0^{1/2} Q_1 \int^{t_1}_0\frac{ds}{\langle t-s\rangle^k}
Q_0^{1/2} Q_1^{m-1}Q_1 ds.
\label{eq:Idef}
\ee
By the mean value theorem, 
$Q_1(s) = Q_1(s) - Q_1(t) + Q_1 (t) = \dot Q_1(\bar s) | t-s| + Q_1(t)$,
where $s\le \bar s\le t$.
Then,
\ba 
I(t) &\le& Q_0^{1\over2}(t)Q^2_1(t) 
 \int^{t_1}_0\frac{ds}{\langle t-s\rangle^k}
Q_0^{1\over2}(s) Q_1^{m-1}(s)\ ds\nn\\
&+& I_1(t),\ {\rm where}\ 
\label{eq:Ibound}\\
I_1(t) &\equiv& c\ Q_0^{1\over2}(t) Q_1(t)
\int^{t_1}_0\frac{ds}{\langle t-s\rangle^{k-1}}Q_0^{1\over2}(s)Q_1^{m-1}(s)
\ Q^{1\over2}_0(\bar s) Q_1(\bar s),
\label{eq:I1def}\ea
where we have used 
\be
\dot Q_1 = \cO\left(Q_0^{1\over2} Q_1\right) + h.o.t. .
\nn\ee
If $\bar s\le t_0$, then using that 
$Q_0^{1\over2}(\bar s)\le \cE_0^\rho \la\bar s\ra^{-1}
 \le \cE_0^\rho\la s\ra^{-1}$,
 we have  the bound 
\be
I_1(t)\ \le\  \cO(\cE_0^\rho)Q_0^{1\over2}(t)Q_1(t)\Jt^{-2}
\le \cO(\cE_0^\rho)\left( Q_0(t)Q_1^2(t) + \Jt^{-4}\right)
\nn\ee
If $\bar s>t_0$, then since $Q_0(s)$ is monotonically increasing
for $s\ge t_0$, and we have  
\be
I_1(t) \le 
\frac{Q_0(t)Q_1(t)}{\la t-t_1\ra}\int_0^{t_1}
\frac{ds}{\langle t-s\rangle^{k-1}}
 Q_0^{1\over2}(s)Q_1^{m-1}(s)\  Q_1(\bar s)
\nn\ee
We now expand the latter factor of $Q_1$ in the integrand using the 
 mean value theorem. Specifically,
  there exists $s'$ with $t_0< \bar s\le s'\le t_1$
such that:
\ba
Q_1(\bar s) &=& Q_1(t_1)+\dot Q_1(s')(s'-t_1)\nn\\
&=&  Q_1(t_1) + \cO\left( Q_0^{1\over2}(s')Q_1(s') \right) |s'-t_1|
\nn\\
&\le& Q_1(t_1) + \cO\left( Q_0^{1\over2}(t_1)Q_1(s') \right) |\bar s-t_1|,
\label{eq:mvt1}\ea
where the last inequality follows by monotonicity of $Q_0$. 

Substitution into integrand in (\ref{eq:mvt1}) 
 gives:
\be
I_1(t) \le
c \cE_0^\rho \frac{Q_0(t)Q_1(t)}{\la t-t_1\ra}
  Q_1^{1\over2}(t_1)\int_0^{t_1}
\frac{ds}{\langle t-s\rangle^{k-3}}
 Q_0^{1\over2}(s)Q_1^{m-1}(s),\label{eq:typ7}
\ee
where we have used (\ref{eq:Q0toQ1}) to replace $Q_0(t_1)$
by $Q_1(t_1)$. 
A higher order term (one proportional to $Q_1(t_1)$) is
subsumed by the constant, $c$.   

From (\ref{eq:Q1lb}) we have
\be
Q_1(t_1)\ \le\   
\left( 1+4(\Gamma'+\delta)Q_1(t_1)Q_0(t_1)(t-t_1)\right)\ Q_1(t)
\nn\ee

\nit {\bf Case 1: $4(\Gamma'+\delta)Q_1(t_1)Q_0(t_1)(t-t_1)\le 100$} Here, 
$Q_1(t_1)\le 101Q(t)$, and therefore
\be
I_1(t) \le
c\cE_0^\rho \frac{Q_0(t)Q_1^{3\over2}(t)}{\la t-t_1\ra}
  \int_0^{t_1}
\frac{ds}{\langle t-s\rangle^{k-3}}
 Q_0^{1\over2}(s)Q_1^{m-1}(s)
\label{eq:typ8}\ee
We now split the integral in (\ref{eq:typ8}) as 
 $\int_0^{t_1}=\int_0^{t_0}+\int_{t_0}^{t_1}$.
Using the $\la s\ra^{-1}$ decay of $Q_0^{1\over2}$ for $s\le t_0$
we have
\be
\bigg|\int_0^{t_0} \cdot\bigg| \le \cE_0^{\rho}\la s\ra^{-1},
\nn\ee
and using the monotonicity of $Q_0$ for $t_1\ge s\ge t_0$ and the relation
$Q_0(t_1)=\cE_0^rQ_1(t_1)\le 101\cE_0^rQ_1(t)$  we have
\be
\bigg|\int_{t_0}^{t_1} \cdot\ ds\bigg| \le \cE_0^\rho 
 Q_0^{1\over2}(t_1) = \sqrt{101}\cE_0^{\rho-{r\over2}}Q_1^{1\over2}(t)
\nn\ee
Therefore, choosing $m$ sufficiently large, we get
\ba
I_1(t)\le c'\left( \cE_0^{\rho_1}Q_0(t)Q_1^{3\over2}(t) \la t\ra^{-1} 
 + \cE_0^{\rho_2}Q_0(t)Q_1^2(t)\right) 
\le c''\cE_0^{\rho_3}\left( Q_0Q_1^2 + \Jt^{-4}\right),\nn\\
&&\label{eq:I1bound1}\ea
where  $\rho_3=\min\{\rho_1,\rho_2\}$.

\nit {\bf Case 2: $4(\Gamma'+\delta)Q_1(t_1)Q_0(t_1)(t-t_1)\ge100$}  
 Then, (\ref{eq:Q1lb}) and monotonicity of $Q_0$ for $t\ge t_0$ 
 implies
\be  
\frac{1}{\la t-t_1\ra}\le 4(\Gamma'+\delta)Q_0(t)Q_1(t).
\nn\ee
This gives
\be
I_1(t)\le C\cE_0^\rho Q_0(t)Q_1^2(t)\nn\ee
We now turn to (\ref{eq:typ4}) for $t\in [t_0,t_1]$.
It suffices to estimate the nonlocal terms 
\be
\cO(\alpha_0|\beta_1|^2)\left\la\chi,\int_0^te^{-i\cH_0(t-s)}
 P_c{\tilde\cH_0}^m
{\tilde\chi}\alpha_0\beta_1^{2m}\ ds\right\ra
\nn\ee
for $t_0\le t\le t_1$.
\ In this region $Q_0$ and $(Q_0/Q_1)$ are increasing functions.
  Also $Q_0(t)
\leq \cE_0^r Q_1 (t)$.
The main difficulty is the need to pull a factor of $Q_1(t)$ out of
the nonlocal term.  To this end we use the following proposition:

\begin{prop}\label{prop:I2}
There exists a constant $\delta>0$ such that for $t\ge t_0$
\be
2\Gamma\int^t_{t_0} Q_0Q_1^m ds \leq \delta 
 Q_0^{1\over2}(t) Q_1(t)^{m-2}\ + \ Q_0(t)Q_1^{m-2}(t)\ +
\ {\rm h.o.t.}
\label{eq:I2int}\ee
\end{prop}

\begin{cor}\label{cor:cor1}
\be
\int^t_{t_0} \frac{ds}{\langle t-s\rangle^{3/2}} 
Q_0^{1/2}(s) Q_1^{m-1}(s)
  \leq
C(1 +\delta)Q_0^{3\over8} Q_1^{m-{1\over2}}.
\nn\ee
\end{cor}
\nit The corollary follows from Proposition \ref{prop:I2} by 
 the H\"older's  inequality. 

{\bf Proof of Proposition \ref{prop:I2}:}
Recall that
\be
\frac{dQ_0}{dt} = 2\Gamma Q_0Q_1^2 + R_0,\ \  
\frac{dQ_1}{dt} \leq 
 -4\Gamma Q_0 Q_1^2 + \cO\left( Q_0^{1\over2} Q_1^m\right)
\label{eq:Q0Q1sys}
\ee

\nit\underline{Claim:}
\be
{1\over Q_0^{1\over2}}\int_{t_0}^t R_0 ds\ \le\
 \cO(\cE_0^\rho)\left[ 1+\int_{t_0}^tQ_0(s)Q_1^{2m}(s)ds\right]
\label{eq:claim}
\ee 
\nit\underline{Proof of claim:}
 The leading order term of $R_0$, in the variable $\alpha_0$,
  which is nonlocal, 
is a term of the form (\ref{eq:nlc}). From this term we have  
after integration by parts to obtain $e^{-i\cH_0(t-s)}\cH_0$ from
 (\ref{eq:nlc}),
\ba
\int^t_{t_0} 
 R_0 &\leq& Q_0^{1\over2}(t)\int^t_{t_0} Q_1 (t') \int^{t'}_{t_0}
\frac{ds}{\langle t-s\rangle^{5/2}} Q_0^{1/2} Q_1^m ds dt'\nn\\
&\leq& Q_0^{1\over2}(t) \int^t_{t_0} Q_1 (t') dt' \left[\int^{t'}_{t_0}
\frac{ds}{\la t-s\ra^{4-\delta}} +\frac{1}{\la t-t'\ra^{1+\delta'}}
 \int^{t'}_{t_0} Q_0 Q_1^{2m}
ds \right]\nn\\
&\leq& c Q_0^{1\over2}(t) \int^{t'}_{t_0} Q_1 (t') dt' \left[ \langle
t-t'\rangle^{-3+\delta'} + \langle t-t_0\rangle^{-3+\delta'}+ \langle
t-t'\rangle^{-1-\delta'}\int^{t'}_{t_0} Q_0Q_1^{2m} ds \right]\nn\\
&\leq& cQ_0^{1\over2}(t) \left[ \cO(\cE_0) (1 + \int^t_{t_0} Q_0
Q_1^{2m}\ ds)\right],
\nn\ea
thus proving the claim.
\bigskip

We first rewrite the right hand side of (\ref{eq:I2int}) and 
use the differential equation for $Q_0$ and the above claim  
to integrate by parts:
\ba
&&2\Gamma\int^t_{t_0} Q_0 Q_1^m ds = \int^t_{t_0} 2\Gamma Q_0 Q^2_1
Q^{m-2}_1 ds\nn\\ 
&=& \int_{t_0}^t Q_1^{m-2}\frac{d}{ds}\left[ Q_0-\int_{t_0}^sR_0 \right]\ ds
\nn\\
&=& Q_1^{m-2}(t')\left[ Q_0-\int_{t_0}^sR_0 \right]\bigg|_{t_0}^t
 -\int_{t_0}^t \frac{d}{ds}Q^{m-2}(s)
 \left[ Q_0(s)-\int_{t_0}^{s}R_0 \right] \ ds\nn\\
&=& Q_1^{m-2}(t)Q_0(t) - Q_1^{m-2}(t_0)Q_0(t_0) - 
Q_1^{m-2}(t)\int_{t_0}^tR_0(s) ds\nn\\
&&\ + (m-2)\int_{t_0}^t dt' Q_1^{m-1}(t')Q_0(t')
       \left[ 4\Gamma Q_0(t')+
 \cO\left(Q_0^{1\over2}(t')Q_1^{m-2}(t')\right)\ \right] \nn\\
&&\ + (m-2)\int_{t_0}^t dt' Q_1^{m-1}(t')
 \left[ 4\Gamma Q_0(t')+\cO\left(Q_0^{1\over2}(t')Q_1^{m-2}(t')\right)\right]
 \ \left( \int_{t_0}^{t'}R_0(s) ds\right)
\nn\ea
The first term on the right hand side is of the form we want (the right
hand side of (\ref{eq:I2int}). The second term is negative so we can drop it.
 The third term is bounded by $\cO(\cE_0)Q_1^{m-2}(t)Q_0^{1\over2}(t)$
 by the claim above, plus $\cO(\int_{t_0}^t\ Q_0Q_1^m\ ds)$, which
 is of the above form   
 and can be absorbed by the left hand by smallness of
 $\cE_0$,  
where we used $t_0<s<t_1,\  Q_0 \leq \cE_0^r Q_1$.
This completes the proof of Proposition \ref{prop:I2}.
\bigskip

Using the proposition and its corollary it is easy to  bound, 
 for $t\ge t_0$,

\be
\cO(\alpha_0 |\beta_1|^2 + |\alpha_0|^2\beta_1)
\ \D^k_t\ \left\la \chi, \int^t_0 e^{-i\cH_0(t-s)}\ P_c\tilde \chi \alpha_0
\beta_1^{2m} ds\ \right\ra
\nn\ee
by $\cO (\cE_0^\rho) \left( Q_0 Q_1^2 + \langle t\rangle^{-3}\right)
+ h.o.t.$
\bigskip

It remains to consider $t\leq t_0$.  In this case we only know that
$|\talpha_0| \le k_0 \langle t\rangle^{-1}$.

The first term of (\ref{eq:nlc})
\be
\bar \alpha_0
  \cO\left(\bar\alpha_0\beta_1 e^{-i\lambda_+t} \right)
 \langle \chi,
\int^t_0 e^{-i\cH_0(t-s)}P_c \alpha_0 \beta_1^{2m} \tilde \chi ds\rangle
\label{eq:nlc1}
\ee
is bounded above by $\cO(\la t\ra^{-3})$,
 due to dispersive estimates and the decay of 
the decay of $Q_0$ for $t\le t_0$.

In the second term of (\ref{eq:nlc}),
\ba
&& \cO (\alpha_0|\beta_1|^2)\partial_t^k\langle \chi, \int_0^t\cdots
\rangle\nn\\
&\sim& 
\cO\left(Q_0^{1\over2}(t)Q_1(t)\right)\ 
\left\la \chi,\int_0^te^{-i\cH_0(t-s)}P_c{\tilde\cH_0}^k{\tilde\chi}
 \talpha_0\tbeta_1^{2m} e^{i\Omega s} \ ds\right\ra,\ \Omega\ne0
\nn\\
&&\label{eq:typ9}
\ea
we need to pull $Q_0^{1\over2}(t)Q_1(t)$ out of the nonlocal (integral) term. 

By dispersive estimates, we have the bound 
\be
\cO\left(Q_0^{1\over2}(t)Q_1(t)\right)
 \int_0^t\frac{1}{\la t-s\ra^{ {3\over2}+k} } |\talpha_0(s)|
Q_1^m(s)\ ds\label{eq:typ10}
\ee
To pull the $Q_1$ term we proceed as earlier.
By the mean value theorem, there exists $\bar s$, with 
$s\le \bar s\le t\le t_0$ such that
\ba
Q_1(s) &=& Q_1(t) + Q_1(s) - Q_1(t) = Q_1 (t) - \dot Q_1 (\bar s) (t-s)\nn\\
 &=& Q_1(t) + \cO\left(Q_0^{1\over2}(\bar s)Q_1(\bar s)\right) (t-s)\nn\\
 &=& Q_1(t) + \cO\left(\cE_0^\rho\la\bar s\ra^{-1}\la t-s\ra\right), 
\nn\ea
where we have used that $\dot Q_1\sim Q_0^{1\over2}Q_1$ for $t\le t_0$.
Therefore, the expression in (\ref{eq:typ10}) satisfies a bound of order 
\ba
&&Q_0^{1\over2}(t)Q_1^2(t)
\int_0^t\frac{1}{\la t-s\ra^{ {3\over2}+k} }|\talpha_0(s)|
Q_1^{m-1}(s)\ ds\nn\\
&& + 
Q_0^{1\over2}(t)Q_1(t)
 \int_0^t\frac{1}{\la t-s\ra^{ {1\over2}+k} }Q_0^{1\over2}(s)
\cO\left(Q_0^{1\over2}(\bar s)Q_1^{m'}(\bar s)\right) Q_1^{m-1}(s)\ ds
\nn\\
&=&Q_0^{1\over2}(t)Q_1^2(t)
\int_0^t\frac{1}{\la t-s\ra^{ {3\over2}+k} }\ Q_0^{1\over2}(s)
Q_1^{m-1}(s)\ ds\nn\\
&+& \cE_0^\rho\ Q_0^{1\over2}(t)Q_1(t) \frac{1}{\la t\ra^2}
\label{eq:typ11}
\ea
where the last term, which is bounded by 
$\cE_0^\rho \left( Q_0Q_1^2 + \la t\ra^{-4}\right)$,
 is obtained using the decay of $Q_0(s)$ for $s\le t_0$. 

It remains to estimate the second to last term in (\ref{eq:typ11}).
Estimating the convolution, using the $\la s\ra^{-1}$ decay of
 $Q_0(s)$, we obtain the bound
$\cO\left(Q_0^{1\over2}(t)Q_1^2(t)\right) \la t\ra^{-1}$ which
is not bounded by the desired 
 $\cO\left(\ \cE_0^\rho (Q_0Q_1^2 + \la t\ra^{-3})\ \right)$.

We will obtain the desired bound  by turning to an earlier expression,
 derivable from 
 (\ref{eq:typ9}). The expression we must consider is 
\be 
\cO\left(Q_0^{1\over2}(t)Q_1^2(t)\right)\
\left\la \chi,\int_0^te^{-i\cH_0(t-s)}P_c{\tilde\cH_0}^k{\tilde\chi}
 \talpha_0(s)\tbeta_1^{2m-2}(s) e^{i\Omega s} \ ds\right\ra,\ \Omega\ne0
\label{eq:typ12}
\ee
We will show that this term is  
of order $\cE^\rho Q_0(t)Q_1^2(t) \la t\ra^{-3/2}$, which implies the 
desired bound.  
We proceed as follows. First, by equation (\ref{eq:normalform})
of  Proposition \ref{prop:normalform} the equation for $\talpha_0$ may
be rewritten as:
\be
i\D_t\ta0(t) = (c+i\Gamma_\omega)|\tb1(T)|^4\ta0(t)\ +\ 
 (c+i\Gamma_\omega)\left( |\tb1(t)|^4 - |\tb1(T)|^4\right)\ta0(t)
 + F_\alpha\label{eq:alphaT}
\ee 
Introducing $\ta0^\#(t)$ via the equation
\be \ta0(t) \equiv e^{-it(c+i\Gamma_\omega)|\tb1(T)|^4t}\ta0^\#(t),
\label{eq:ta0sharpdef}
\ee
we have
\be
i\D_t\ta0^\#(t) = (c+i\Gamma_\omega)\left( |\tb1(t)|^4-|\tb1(T)|^4\right)
 \ta0^\#(t)  +   e^{it(c+i\Gamma_\omega)|\tb1(T)|^4}F_\alpha
\label{eq:ta0sharp-eqn}
\ee
The integral in (\ref{eq:typ12}) can be written, keeping only leading order
terms, as  
\ba
&&\left\la \chi,\int_0^te^{-i\cH_0(t-s)}P_c{\tilde\cH_0}^k{\tilde\chi}
 \ta0^\#(s)\tb1^{2m-2}(s) e^{i\tOmega s} \ ds\right\ra\nn\\
&=&\ \left\la \chi,\int_0^te^{-i\cH_0(t-s)}P_c{\tilde\cH_0}^k{\tilde\chi_1}
 \dot{\ta0}^\#(s)\tb1^{2m-2}(s) e^{i\tOmega s} \ ds\ +\ \dots\right\ra\nn\\
&=&\ \left\la \chi,\int_0^te^{-i\cH_0(t-s)}P_c{\tilde\cH_0}^k{\tilde\chi_1}
 \left(|\tb1(t)|^4-|\tb1(T)|^4
   \right)\ta0^\#(s)\tb1^{2m-2}(s) e^{i\tOmega s} \ ds\right\ra\nn\\
&\le&
 \cO(\cE_0^2)\int_0^t\frac{ds}{\la t-s\ra^{{3\over2}+k}\la s\ra}
                      \left(|\tb1(t)|^4-|\tb1(T)|^4 \right)\ +\ \dots
\label{eq:typ13}
\ea
It suffices to show
\begin{prop}\label{prop:P1diff}
\be
|P_1(t)-P_1(s)|\ \le\ \frac{C}{\la s\ra^{1\over2}},\ 0\le s\le t
\label{eq:P1est}
\ee
\end{prop}
For the proof we turn to the $P_1$ equation (\ref{eq:P1eqn}):
\ba
{dP_1\over dt} &=& -4\Gamma P_0P_1^2 \ +\ R_1\nn\\
R_1 &=& R_1[\ta0,\tb1,\eta,t]=  2\Im (\ol{\tb1}F_\beta).
\label{eq:P1eqna}
\ea
The key term in $\Im (\ol{\tb1}F_\beta)$ is the form
$\tb1^m\ta0^\#e^{i\Omega t},\ \Omega\ne0$. We have, after two integrations
by parts,  
\ba
&&\left| P_1(t)-P_1(s)\right|\ \le\ \int_s^t \tb1^m(s)
             \ta0^\#(s')e^{i\Omega s'}ds'
\ +\ \dots\nn\\
&=&\ \tb1^m(s)\ta0^\#(s'){1\over i\Omega}e^{i\Omega s'}\left.\right|_s^t
 \ -\ \int_s^t \frac{d}{ds'}\left(\tb1^m(s')\ta0^\#(s')\right)
 {1\over i\Omega}e^{i\Omega s'}\ ds' +\ dots\nn\\
&=&\ \cO\left(\frac{1}{\la s\ra}\right) + \cO\left(\frac{\log s}{\la s\ra}\right) +
 \cO(\cE_0)\int_s^t\ \left|
 \left(P_1(t)-P_1(s')\right)^2\ta0^\#(s')e^{i\tOmega s'} ds'\right|\nn\\
&\le& \cO\left(\frac{1}{\la s\ra}\right) + 
 \cO\left(\frac{\log s}{\la s\ra}\right) +
 \cO(\cE_0)\sup_{s''}\la s''\ra^{1\over2}|P(t)-P(s'')|
 \int_s^t \frac{1}{\la s'\ra^2} ds'\nn\\
&=& \cO\left(\frac{1}{\la s\ra}\right) + 
 \cO\left(\frac{\log s}{\la s\ra}\right) + 
 \cO\left(\frac{\cE_0}{\la s\ra}\right)
 \sup_{s''}\la s''\ra^{1\over2}|P(t)-P(s'')|. 
\label{eq:P1diff}
\ea
Multiplication by $\la s\ra^{1\over2}$ and taking the supremum over
  $s\le t$ implies Proposition \ref{prop:P1diff} and therewith
the proof of part (a) of Proposition \ref{prop:ODE}.
\bigskip

\subsection{Proof of part (2) of Proposition \ref{prop:ODE}}

The proof is similar to that of part (a) but simpler, 
 since we can allow for the nonlocal terms to be controlled, in addition,  by 
terms of order $Q_0^{1\over2}Q_1^m$.

The leading contributions are again nonlocal, linear $\tetab$
contributions:

From (\ref{eq:beta_eqn3}) we read 
\ba 
i\partial_t \beta_1 &\sim& \cdots + 2 \lambda \langle \chi, \bar
\Psi_1\tetab + \Psi_1 {\overline{\tetab}}\rangle e^{i\lambda_+ t}
\alpha_0\nn\\
&+& \lambda e^{i\lambda_+ t} 2 \langle \chi, \Psi_1 \tetab\rangle
\bar \alpha_0\nn\\
&&
+\lambda e^{i\lambda_+ t} \langle \chi, \Psi^2_1 {\ol{\tetab}}
+ 2|\Psi_1|^2 \tetab\rangle + h.o.t.\\
&+& X^{-1} (t) (A(t) - A(T)) X(t) \beta_1.\label{eq:beta4}
\ea
The first two terms on the righ hand side of (\ref{eq:beta4}) 
 are easily seen to be of 
order 
\noindent $O(\alpha_0) P_1^{2m}$ or $\cO(\cE_0^\rho) P_0P_1^2$, 
 by integration
by parts over the ODE source terms in $\tetab$, as before.

The third term contributes to the $P_1$ equations, after normal form
transformations and remaining resonant terms:

$$\Im\bar \beta_1 e^{i\lambda_+ t} \partial^n_t \langle \chi,
|\beta_1|^2\int^t_0 e^{-i \cH_0(t-s)}P_c \alpha_0 |\beta_1 |^2 \tilde \chi
ds\rangle
$$
and higher order/similar terms.
\bigskip

The leading term, after integration by parts of the integral term
(note that $H^{-1}P_c$ is bounded):

$$
\bigg|\Im\bar \beta_1 e^{i\lambda_+ t}|\beta_1|^2\langle \chi,
\int^t_0 e^{-i\cH_0(t-s)} P_c \cH_0^n\talpha_0\tbeta_1^{2m} 
 \tilde \chi ds\bigg|
$$
$$
\leq c |\beta_1|^3 \int^t_0 \frac{ds}{\langle
t-s\rangle^k}Q_0^{1/2}(s) Q_1^m.
$$
To this end we repeat the argument of part (a) to estimate the above
integral by
$$
Q_0^{1\over2}(t) Q_1^{m-1}(t) + \cO(t^{-3}) + h.o.t. (ODE).
$$
The main new type of term we need to control comes from the last term
on the term $\cR_\beta$ in (\ref{eq:beta_eqn2}-\ref{eq:Rbeta_def}), 
coming from the difference $\bA(t) - \bA(T)$.

This term contributes to the $P_1$ equation terms like 
\be
\Im \left(\ |\beta_1|^2 \left[\alpha_0^2(t) -
\alpha_0^2(T)\ \right] {\bar \alpha}_0^2(t)\ \right), \Im \left[
e^{i\omega t } \beta_1^2\ (P_0(t) - P_0(T)) \cO(\alpha^2_0(T))\right].
\nn\ee

The term with phase $e^{i\omega t}$ can be integrated by parts, and
gives higher order terms;

The term without a phase requires the estimate of 
$$
\alpha_0 (t)^2 - \alpha_0(T)^2 = 2\int^T_t\alpha_0
\frac{d\alpha_0}{ds} ds.
$$

Using that 
$$
\frac{d\alpha_0}{ds} = c\bar \alpha_0\beta^2_1 e^{-2i\lambda_+ t} + h.o.t.
$$
we can repeatedly integrate by parts to get
$$
\alpha_0^2 (t) - \alpha^2_0 (T) = \ {\rm local\ terms }\ + \int^T_t 
\cO(\alpha_0^2) \beta_1^{2m} ds + h.o.t.
$$
where local terms = $\cO(\bar \alpha_0 P_1)$ and higher.

clearly, then
$$
|\beta_1|^2 \bar \alpha_0^2 (t) \left[\ { local\ terms }\ +
h.o.t.\right] \leq \cO(\cE_0) \left[ P_0 P_1^2 + \langle t
\rangle^{-3}\right].
$$
So, we need to estimate
\be
|\beta_1|^2 \bar\alpha_0^2 (t) \int^T_t \cO(\alpha_0^2) \beta_1^{2m} ds
\ {\rm by}\ \cO(\cE_0^\rho)Q_0Q_1^2\ +\ {\rm h.o.t.}
\nn\ee
For $t\geq t_1 \quad  |\beta_1|^2 \leq Q_1$ is monotonic decreasing.
  Hence 
$$
|\int^T_t \cO(\alpha_0^2) \beta_1^{2m} ds | \leq |\beta_1(t)|^2
\int^T_t Q_0Q_1^{m-1} ds.
$$

For $t> t_1, Q_1\downarrow$ is bounded by 
$$
Q_1(t) \leq Q_1(t_1) (1+2\Gamma Q_0(t_1) Q_1 (t_1) (t-t_1))^{-1} \leq c
\langle t-t_1\rangle^{-1} Q_0 (t_1)^{-1}
$$
$$
= cQ_1 (t_1)^{-1} \cE_0^{-r} \langle t-t_1\rangle^{-1}.
$$
since $Q_0(t_1) = \cE_0^rQ_1 (t_1)$.

So,
$$
\left[Q_1(t)^{1+r} Q_1(t)\right]^k \leq \cO(\cE_0) \langle
t-t_1\rangle^{-k}.
$$
Hence, for $k>1$, or for $m> 2 + r + 1 = r+3$,
$$
\int^T_t Q_0 Q_1^{m-1} \leq \cO(\cE_0)
$$so 
$$
\big|\int^T_t \cO(\alpha_0^2) \beta_1^{2m} \big| \leq \cO(\cE_0)
Q_0Q_1^2.
$$

If $t>t_0, t < t_1, Q_0 (t) \leq \cE_0^rQ_1 (t); \
 {\rm  therefore }$
$$
\int^T_t Q_0 Q_1^{m-1} ds = \int^T_t Q_0Q_1^2 Q_1^{m-3} ds =
\left(\frac{1}{2\Gamma} Q_0 (s) + R_0(s)\right) Q_1^{m-3} (s)
\big|^T_{s=t}
$$
$$
-\int^T_t \left(\ 
 \frac{1}{2\Gamma} Q_0 + \int_{t_0}^sR_0(s)ds\ \right) Q_1^{m-4} (m-3)
\frac{dQ_1}{ds} ds.
$$
The first term on the right hand side is local and high order.
The second term on the right hand side  is bounded by 
\ba
c\int^{t_1}_t Q_0^2 Q_1^{m-2} ds + \int^T_{t_1} \left(\cdots\right) +
\ {\rm h.o.t.}\ &\le& \cO(\cE_0) Q_1^2 +\cO(\cE_0^\rho)\la t\ra^{-3} + 
 c\int^{t_1}_t Q_0^2 Q_1^{m-2} ds\nn\\
&\le& \cO (\cE_0) \left[ Q_1^2 + \langle t
\rangle^{-3}\right] + c \cE_0^r\int^{t_1}_t Q_0Q_1^{m-1} ds.
\nn\ea

Hence,
\be
\int^T_t Q_0Q_1^{m-1} ds \leq \cO(\cE_0) \left[ Q_1^2 +
\langle t\rangle^{-3}\right] + \cO(\cE_0) \int^T_t Q_0Q_1^{m-1} ds
\nn\ee
which implies that 
\be
\int^T_t Q_0 Q_1^{m-1} ds \leq \cO(\cE_0) \left[ Q_1^2 +
\langle t \rangle^{-3}\right]
\nn\ee
for all $t\geq t_0$.

For $0< t < t_0$, we need to estimate 
$$
|\beta_1 |^2 \bar\alpha^2_0(t) \int^{t_0}_t O (\alpha_0^2) \beta_1^{2m}
ds.
$$
Using that for $t< t_0, | \alpha_0 (t)|^2\leq k_0 \la t\ra^{-1}
 \leq \cE_0 \langle t_0\rangle^{-1} \langle t\rangle^{-1}$ the 
 above
expression is bounded by 
$$
\cO(\cE_0) k_0(t_0) \langle t\rangle^{-1} (\ln
\frac{t_0}{\langle t\rangle}) k_0 (t_0) \leq \cO(\cE_0)
k_0(t_0) \langle t \rangle^{-1} \left(\frac{\langle t\rangle}{\langle
t_0\rangle} \ln\frac{t_0}{\langle t\rangle}\right) \langle t
\rangle^{-1}
$$
$$
\leq \cO(\cE_0) k_0 (t_0) \langle t\rangle^{-2}.
$$

This completes the proof of Proposition \ref{prop:ODE}. \bigskip

\section{$R_1^{(j)}(\tetab)$ terms of  Proposition \ref{prop:Master}}
\label{sec:Rj1teta}

\begin{prop}\label{prop:Riteta} 
Assume either $t\le t_0$ or Montonicity Property {\bf Q} on 
 $[t_0,t_0+\delta_*]$. Then, the terms $R_1^{(i)}[\tetab],\ i=0,1$ 
 in (\ref{eq:P0P1}) satisfy the estimates:
\be
\left|R_1^{(i)}[\tetab]\right| \leq \cO(\cE_0) \left[2\Gamma Q_0 Q_1^2 +
\langle t_0\rangle^{-1} \langle t\rangle^{-2} \left(Poly \left[P_0(0), P_1(0),
P_0(T), P_1 (T)\right]   \right)   \right]
\label{eq:Rjitetabound}
\ee
where Poly $[\cdots]$ stands
for polynomial in the bracketed variables.
\end{prop}

{\bf Proof:}  The contributions to $R_1^{(i)}(\tetab)$ comes
form linear and nonlinear terms in $\tetab$ in the $P_i$
equations.

Consider first the nonlinear contributions:

In the $P_0$ equations we have terms like (7.41)
\ba
&&\bar \alpha_0^2\la \chi, \tetab^2\ra,\ \ 
 \bar \alpha_0 \bar\beta_1 e^{i\lambda_+ t} \la \chi, \tetab^2\ra,\nn\\
&&e^{i\lambda_+ t}\bar \alpha_0 \beta_1 \la\chi, |\tetab|^2\ra,\ \ 
 \langle \chi, |\tetab|^2 \tetab\rangle\bar \alpha_0.
\label{eq:termss}
\ea
Since for $t\in I_0, I_2$ we have time decay of either $\alpha_0$ or
$\beta_1$ respectively, the main contribution is when $t\in I_1= [t_0,
t_1]$.

For $t\in I_2$ the bound on $\tetab$ we have is
$$
\|\tetab(t)\|_{H^k} \leq c Q_0^{1/2}(t) \int^t_0 \frac{ds}{\langle
t-s\rangle^{3/2}} Q_1^m (s).
$$
The second and third terms can be integrated by parts leaving terms of
the type
$$
\cO(\beta_1^2)\ \cO(\bar\alpha_0)\ \partial^m_t\langle \chi, |\tetab|^2 + 
\tetab^2  \rangle.
$$
We also need to integrate by parts the two other terms.  For this we
need to pull out a phase factor from the leading nonlocal.

Pulling a phase as in the proof of Proposition \ref{prop:ODE}
  we are left with
estimating term of the type
$$
O(\alpha_0^2 + \alpha_0\beta_1) \langle \chi, 
\tetab\partial^k_t\tetab\rangle + \langle \chi, |\tetab|^2
\partial_t^k \tetab\rangle O(\alpha_0).
$$  
As in the estimates of Proposition \ref{cor:Corollary 3} , for $t\in I_1,$
$$
\|\partial^k_t\tetab\|_{H^s} \leq C
Q_0^{1/2}(t)\int^t_0\frac{ds}{\langle t-s\rangle^{k'}}Q_1^m (s).
$$
with $k'$ large for $k$ large.

To this end we use the following.

\begin{prop}
\label{prop:Lemma}

For $t\in I_1$:
$$
\int^t_{t_0}\frac{ds}{\langle t-s\rangle^{k'}} Q_1^{\bar m} (s) \leq
\cO(\cE_0) Q^2_1 (t).
$$
\end{prop}

\nit{\bf Proof of Proposition \ref{prop:Lemma}:}
$$
\int^t_{t_0} \frac{ds}{\langle t-s\rangle^{k'}}Q_1^{\bar m} (s) =
Q_1(t) \int^t_{t_0} \frac{ds}{\langle t-s\rangle^{k'}} Q_1^{\bar m -1}
+ \int^t_{t_0} \frac{ds}{\langle t-s\rangle^{k'}} (Q_1 (s) - Q_1(t))
Q_1^{\bar m -1}.
$$
The second term on the right hand side can be estimated, using that
$\dot Q_1=-\Gamma Q_0Q_1^2+\cO(Q_0^{1\over2}Q_1^m)$ and monotonicity of $Q_0$,
 as follows:
 $(s\leq \xi\leq t)$
\ba
&\leq& C \int^t_{t_0} \frac{ds}{\langle t-s\rangle^{k' -1}} \dot Q_1
(\xi) Q_1^{\bar m-1}(s) \leq C Q_0 (t) \int^t_{t_0}
\frac{ds}{\langle t-s\rangle^{k'-1}} Q_1^2 (\xi) Q_1^{\bar m-1}(s)\nn\\
&+&CQ_0^{1/2}(t) \int^t_{t_0} \frac{ds}{\langle t-s\rangle^{k'-1}} Q_1^m
(\xi) Q_1^{\bar m-1}(s)\nn\\
&&
\leq \cO(\cE_0)Q_1(t) \int^t_{t_0} \frac{ds}{\langle
t-s\rangle^{k'-1}} Q_1^2 (\xi) Q_1^{\bar m-1} \nn\\
&+& \cO
(\cE_0)Q_1^{1/2} (t) \int^t_{t_0} \frac{ds}{\langle
t-s\rangle^{k'-1}} Q_1^{m-1} (\xi) Q_1^{\bar m-1}.
\nn\ea
Repeating this argument, we have
\ba
\int^t_{t_0} \frac{ds}{\langle t-s\rangle^{k'}} Q_1^{\bar m} (s) &\leq&
\cO(\cE_0) Q_0^j (t) \int^t_{t_0} \frac{ds}{\langle
t-s\rangle^{k'-2j}} Q_1^{\bar m - 2j}\nn\\
&\leq& \cO(\cE_0)Q_1^j(t).
\nn\ea
for $k' - 2j> 1$, which proves the Proposition \ref{prop:Lemma}. 

This Proposition together with the estimate 
\be
\| \tetab\|_{W^{k,\infty}} \leq \cO(\cE_0) Q_0^{1/2}
\nn\ee
implies that 
\be
R_1^{(0)} \leq \cO(\cE_0) 
 \left[\ Q_0 Q_1^2 + {\rm higher\ order\ terms}\ \right].
\nn\ee
The estimates of $R_1^{(1)}$ are similar.

It remains to estimate the linear $\tetab$ terms in the $P_0,
P_1$ eqs.

The leading order source term of $\tetab$ was estimated in
Proposition \ref{prop:ODE}.  It remains to estimate the higher order
corrections.

To this end we need to estimate terms of the following 
 type appearing in the $P_0$
equation, (\ref{eq:P0P1}):
\ba
&&\bar\alpha_0\left(|\beta_1|^2 + \alpha_0 \beta_1\right)
\left\la \chi, \int^t_0 e^{-i\cH_0(t-s)} P_c \psi_0 \tetab^2 ds\ \right\ra \ 
 {\rm and}\nn\\
&&\left(|\beta_1|^2 \alpha_0 + \bar \alpha_0
\alpha_0\beta_1\right)\left\la \chi, \int^t_0 e^{-i\cH_0(t-s)}
\ P_c|\tetab|^2 \tetab ds\ \right\ra,
\nn\ea
and similar terms in the $P_1$ equation.

Again we focus on $t\in I_1$.
Since for $0\leq s \leq t_0$ $\alpha_0$ and $\tetab$ are of order
$\frac{\cO(\cE_0)}{\langle s \rangle}$,
clearly these contributions are of order
$$
\cO(\cE_0) \left[ Q_0Q_1^2 + \langle t_0\rangle^{-3} \right],
$$
so it remains to estimate the $s$-integrals above on $I_1$.

But on $I_1, \|\tetab\| \leq \cO(\cE_0)Q_0^{1\over2}(t) \leq
\cO(\cE_0) Q_1^{1\over2}(t)$
 and since $Q_0$ is monotonic increasing on $I_1$, the above nonlocal
term are bounded by 
$$
\cO(\cE_0)Q_0^{3\over2}(t).
$$
So,
$$
\cO(|\beta_1|^2\bar\alpha_0) \cO(\cE_0) Q_0^{3\over2}(t)\leq
\cO(\cE_0) Q_1^2 Q_0
$$
$$
O(|\alpha_0|^2\beta_1) \cO(\cE_0)Q_0^{3\over2} (t) \leq 
\cO(\cE_0) Q_1^2 Q_0
$$ 
since $Q_0 \leq \cO(\cE_0) Q_1$ on $I_1$.
\section{Bootstrapping it all}\label{sec:bootstrap}

We assume that $t_0<\infty$, where $t_0$ is given by (\ref{eq:t0def}). 
Consider the equations for $P_0$ and $P_1$, (\ref{eq:P0P1}) 
displayed in Proposition \ref{prop:Master}. Explicit in (\ref{eq:P0P1})
are terms which\ 

\nit (1)\ are driven by the dispersive part of the 
initial data: $R_0^0[\eta_0]$ and $R_0^1[\eta_0]$

\nit (2)\ encompass interactions of the two bound
states and dispersive waves: $R_1^0[\tetab]$\ and 

\nit (3) encompass interactions between bound states: $R_2^0[P_0,P_1]$. 

\nit  By Proposition \ref{prop:biR0i} the 
 $R_0^j[\eta_0],\ j=0,1$ terms satisfy:
\be
R_0^j[\eta_0]\ =\ \cO\left(b_j(t_0,[\eta_0]_X,\cE_0\right)
 \la t\ra^{-2}\ +\ \cO(\cE_0) 
 2\Gamma P_0P_1^2
\label{eq:R0jest}
\ee
Therefore, by Proposition \ref{prop:propB} and its proof
 (see eqn. (\ref{eq:k0_def}) ), it is natural to introduce the functions: 
\be
Q_0(t)\ =\ P_0-\frac{k_0}{\la t\ra},\ \ Q_1(t)\ =\ P_1+\frac{k_1}{\la t\ra}
\label{eq:Q0Q1}
\ee
where
\ba
k_0\ &=&\ b_0\ +\ \cO(\cE_0)b_1^2\ ,
 \ \ \ k_1\ =\ 10b_1\label{eq:k0k1def}\\
b_0\ &=&\ \la t_0\ra^{-1}\left(\ [\eta_0]_X\ +\ c_*\cE_0^2\ \right),
\label{eq:b0a}\\ 
\ b_1\ &=&\ \la t_0\ra^{-{1\over2}}
 \left(\ [\eta_0]_X^{2\over3}\ +\ d_*\cE_0^2\ \right), \label{eq:b1a} 
\ea
where the constants $c_*$ and $d_*$ are to be chosen.
We find, for any $m\ge4$ and all  $t\ge0$: 
\ba
\frac{dQ_0}{dt} &\ge& 2\Gamma Q_0 Q_1^2 + R_1^{0,\#}[\tetab] + R_2^{0,\#}[Q_0,Q_1]
 \ +\ \frac{\cE_0^2c_*}{\la t_0\ra\ \la t\ra^2}
 \nn\\
\frac{d Q_1}{dt}&\le& -4\Gamma Q_0 Q_1^2 + R_1^{1,\#}[\tetab] + R_2^{1,\#}[Q_0,Q_1]
 + \cO(\cE_0;m)\sqrt{Q_0}Q_1^m\ -\ 
 \frac{\cE_0^2d_*}{\la t_0\ra^{1\over2}\ \la t\ra^2}.\nn\\  
&&\label{eq:RRSS}
\ea
The analogous reverse inequalities hold as well with slightly different constants.

By the definition of $t_0$,  $Q_0(t_0)>0$.
Furthermore, using the energy estimate on the bound state amplitudes
 and (\ref{eq:Q0Q1}) of section \ref{sec:bounds}, we have
\be  Q_0(t)+ Q_1(t)\leq C\cE_0,\ \ t>0.\label{eq:Q0Q1ebound}
\ee

\bigskip

\bigskip

We now introduce a set of norms. The norm of
$q(t)\equiv \left(\ Q_0(t),Q_1(t),\tetab(t)\ \right)$
is defined as
\be
\|q(t)\|_\bY =  |Q_0 |_{y_0(t)} +
 |Q_1|_{y_1(t)}  + \|\tetab\|_{y_2(t)}.
\label{eq:bYdef}
\ee
The norm, $\|q(t)\|_\bY$, 
 encodes all the estimates for $Q_0,Q_1$ and $\tetab$
 in the intervals $I_0,I_1$ and $I_2$ through
the following: 
\ba
|Q_0|_{y_0(t)} &\equiv& \sup_{0\le s\le t} |Q_0(s)|
       \ +\  \sup_{0\le s\le \min\{t,t_0\} }\la t_0\ra\la s\ra |Q_0(s)|
\label{eq:y0norm}\\
|Q_1|_{y_1(t)} &\equiv& \sup_{0\le s\le t} |Q_1(s)|
       \ +\  \sup_{t_1\le s\le t}|s-t_1|Q_1(s)\Gamma'Q_0(t_1)Q_1(t_1)
\label{eq:y1norm}\\
\|\tetab\|_{y_2(t)} &\equiv& \sup_{0\le s\le \min\{t,t_0\} }
                        \la s\ra \|\tetab(s)\|_{W^{k,\infty}}
  \ +\  \sup_{t_0\le s\le \min\{t,t_1\} } \|\tetab(s)\|_{W^{k,\infty}}\nn\\
    &&\ \ +\ \sup_{t_1\le s\le t}\la s-t_1\ra^{1\over2}
                         \|\tetab(s)\|_{W^{k,\infty}}
\label{eq:y2norm}
\ea
{\it In these definitions we use the convention that terms for which the $s$-
  range
 empty are set to zero.} 
\bigskip 

 By the $H^1$ {\it \'a priori} bounds
\be
\sup_{0\le s\le t} |Q_0(s)| \le \cE_0\left( 1+\cE_0^2\|\tetab\|_\infty\right)
\nn\ee
and by definition of $t_0$, (\ref{eq:t0def}),  for $t_0<\infty$,
\be
\sup_{0\le s\le t_0}\la t_0\ra\la s\ra |Q_0(s)|
 \le {1\over2}\left(\ \cE_0+[\eta_0]_X\ \right).
\nn\ee

In terms of these norms, we have bounds on $R_i^{j,\#}$. By Proposition
 \ref{prop:Riteta}
\be
|R_1^{0,\#}[\tetab]|+|R_1^{1,\#}[\tetab]|\ \le\ 
 \cO(\cE_0^\rho)\|q(t)\|_Y^{l_1}Q_0Q_1^2 +
  C\ \frac{\cE_0^\rho\|q(t)\|^{l_2}_\bY}{\la t_0\ra\la t\ra^2}.
 \label{eq:RRj}
\ee
By Proposition
 \ref{prop:ODE}
\ba
|R_2^{0,\#}[Q_0,Q_1]|&\le&\cO(\cE_0^\rho)Q_0Q_1^2\ +\ 
 C\ \frac{\cE_0^\rho+\|q(t)\|^l_\bY}{\la t_0\ra \la t\ra^2}\label{eq:RR2}\\
|R_2^{1,\#}[Q_0,Q_1]|&\le&\cO(\cE_0^\rho)Q_0Q_1^2\ +\ 
C\ 
\frac{\cE_0^\rho+\|q(t)\|^l_\bY}{\la t_0\ra^{1\over2} \la t\ra^2}\label{eq:SS2}
\ea
where $l\ge2$.
\bigskip

By the definition of $I_0,\ (0\le t\le t_0)$ and Propositions \ref{prop:biR0i}, 
 \ref{prop:ODE} and \ref{prop:Riteta}, we have estimates
(\ref{eq:RRj},\ref{eq:RR2},\ref{eq:SS2}).
Therefore, for an appropriate choice of $c_*$ and $d_*$ 
we have for $0\le t\le t_0$
\ba
\frac{dQ_0}{dt}\ &\ge&\ 2\Gamma' Q_0Q_1^2\ +\ 
 \frac{c_*}{2}\frac{\cE_0^2}{\la t_0\ra\la t\ra^2}\nn\\
\frac{dQ_1}{dt}\ &\le&\ -4\Gamma' Q_0Q_1^2\ +\ \cO(\cE_0;m)\sqrt{Q_0}Q_1^m\ -\ 
 \frac{d_*}{2} \frac{\cE_0^2}{\la t_0\ra^{1\over2}\la t\ra^2},
\label{eq:shp1}
\ea
where $m\ge4$.
Note that by definition of $I_0$, $Q_0(t)<0$ for $t\in I_0$.

By continuity, (\ref{eq:shp1}) holds for $t_0\le t\le t_0+\delta$, for some 
 $\delta>0$. It follows, using that $Q_0(t_0)>0$ and 
  Propositions \label{prop:B.5} and \label{prop:C},
  that (\ref{eq:PropertyQ}) ({\bf Monotonicity Property Q}) holds
 on $t_0\le t\le t_0+\delta$. 
  Therefore, by Propositions \ref{prop:biR0i}, \ref{prop:ODE} and 
 \ref{prop:Riteta} the terms $J_0$ and $J_1$ in (\ref{eq:Q0Q1b})
both satisfy the bound
\be
|J_k|\ \le \frac{\cO(\cE_0^{2+\rho})}{\la t_0\ra\la t\ra^2}
 \ +\ \cE_0 Q_0Q_1^2,\ \ \rho>0,\ k=0,1.
\label{eq:Jbounds}
\ee
Therefore,  for $\cE_0$ sufficiently small (\ref{eq:shp1}) holds
with $c_*,d_*$ replaced by $c_*/2, d_*/2$.

Define
\be
T_*\ =\ \sup\{t\ge t_0:\ (\ref{eq:shp1})\ {\rm holds\ for\ some}\  c_*>0\
 {\rm and}\ 
  d_*>0\}
\label{eq:T*def}\ee 
For $t\in [0,T_*)$, $\|q(t)\|_{\bf Y}$ is small.
We claim that $ T_*=\infty$. Suppose $t_0\le T_*<\infty$. Then, 
for $t\in [t_0,T_*)$ we have, by Propositions  \ref{prop:B.5} and \ref{prop:C},
 that the monotonicity property (\ref{eq:PropertyQ}) holds 
 at $t=T_*$ and slightly
beyond. Thus, the {\'a priori} bounds on $J_0,J_1$ of 
Propositions \ref{prop:biR0i}, \ref{prop:ODE} and
 \ref{prop:Riteta}, the $\tetab$ - bounds of Propositions \ref{prop:etabonI0},
 \ref{prop:etabonI1} and \ref{prop:etabonI2}, and the smallness of
   $Q_0$ and $Q_1$
imply persistence of the inequalities (\ref{eq:shp1}), with perhaps a slightly
smaller choice of positive constants $c_*$ and $d_*$. 
This implies that $T_*=\infty$.   
\bigskip

\section{Nongeneric behavior}\label{sec:nongeneric} 

Recall that $t_0$ is defined by (\ref{eq:t0def}).
and consider the case where $t_0=\infty$ .
We would like to show that 
\ba
P_0(t) &\to& 0 \ {\rm as }\ t \to \infty\nn\\
P_1(t) \ && {\rm has \ a\ limit}.
\nn\ea

The following is a consequence of the definition of $t_0$.
\begin{prop}\label{prop:NG1}
Assume $t_0=\infty$. Then, 
$P_0(t) = \cO\left([\eta_0]_X \la t\ra^{-2}\right).$
Therefore, $\alpha_0\to0$ and the ground state decays.
\end{prop}
\bigskip

\begin{prop}\label{prop:NG2} 
 Let  $t_0=\infty$. Then, $\beta_1$ has a limit as $t\to\infty$.
\end{prop}
States with this (nongeneric) behavior were constructed in \cite{TY01b}. 

\nit{\bf Proof of Proposition \ref{prop:NG2}:}
The equation (\ref{eq:P1eqn}), together with the above estimate 
\be P_0(t) = \cO(\ol{\eta_0})\la t\ra^{-2}
\nn\ee
implies
\ba
\frac{dP_1}{dt} &=& (-4\Gamma P_0 P_1^2 + \cO(\ol{\eta_0})\la
t\ra^{-3/2})(1+\cO(P_0) + \cO(P_1))\nn\\
&+& \Re (c e^{i\lambda_+ t} |\beta_1|^2\ol{\beta_1} \alpha_0) + \
 {\rm h.o.t.}
\ea
To show that $P_1$ has a limit, we show that $\int_0^t\D_sP_2(s) ds$
has a limit. 
All terms other than the $\cO(\alpha_0)$ term, 
 on the right hand side are absolutely integrable since
 $P_0 = \cO(\la t\ra^{-2})$. 

It is left to integrate the $\cO(|\beta_1|^2 \bar\beta_1 \alpha_0)$ term.
For $T$ given, let $\beta_T^2 \equiv \beta_1(T)^2$.  
 Then, equation (\ref{eq:alpha0_eqn2}) reads
\be 
2i\partial_t \alpha_0 = \lambda \la\psi_{0*},\Psi_1 (t)^2\ra\ol{\alpha_0}\ 
  + \ {\rm integrable\ in\ t }.
\nn\ee
 Using 
the expression for $\Psi_1(t)=\alpha_1\psi_1(\cdot,|\alpha_1|^2)$:
\ba
2i\D_t\alpha_0&=& \lambda \la\psi_{0*}, \alpha_1 (t)^2 \psi^2_{1*}\ra
\bar\alpha_0 + {\rm h.o.t.}\nn\\
&=&\lambda \langle \psi_{0*}, \psi^2_{1*}\rangle e^{-2i\lambda_+
t}\beta_1(t)^2\bar\alpha_0 (t) + O(\alpha^2_0 (T)\bar\beta_1) + {
\rm h.o.t.}\nn\\
&\equiv& \tilde\lambda e^{-2i\lambda_+ t} \beta_T^2 \bar\alpha_0 (t) +
\tilde \lambda e^{-2i\lambda_+t}[\beta^2_1 (t) -
\beta^2_1(T)]\bar\alpha_0(t) + O (\frac{1}{T^2})\nn\\
&+& \ \ \ {\rm h.o.t.}
\label{eq:inhom}\ea
where $\tilde\lambda\equiv\lambda\langle \psi_{0*}
\psi^2_{1*}\ra \ {\rm and}\    
\beta_T^2 \equiv\beta^2_1 (T)$.

Solving the homogeneous part of (\ref{eq:inhom}):
\be
2i\partial_t\hat\alpha_0 = \tilde\lambda e^{-2i\lambda_+t}
\beta_T^2 \overline{\hat \alpha}(t)
\label{eq:hom}\ee
we have, using the Ansatz 
$$
\alpha_0(t) = A(t) e^{i(\lambda -a)t} + B(t) e^{i(\lambda + a) t}
$$
with
$$
\dot A \sim \tilde \lambda e^{-2i\lambda_+ t} [\beta^2_1 (t) -
\beta^2_1(T)]\bar A e^{i\theta(t)}+ \ {\rm h.o.t.}.
$$
and a similar equation for $B(t)$.

We have
$$
\frac{dP_1}{dt} =- 4\Gamma P_0P_1^2 + \Re \left(c
e^{i\theta_A t }|\beta_1|^2\bar\beta_1 A 
 + e^{i\theta_B t }|\beta_1|^2\bar\beta_1 B\ \right) + \ {\rm h.o.t.},
$$
$\theta_A, \theta_B\ne0$.
Integration  of the above equation, 
integration by parts (twice) of the $A$ and $B$, implies 
\ba
&&
\langle t\rangle^{1\over2} |P_1(t) - P_1(T) | \leq C\langle
t\rangle^{1\over2}\ \int^T_t |\beta_1|^2\bar\beta_1\langle
t'\rangle^{-1}\langle t'\rangle [\beta^2_1 (t') -
\beta^2_1(T)]^2\bar A(t') dt' + \ {\rm h.o.t. } \langle t\rangle^{1\over2}
\nn\\
&\leq& C\cE_0^m\left(\int^T_t \langle t'\rangle^{-{1\over2} - 1}
dt'\right) \left( \sup_{0\leq t' \leq T} \langle t^{1\over2}\rangle
|P_1(t') - P_1 (T) | \right)^2 + \sup_t\langle
t\rangle^{1\over2}\ {\rm h.o.t. }\nn\\
&&\Rightarrow | P_1(t) - P_1(T) |  \leq C\langle t\rangle^{-{1\over2}}
\nn\ea
which implies integrability of $\dot A(t)$ and limit of $P_1(t)$.

\input mbs-appA 
\input mbs-appB 
\input mbs-appC 
\input mbs-bib
\end{document}

%% file: mbs-appA.tex
\section{Notation}\label{sec:notation}

\noindent $\Re z$ and $\Im z$\ denote the real and  imaginary parts 
 of a  complex number $z$

\noindent $\overline{z}$ denotes the complex conjugate of $z$ 

\noindent $\la x \ra\ = \sqrt{1 + |x|^2},\ t\ge0$

\noindent 
$ P_{c*}$\ -\ projection onto the continous spectral part of the self-adjoint 
 operator, $H$. 


\noindent $\la f,g\ra = \int \ol{f}\ g$


\be \left(\begin{array}{c} a \\ c.c.\end{array}\right) = 
 \left(\begin{array}{c} a \\ \ol{a}\end{array}\right)\nn\ee

\noindent For $j=1,2$, let $\pi_j:\C^2\to\C^1$  be defined by:

\noindent $\pi_1\left(\begin{array}{c} z \\ w\end{array}\right) = z, \ \ 
\pi_2\left(\begin{array}{c} z \\ w\end{array}\right) = w$

\be \sigma_1 = \left(\begin{array}{cc} 0 & 1 \\1 & 0\end{array}\right),\ \ 
 \sigma_3 = \left(\begin{array}{cc} 1 & 0 \\0 & -1\end{array}\right)
\nn\ee

\noindent \underline{Plemelj identities}
\be (x\mp i0)^{-1}\ = \ 
 {\rm P.V.}\ x^{-1}\ \pm\ i\pi\ \delta(x)
\label{eq:Plemelj}\ee 

\noindent
$\vec\nabla_j\ =\ \left( \D_{\alpha_j}, \D_{\overline{\alpha}_j}\right)$

\noindent
$\chi(x,p)$ denotes are real-valued localized function of $x$ which depends
smoothly on a parameter, $p$.

\nit $\chi_k^{(j)}$ denotes a spatially localized  function of order $|\alpha_j|^k$,
 as $|\alpha_j|\to0$.

\nit $\cO_k^{(j)}$ denotes a quantity which is of order
 $|\alpha_j|^k$ as $|\alpha_j|\to0$.  Both $\chi_k^{(j)}$  and $\cO_k^{(j)}$
 are invariant under the map
$\alpha_j\mapsto\alpha_je^{i\gamma}$.

\nit $\cO_k^{(0,1)} = \cO_{k_1}^{(0)}\cO_{k_2}^{(1)},\ \ k=k_1+k_2$.

%% file: mbs-appB.tex
\section{Appendix B - 
Proof of Proposition \ref{prop:Hspectrum}}\label{sec:appB}

 Parts (1), (3) and (5) of Proposition \ref{prop:Hspectrum}
 follow from \cite{MIW85}.
We now prove parts (2) and (4) by a perturbation argument  
about the case $\alpha_0=0$. 

Consider the eigenvalue problem ${\cal H}_0\vec f=\mu\vec f$. Since $\alpha_0$
is assumed small it is natural to make explicit the leading order and 
perturbation terms. Thus we have
\be
{\cal H}_0\vec f = \sigma_3 \left[\ (H-E_{0*}) -E_0^{(1)}|\alpha_0|^2 I +
 \psi_0^2\left(\begin{array}{cc} 2|\alpha_0|^2 & \alpha_0^2\\
                    \overline{\alpha_0}^2 & 2|\alpha_0|^2 
               \end{array}\right)\ \right] \vec f = \mu \vec f
\label{eq:H0evp}
\ee
Recall that $E_0^{(1)}$ and $\psi_0$ are defined in Proposition 
 \ref{prop:bifurcate}.

The zeroth order problem ($\alpha_0=0$) is 
\be \sigma_3(H-E_{0*})\vec f_0 = \mu_0 \vec f_0,
\label{eq:0thorder}
\ee
which has two linearly independent solutions:
\ba \mu_0 &=& E_{*1}-E_{0*},\ \ 
 \vec f_0 = \left(\begin{array}{c} 1 \\ 0\end{array}\right)\psi_{1*}
\label{eq:0thorder1}\\
\mu_0 &=& -( E_{*1}-E_{0*}), \ \ \overline{\sigma_1\vec f_0}.
\label{eq:0thorder2}
\ea
We develop the perturbation theory of (\ref{eq:0thorder1}). 
 That of the second 
is completely analogous.

For $\alpha_0$ and small we define the perturbations about the zeroth
order eigenstates via: 
\ba \vec f &=& \vec f_0 + \vec f_1\label{eq:fexpand}\\
    \mu    &=& E_{1*}-E_{0*} + \mu_1\label{eq:muexpand1}
\ea
Substitution into (\ref{eq:H0evp}) yields:
\ba
&&\left[\ \sigma_3(H-E_{0*}) - (E_{1*}-E_{0*})I\ \right] \vec f_1\nn\\
&=&\ \ 
 |\alpha_0|^2 E_0^{(1)}\sigma_3\vec f_0 
 -\lambda\psi_0^2\sigma_3\left(\begin{array}{cc} 2|\alpha_0|^2 & \alpha_0^2\\
                      \overline{\alpha_0}^2 & 2|\alpha_0|^2
          \end{array}\right)\ \vec f_0\ + \mu_1\vec f_0\nn\\
&&\ \ + |\alpha_0|^2 E_0^{(1)}\sigma_3\vec f_1
 -\lambda\psi_0^2\sigma_3\left(\begin{array}{cc} 2|\alpha_0|^2 & \alpha_0^2\\ 
                      \overline{\alpha_0}^2 & 2|\alpha_0|^2 
          \end{array}\right)\ \vec f_1\ + \mu_1\vec f_1
\label{eq:H0evp1}
\ea 
We consider, individually, the first and second equations
of the system (\ref{eq:H0evp1}), governing $f_{1j}=\pi_j\vec f_1,\ j=1,2$.
The first component of (\ref{eq:H0evp1}) is:
\ba
\left(H-E_{1*}\right) f_{11} &=&
 |\alpha_0|^2\left( E_0^{(1)}-2\lambda\psi_0^2\right)\psi_{1*}
 + \mu_1\psi_{1*}\nn\\
 &+& |\alpha_0|^2\left(E_0^{(1)} - 2\lambda\psi_0^2\right)f_{11} 
 -\lambda\alpha_0^2\psi_0^2f_{12} +\mu_1 f_{11}\label{eq:f11eqn}
\ea 
Let $\nu_*=2E_{0*}-E_{1*}$.
 The second component of (\ref{eq:H0evp1}) is:
\ba
(H-\nu_*)f_{12} &=& 
 -\lambda\overline{\alpha_0}^2\psi_0^2\psi_{1*} + 
 |\alpha_0|^2E_0^{(1)}f_{12}\nn\\
&& - \lambda\psi_0^2\left( \overline{\alpha_0}^2f_{11}+2|\alpha_0|^2f_{12}\right)
 - \mu_1f_{12}
\label{eq:f12eqn}
\ea
We wish to make the dependence of $\vec f_1$ on $\alpha_0$ and 
 $\overline{\alpha_0}$ explicit. Define
\be \mu_1=|\alpha_0|^2\tmu_1,\ f_{11}=|\alpha_0|^2\tf_{11}, \ 
 f_{12}=\overline{\alpha_0}^2\tf_{12}\label{eq:tdefs}\ee
Equations (\ref{eq:f11eqn}) and (\ref{eq:f12eqn}) reduce to the following
system for $\tf_{11}$ and $\tf_{12}$:
\ba
(H-E_{1*})\tf_{11} &=& \left( E_0^{(1)}-2\lambda\psi_0^2\right)\psi_{1*}
+ \tmu_1\psi_{1*}\nn\\ 
&+& |\alpha_0|^2\left(E_0^{(1)} - 2\lambda\psi_0^2\right)\tf_{11} 
 -\lambda|\alpha_0|^2\psi_0^2\tf_{12} + |\alpha_0|^2 \tmu_1 \tf_{11}
\label{eq:tf11eqn}\\
(H-\nu_*)\tf_{12} &=& 
 -\lambda\psi_0^2\psi_{1*} + 
 |\alpha_0|^2E_0^{(1)}\tf_{12}\nn\\
&& - \lambda|\alpha_0|^2 
 \psi_0^2\left( 2\tf_{11}+2\tf_{12}\right)
 - |\alpha_0|^2\tmu_1\tf_{12}
\label{eq:tf12eqn}
\ea
We seek a solution to the system (\ref{eq:tf11eqn}), (\ref{eq:tf12eqn}):
\be
|\alpha_0|^2\mapsto 
 \left(\tf_{11}(|\alpha_0|^2), \tf_{12}(|\alpha_0|^2), \tmu(|\alpha_0|^2)\right)
\ \in L^2\times L^2\times \R
\label{eq:sol}\ee
defined in a neighborhood of $\alpha_0=0$. 

For $\alpha_0=0$ the  system (\ref{eq:tf11eqn}), (\ref{eq:tf12eqn})
reduces to:
\ba
(H-E_{1*})\tf_{11}^0 &=& 
 \left( E_0^{(1)}(0)-2\lambda\psi_{0*}^2\right)\psi_{1*}
+ \tmu_1\psi_{1*}
\label{eq:tf11eqn0}\\
(H-\nu_*)\tf_{12}^0 &=&
 -\lambda\psi_{0*}^2\psi_{1*} 
\label{eq:tf12eqn0}
\ea
Note that  $\nu_*<0$ is not in the spectrum of
$H$. Therefore, (\ref{eq:tf12eqn0}) is solvable for $\tf_{12}^0$
and we have:
\be
\tf_{12}^0 = -\lambda(H-\nu_*)^{-1}\psi_{0*}^2\psi_{1*}.
\label{eq:tf120}
\ee
Since $(H-E_{1*})\psi_{1*}=0$, (\ref{eq:tf11eqn0}) is solvable if and only
if its right hand side is orthogonal to $\psi_{1*}$. This determines
$\mu_1(0)$:
\be
\tmu_1(0) = -E_0^{(1)}(0) + 2\lambda\la\psi_{0*}^2,\psi_{1*}^2\ra,
\label{eq:tmu1of0}
\ee
and now (\ref{eq:tf11eqn0}) can be solved for $\tf_{11}^0$.

To solve in a neighborhood of $\alpha_0=0$ we proceed as follows.
Rewrite (\ref{eq:tf12eqn}) as 
\be
(H + |\alpha_0|^2W_{12} -\nu_*)\tf_{12} =
 -\lambda\psi_0^2\psi_{1*} - \lambda|\alpha_0|^2 \psi_0^2\tf_{11}, 
\label{eq:tf12eqn1}
\ee
where $W_{12}$ is a multiplication operator defined by
\be
 W_{12}(|\alpha_0|^2,\tmu_1) = 
 -E_0^{(1)}+2\lambda\psi_0^2+\tmu_1.\label{eq:W12def}
\ee
For $\tmu_1$ in a fixed compact set and $\alpha_0$ sufficiently small,
 the operator $H + |\alpha_0|^2W_{12} -\nu_*$
has a bounded inverse, $B(|\alpha_0|^2)$. Thus,
\ba
\tf_{12} &\equiv& \tf_{12}[\tf_{11},|\alpha_0|^2]\nn\\
         &=& 
        -\lambda B(|\alpha_0|^2)\psi_0^2\psi_{1*} 
           -\lambda|\alpha_0|^2 B(|\alpha_0|^2) \psi_0^2\tf_{11}
\label{eq:tf12}
\ea
Substitution of (\ref{eq:tf12}) into (\ref{eq:tf11eqn}) yields the following
closed equation for $\tf_{11}$:
\be
(H-E_{1*})\tf_{11}
 = \left( E_0^{(1)}-2\lambda\psi_0^2\right) \psi_{1*}
+ \tmu_1\psi_{1*} +  |\alpha_0|^2W_{11}\tf_{11},\label{eq:tf11eqn1}
\ee
where the operator $W_{11}$ is defined by 
\be
W_{11}(|\alpha_0|^2,\tmu_1) =
 (E_0^{(1)}-2\lambda\psi_0^2)+\tmu_1+\psi_0^2\tf_{12}[\cdot,|\alpha_0|^2].
\label{eq:W11def}
\ee
Setting the inner product of the right hand side of (\ref{eq:tf11eqn1})
equal to zero, gives the solvability condition for (\ref{eq:tf11eqn1}):
\be
\tmu_1 = 2\lambda\la\psi_0^2,\psi_{1*}^2\ra - E_0^{(1)} - 
 |\alpha_0|^2\la \psi_{1*},W_{11}\tf_{11}\ra
\label{eq:solvability}
\ee
The system (\ref{eq:tf11eqn1}), (\ref{eq:solvability}) is of the form:
\be F(\tf_{11},\tmu_1,s) = 0, \label{eq:Feq0}\ee 
with the solution $\tf_{11}=\tf_{11}^0, \tmu_1=\tmu_1(0), s=0$ defined 
by (\ref{eq:tmu1of0}). Furthermore, the Jacobian of $F(\tf_{11},\tmu_1,s)$
with respect to $(\tf_{11},\tmu_1)$ evaluated at $(\tf_{11}^0,\tmu_1(0),0)$
is given by:
\be
\left(\begin{array}{cc} H-E_{1*} & -\psi_{1*}\\
                        0        & I \end{array}\right)
\label{eq:jacobian}
\ee
which maps $H^2\times\R$ one to one and onto 
 $L^2\times\{\la g,\psi_{1*}\ra:g\in L^2\}$.
 Therefore, by the implicit function theorem \cite{Nirenberg},
  we have a real analytic curve
of solutions   
$ s\mapsto (\tf_{11}(s),\tmu_1(s),s)$, defined in a neighborhood of $s=0$
 and coinciding with 
 $(\tf_{11}^0,\tmu_1(0),0)$ for $s=0$. The family of solutions we seek
is obtained by restriction to $s=|\alpha_0|^2\ge0$.
This completes the proof of Proposition \ref{prop:Hspectrum}. 

%% file: mbs-appC.tex
\section{Appendix C: A commutator term}\label{sec:commutator} 

In this section we record a calculation of a "commutator term"
appearing in the modulation equations of section \ref{sec:dec-mod}.

\begin{prop}\label{prop:commutators}
\ba
i\la\D_t(\sigma_3 \xi_{01}),\Phi_2\ra &=&
i\D_t(|\alpha_0|^2)\
 \left\la F_0'G_0\left(\begin{array}{c}1 \\ 1\end{array}\right)\ ,\ \Phi_2
 \right\ra\nn\\
&+& \D_t\gamma_0\ |\alpha_0|^2
 \left\la  \chi G_0\left(\begin{array}{c}1\\ 1\end{array}
 \right)\ ,\ \Phi_2\right\ra\label{eq:com1}\\
i\la\D_t(\sigma_3 \xi_{02}),\Phi_2\ra &=&
i\D_t(|\alpha_0|^2)\
 \left\la F_0''\sigma_3 G_0\left(\begin{array}{c}1 \\ 1\end{array}\right)
 \ ,\ \Phi_2 \right\ra\nn\\
&+& (\D_t\gamma_0)\
 |\alpha_0|^2\left\la \chi\sigma_3G_0
\left(\begin{array}{c}1\\ 1\end{array}\right) \ ,\ \Phi_2
\right\ra\label{eq:com2}
\ea
\end{prop}

\noindent{\bf proof:}
By direct computation from (\ref{eq:G0_def})
\be \D_t G_0(t) =  i(\D_t\gamma_0)\ \sigma_3 G_0(t).\label{eq:DtG0}\ee
Note also that by (\ref{eq:useful})
\ba
\sigma_3G_0\left(\begin{array}{c}1 \\ 1\end{array}\right) F_0 &=& \xi_{01}
 =  2\zeta_{01} +
 |\alpha_0|^2 \sigma_3G_0\left(\begin{array}{c}1 \\ 1\end{array}\right)
 \chi(x;|\alpha_0|^2)\nn\\
G_0\left(\begin{array}{c}1 \\ 1\end{array}\right) F_0' &=& \xi_{02} =
{1\over2} \zeta_{02} + |\alpha_0|^2
 G_0\left(\begin{array}{c}1 \\ 1\end{array}\right)
 \chi(x;|\alpha_0|^2)
\nn\ea
Using these relations we have
for  $j=1$
that
\ba \D_t( \sigma_3\xi_{01}(t) ) &=& \D_t(|\alpha_0|^2)
 G_0 \left(\begin{array}{c}1 \\ 1\end{array}\right) F_0'(|\alpha_0|^2)\nn\\
&+& i(\D_t\gamma_0) \sigma_3 G_0\left(\begin{array}{c}1 \\ 1\end{array}\right)
 F_0(|\alpha_0|^2)\nn\\
&=& \D_t(|\alpha_0|^2) G_0 \left(\begin{array}{c}1 \\ 1\end{array}\right)
 F_0'(|\alpha_0|^2) + 2i(\D_t\gamma_0) \zeta_{01}(t) +
 (\D_t\gamma_0)\ |\alpha_0|^2\chi(x;|\alpha_0|^2)G_0
    \left(\begin{array}{c}1 \\ 1\end{array}\right).\nn
\ea
Substitution into the inner product
 $\la \D_t( \sigma_3\xi_{01}(t) ) , \Phi_2\ra$ and using the constraint
$\la \zeta_{01}(t),\Phi_2\ra=0$ yields the result for $j=1$.

For $j=2$
\ba \D_t( \sigma_3\xi_{02}(t) ) &=& \D_t(|\alpha_0|^2)
\sigma_3
G_0 \left(\begin{array}{c}1 \\ 1\end{array}\right) F_0''(|\alpha_0|^2)\nn\\
&+& i(\D_t\gamma_0) G_0\left(\begin{array}{c}1 \\ 1\end{array}\right)
 F_0'(|\alpha_0|^2)\nn\\
&=& \D_t(|\alpha_0|^2) \sigma_3 G_0
 \left(\begin{array}{c}1 \\ 1\end{array}\right)
 F_0'(|\alpha_0|^2) + {i\over2}(\D_t\gamma_0) \zeta_{02}(t) +
 \D_t\gamma_0|\alpha_0|^2\chi(x;|\alpha_0|^2)\sigma_3G_0
    \left(\begin{array}{c}1 \\ 1\end{array}\right).\nn
\ea
Substitution into the inner product
 $\la \D_t( \sigma_3\xi_{02}(t) ) , \Phi_2\ra$ and using the constraint
$\la \zeta_{02}(t),\Phi_2\ra=0$ yields the result for $j=2$.
This completes the proof of Proposition \ref{prop:commutators}.

%% file: mbs-bib.tex

%